\documentclass[12pt, oneside]{article}   	

\usepackage{jheppub}

\usepackage{amsmath}
\usepackage{amssymb}
\usepackage{amsfonts}
\usepackage{amsthm}
\usepackage{amsbsy}
\usepackage{verbatim}
\usepackage{array}
\usepackage{dsfont}
\usepackage[linesnumbered,lined,boxed,commentsnumbered]{algorithm2e}

\usepackage[OT2,OT1]{fontenc}
\newcommand\cyr
{
\renewcommand\rmdefault{wncyr}
\renewcommand\sfdefault{wncyss}
\renewcommand\encodingdefault{OT2}
\normalfont
\selectfont
}
\DeclareTextFontCommand{\textcyr}{\cyr}

\usepackage{mathtools}
\usepackage[usenames,dvipsnames]{xcolor}

\usepackage{subcaption}

\usepackage{dsdshorthand}
\usepackage{tikz}

\usepackage[vcentermath]{youngtab}

\def\ben{\begin{equation}}
\def\een{\end{equation}}
\def\bea{\begin{eqnarray}}
\def\eea{\end{eqnarray}}

\newcommand{\el}{\in}

\newcommand{\la}{\langle}
\newcommand{\ra}{\rangle}
\let\a=\alpha \let\b=\beta \let\c=\chi \let\d=\delta \let\e=\epsilon \let\ve=\varepsilon \let\g=\gamma \let\h=\eta \let\k=\kappa \let\l=\lambda  
 \let\p=\phi \let\r=\rho  \let\t=\tau \let\th=\theta  \let\vp=\varphi \let\x=\xi  \let\z=\zeta
 \let\D=\Delta   \let\O=\Omega \let\P=\Phi     
\def\nn{\nonumber}
\def\inf{\infty}

\def\mbK{\mathbb{K}}

\makeatletter
\def\@fpheader{\ }
\makeatother

\title{Jackiw-Teitelboim quantum gravity with defects and the Aharonov-Bohm effect
}

\author{Eric Mefford and Kenta Suzuki}
\affiliation{CPHT, CNRS, Ecole Polytechnique, Institut Polytechnique de Paris, Route de Saclay,
91128 PALAISEAU, France}

\emailAdd{eric.mefford@polytechnique.edu}
\emailAdd{kenta.suzuki@polytechnique.edu}

\preprint{CPHT-RR085.112020}

\abstract{
We study the theory of Jackiw-Teitelboim gravity with generalized dilaton potential on Euclidean two-dimensional negatively curved backgrounds. The effect of the generalized dilaton potential is to induce a conical defect on the two-dimensional manifold. We show that this theory can be written as the ordinary quantum mechanics of a charged particle on a hyperbolic disk in the presence of a constant background magnetic field plus a pure gauge Aharonov-Bohm field.
This picture allows us to exactly calculate the wavefunctions and propagators of the corresponding gravitational dynamics.
With this method we are able to reproduce the gravitational density of states as well as compute the R\'eyni and entanglement entropies for the Hartle-Hawking state.
While we reproduce the classical entropy at high temperature, we also find an extra topological contribution that becomes dominant at low temperatures. We then show how the presence of defects modify correlation functions, including the out-of-time-ordered correlation, and decrease the Lyapunov exponent.
This is achieved two ways: by directly quantizing the boundary Schwarzian theory and by dimensionally reducing $SL(2,\mathbb{Z})$ black holes.}

\date{}

\begin{document}
\maketitle

\section{Introduction}
\label{sec:int}

Jackiw-Teitelboim (JT) gravity is a two-dimensional theory of quantum gravity coupled to a real scalar field \cite{Jackiw:1984je, Teitelboim:1983ux}.
The theory is remarkable for many reasons, especially when it is defined on a Euclidean negatively curved backgrounds. Among these, two stand out.
The first is that since there is no propagating degrees of freedom,
all gravitational dynamics is reduced to a theory of boundary coordinate reparametrizations which is described by a one-dimensional Schwarzian action
\cite{Almheiri:2014cka, Maldacena:2016upp,Jensen:2016pah, Engelsoy:2016xyb}.
This action also describes the low energy dynamics of the Sachdev-Ye-Kitaev (SYK) model \cite{Sachdev:1992fk,Polchinski:2016xgd,Maldacena:2016hyu,Jevicki:2016bwu,Jevicki:2016ito}. Both these theories have relevance to the study of quantum chaos and random matrices.
Furthermore, there is strong evidence that JT gravity is a special kind of holographic theory whose dual is not a specific quantum mechanical system living on the boundary of some AdS spacetime but one whose dual is a random ensemble of quantum mechanical systems \cite{Saad:2019lba, Stanford:2019vob, Maxfield:2020ale, Witten:2020wvy}. But perhaps the most surprising feature of the Schwarzian sector of JT gravity and of the SYK model is that it is one-loop exact. In other words, it is an exactly solvable theory of quantum gravity \cite{Bagrets:2016cdf, Stanford:2017thb, Mertens:2017mtv}.
Furthermore, this theory is universal in the sense that it arises in the near-horizon and near-extremal limit of many higher dimensional black holes discussed for instance in \cite{Bardeen:1999px, NavarroSalas:1999up,Nayak:2018qej,Sachdev:2019bjn}. Closely related to this, a universal Schwarzian sector also appears in higher dimensional quantum systems including two-dimensional conformal field theories \cite{Ghosh:2019rcj}. The fact that one can exactly solve a system of quantum gravity has recently allowed for a major breakthrough in the black hole information problem \cite{Penington:2019npb, Almheiri:2019psf}.

Given the absence of propagating degrees of freedom, two-dimensional gravity is purely topological and has constant (local) curvature \cite{Jackiw:1984je}.
When the geometry is Euclidean with an asymptotically AdS boundary, the theory is equivalent to the study of Riemann surfaces with boundaries. A major goal of recent work on the subject is to understand the theory of JT gravity on arbitrary genus Riemann surfaces both in the Schwarzian limit and beyond.
These surfaces can be constructed via cut-and-sew procedures detailed in \cite{Saad:2019lba, Stanford:2019vob, Maxfield:2020ale, Witten:2020wvy} whereby one assembles the Riemann surface via smaller building blocks: an asymptotically AdS trumpet with a geodesic boundary and a three-holed sphere with geodesic boundaries. If one allows for conical deficits, then one must also include the same building blocks where one or more of the geodesic boundaries is replaced by a conical singularity. A first step to understand the theory on general Riemann surfaces is then to understand the theory on the hyperbolic trumpet with one geodesic boundary and on the related surface where this geodesic boundary is replaced by a conical singularity, which we will occasionally call the punctured disk. In \cite{Maxfield:2020ale, Witten:2020wvy}, it was shown that Riemann surfaces with conical singularities naturally arise when we generalize the dilaton potential of the JT gravity to have exponential form.

The theory of JT gravity on the disk, which arises as the leading term in the genus expansion, has been understood via many different routes.
A very formal approach is to study the intersection theory on the moduli space of Riemann surfaces \cite{Witten:1990hr} (also see references in \cite{Witten:2020wvy}).
Another formal method exploits the topological nature of two-dimensional gravity via a reframing of the first-order formalism as a topological gauge theory \cite{Jackiw:1992bw, Mertens:2018fds,Blommaert:2018oro}, often called the BF theory, some details of which we review in appendix~\ref{app:monodromiesinBFtheory}.
Another yet related route is to reframe the theory in terms of a Liouville CFT \cite{Mertens:2017mtv}.
A different route is through the dimensional reduction of the near-extremal BTZ black hole and its CFT$_2$ dual \cite{Ghosh:2019rcj, Maxfield:2020ale}.
In the Schwarzian limit, the boundary time reparametrizations can be directly quantized and the theory can be solved via a path integral whose measure is derived from the Weil-Petersson symplectic form \cite{Saad:2019lba, Cotler:2019nbi}. 
Finally, the theory can be rephrased as the ordinary quantum mechanics of a charged particle propagating on a hyperbolic disk in the presence of a constant background magnetic field \cite{Yang:2018gdb, Kitaev:2018wpr}. 

Each of these methods brings different perspectives to the problem and all but the last,
the quantum mechanics of a charged particle, have been directly applied to the problem of deformed JT gravity.
This is unfortunate because we feel that  the charged particle picture with dimensional reduction of asymptotically AdS$_3$ black holes give perhaps the most direct and intuitive understanding of quantum gravity in two-dimensional AdS space.
Therefore, in the present paper, we study the charged particle picture for the punctured disk and trumpet.
In particular, the wavefunctions and propagators for the theory are exactly solved.
This allows us, for instance, to construct the bulk-boundary propagator of a specific boundary reparametrization, to calculate correlation functions, to derive the density of states at a particular energy, and the calculate the entanglement entropy beyond the semi-classical area term. 

The rest of the paper is organized as follows.
First in section~\ref{sec:jt}, we summarize how the generalized dilaton potential of the JT gravity leads to a conical singularity \cite{Maxfield:2020ale, Witten:2020wvy}.
In section~\ref{sec:charged}, we start our study of the charged particle picture for the punctured disk and trumpet.
In particular, we show that these two geometries require the inclusion of a purely topological component of the magnetic field, an Aharonov-Bohm (AB) field,
whose flux is directly related to the conical deficit angle and geodesic length, respectively.
The trumpet and punctured disk results are related by a simple analytic continuation and our derivation focuses predominantly on the punctured disk. After solving this problem, we calculate the R\'enyi and von Neumann entropies. We find that, in contrast to the theory on the hyperbolic disk, the von Neumann entropy of the punctured disk, and hence the trumpet as well, contains a purely topological piece. At low temperatures, this piece will dominate the entropy.

In section~\ref{sec:sl(2,z) bh}, we then reproduce many of our results by considering the dimensional reduction of $SL(2,\mathbb{Z})$ black holes.
We also discuss some consequences of the conical deficit on correlation functions.
In particular, we calculate the out-of-time-ordered correlation function (OTOC) of the boundary operators in the $SL(2,\mathbb{Z})$ black hole backgrounds
and show that the Lyupanov exponent is decreased $\lambda_L = (2\pi - \alpha)/\beta$ for a conical deficit $\alpha$.
In section~\ref{sec:schwarzian}, we also reproduce this result by directly computing the OTOC of external conformal operators in the Schwarzian theory. 
Finally our conclusion and discussion are given in section~\ref{sec:conclusion}.
A series of appendices give further insights and computational details to some of the problems discussed in the main text.

\section{Deformed JT Gravity}
\label{sec:jt}
We study the generalized JT gravity in Euclidean signature described by the action 
	\begin{align}
		I \, = \, \underbrace{ - \, \frac{\p_0}{2} \left( \int_M dx^2 \sqrt{g} R \, + \, 2 \int_{\partial M} \sqrt{h} K \right) }_{\text{Einstein-Hilbert Action}}
		\, \underbrace{ - \, \frac{1}{2} \left( \int_M dx^2 \sqrt{g} \Big(\p R +W(\p) \Big) \, + \, 2 \int_{\partial M} \sqrt{h} \p_b K \right) }_{\text{Modified JT Action}} \, ,
	\end{align}
where $h$ and $K$ are the induced metric and extrinsic curvature on the boundary $\partial M$.
For the dilaton field $\p$, we have separated out a background contribution $\p_0$ and we assume $\p_0\gg \p$.
For the second term, we included a general potential for the dilation field and $\p_b$ represents the boundary value of $\p$. 
As in \cite{Maxfield:2020ale, Witten:2020wvy}, we consider the following form of the potential
	\begin{align}
		W(\phi) \, = \, 2\phi \, + \, 2\epsilon \, e^{- \alpha \phi} \, .
	\end{align}
Expanding in $\e$ for the partition function, we have
	\begin{align}
		\exp\big( -I_{(0)} \big) \left( 1 \, + \, \e \int d^2x_1 \, \sqrt{g(x_1)} \, e^{-\a \p(x_1)} \, + \, \mathcal{O}(\e^2) \right) \, .
	\label{expanded-partition-func}
	\end{align}
The $\mathcal{O}(\epsilon^0)$ corresponds to the original JT gravity, and we consider the zero-th and first order contributions.
We first consider the $\mathcal{O}(\epsilon^0)$ contribution.
Since the variation of $\phi$ leads to $R=-2$, the Einstein-Hilbert Action is reduced to the Euler characteristic $\chi$:
	\begin{align}
		I_{(0)} \, = \, - \, 2\pi \phi_0 \chi(M) \, - \, \phi_b \int_{\partial M} K \, .
	\label{on-shell action}
	\end{align}
The two simplest geometries which solve the equations of motion are the disk $D$ and the trumpet $T(b)$ with inner boundary geodesic length $b$ \cite{Saad:2019lba}.
For these geometries, the Euler characteristic is given by $\chi(D)=1$ and $\chi(T)=0$, respectively.
They can be labeled by metrics 
	\begin{align}
		ds_D^2 \, &= \, d\rho^2 \, + \, \sinh^2 \rho \, d\varphi^2 \, , \qquad (0 \, \le \, \varphi \, < \, 2\pi) \, , \\
		ds_T^2 \, &= \, d\rho^2 \, + \, \cosh^2 \rho \, d\varphi^2 \, , \qquad (0 \, \le \, \varphi \, < \, b) \, .
	\end{align}
The solution of the dilaton field for each geometry is given by
	\begin{align}
		\phi_D \, = \, \g_D \cosh \rho \, , \qquad \ \phi_T \, = \, \g_T \sinh \rho \, ,
	\label{phi_D}
	\end{align}
with some constants $\g$.

For the order $\mathcal{O}(\epsilon)$ contribution, we pull the integral over $x_1$ outside of the path integral,
and write the corresponding action as  \cite{Maxfield:2020ale,Witten:2020wvy}
	\begin{align}
		I_{(1)} \, = \, - \, \frac{\phi_0}{2} \left( \int_M R + 2 \int_{\partial M} K \right)
		\, - \, \left( \frac{1}{2} \int_M \phi \big( R + 2 \big) - \alpha \phi(x_1) + \int_{\partial M} \phi_b K \right) \, .
	\end{align}
Now the variation of $\phi$ leads to $R(x)+2=2\alpha \delta^2(x-x_1)$.
Besides the point $x=x_1$, this is still described by (Euclidean) AdS$_2$, but it has a conical singularity at $x=x_1$. We are free to put the conical singularity at the origin, in which case the metric is still given by
	\begin{align}
		ds_{D_\a}^2 \, = \, d\rho^2 \, + \, \sinh^2 \rho \, d\varphi^2 \, , \qquad (0 \, \le \, \varphi \, < \, 2\pi - \alpha) \, ,
		\label{conegauge1}
	\end{align}
but the periodicity of $\varphi$ is changed by the conical singularity.
The dilaton solution is same as $\phi_D$ in (\ref{phi_D}).

For these geometries, by computing the Schwarzian path integrals, the gravitation partition functions are computed in the Schwarzianlimit as \cite{Stanford:2017thb,Saad:2019lba,Mertens:2019tcm}
	\begin{align}
		Z_D(\beta) \, = \, \frac{\exp\left( \frac{2\pi^2}{\beta} \right)}{\sqrt{2\pi} \, \beta^{3/2}} \, , \ \ \quad
		Z_T(\beta) \, = \, \frac{\exp\left( - \frac{b^2}{2\beta} \right)}{\sqrt{2\pi} \, \beta^{1/2}} \, , \ \ \quad
		Z_{D_\a}(\beta) \, = \, \frac{\exp\left( \frac{(2\pi - \alpha)^2}{2\beta} \right)}{\sqrt{2\pi} \, \beta^{1/2}} \, ,
	\label{Z}
	\end{align}
where $\beta$ is the inverse temperature defined in (\ref{betadef}).
Defining the density of states by
	\begin{align}
		Z(\b) \, = \, \int_0^{\infty} ds \, e^{- \beta \frac{s^2}{2}} \rho(s) \, , \ \ \qquad E \, = \, \frac{s^2}{2} \, ,
	\end{align}
they are given by
	\begin{align}
		\rho_D(s) \, = \, \frac{s}{2\pi^2} \, \sinh(2\pi s) \, , \ \quad
		\rho_T(s) \, = \, \frac{1}{\pi} \cos(b s) \, , \ \quad
		\rho_{D_\a}(s) \, = \, \frac{1}{\pi} \cosh\big( 2\pi \z s\big) \, ,
	\label{rho}
	\end{align}	
where we defined $2\pi \z  \equiv 2\pi -\alpha$.
The disk partition function $Z_D(\beta)$ was also reproduced by the charged particle picture in \cite{Yang:2018gdb,Kitaev:2018wpr}
\footnote{This is also related to the Liouville quantum mechanics description of the Schwarzian theory discussed in \cite{Bagrets:2016cdf,Mertens:2017mtv}.}.

In terms of the Schwarzian theory, the different powers of $\beta$ in $Z_D$ and $Z_{D_\alpha}$ are explained by the different stabilizer group $H$ \cite{Stanford:2017thb}.
For $D_\a$, we have $H=U(1)$, but the symmetry is enhanced for $D$ as $H= SL(2, \mathbb{R})$.
Therefore, $\lim_{\alpha \to 0} Z_{D_\alpha}$ does not agree with $Z_D$ \cite{Mertens:2019tcm}.
We would like to understand this discontinuity from the charged particle picture as well as reproducing $Z_T$.

\section{Charged Particle Picture}
\label{sec:charged}
For all three types of the geometries, the on-shell action is reduced to the form of (\ref{on-shell action}), where $\chi(D)=\chi(D_\alpha)=1$ and $\chi(T)=0$.
We introduce the boundary cutoff at $\rho = \rho_0$ and on that we fix the metric along the boundary and the value of the dilaton field \cite{Maldacena:2016upp}:
	\begin{align}
		\varphi\big|_{\rho_0} \, = \, \phi_b \, u \, , \ \qquad {\rm with} \ \qquad \phi_b \, = \, \g_D \cosh \r_0 \ \quad {\rm or} \ \quad \phi_b \, = \, \g_T \sinh \r_0 \, ,
	\label{u}
	\end{align}
and we think of $u$ as the time of the boundary theory, which has the periodicity $u \cong u + \beta$.
Introducing a counterterm for the extrinsic curvature as in \cite{Yang:2018gdb}, besides the topological term proportional to $\c$, we have
	\begin{align}
		I \, = \, - \, \phi_b \int_{\partial M} d\vp \sqrt{g} \, \big( K - 1 \big) \, .
	\label{I_eff}
	\end{align}
If we regard the bulk angular coordinate as a function of the boundary time $\vp = \vp(u)$, evaluating the extrinsic curvature in the $\r_0 \to \infty$ limit,
one gets the Schwarzian action \cite{Maldacena:2016upp}
	\begin{align}
		S[f] \, = \, - \, C \int_0^\b du \, \left\{\tan\left( \frac{\vp(u)}{2} \right) , \, u \right\} \, ,
	\label{S[f]}
	\end{align}
where $C$ is the regularized coupling defined in (\ref{betadef}).
For the $M=\{D, D(\a)\}$ cases, if we redefine $\vp(u) = \{ 2\pi f(u)/\b \, , \, 2\pi \z f(u)/\b \}$, the periodicity of $f(u)$ becomes $\b$.

On the other hand, one can also rewrite the effective action (\ref{I_eff}) in the following form.
The counterterm is the circumference at $\r= \r_0$, which have the following expressions:
	\begin{gather}
		L_M \, \equiv \, \int_{\partial M} d\varphi \sqrt{g} \, , \\[2pt]
		L_D \, = \, 2\pi \sinh \rho_0 \, , \qquad  L_T \, = \, b \cosh \rho_0 \, , \qquad L_{D_\a} \, = \, (2\pi - \alpha) \sinh \rho_0 \, . \label{L}
	\end{gather}
The inverse temperature is defined by
\footnote{In this paper, we follow the convention of the temperature used in \cite{Yang:2018gdb,Kitaev:2018wpr} instead of \cite{Saad:2019lba,Witten:2020wvy}.
They differ by $\beta_{\rm Yang}=\b_{\rm KS}=2\beta_{\rm Witten}=2\beta_{\rm SSS}$.}
	\begin{equation}
		\beta \, = \, C \lim_{\rho_0 \to \infty} \, \frac{L_M}{\phi_b} \, ,
	\label{betadef}
	\end{equation}
where $C$ is a proportionality constant. Therefore for each geometry, we have
	\begin{equation}
		\beta \, = \, \left\{ \frac{2\pi C}{\g_D} \, , \ \frac{b C}{\g_T} \, , \ \frac{(2\pi - \a) C}{\g_{D_\a}} \right\} \, .
		\label{beta}
	\end{equation}
For the extrinsic curvature, we used the Gauss-Bonnet theorem as in \cite{Yang:2018gdb}:
	\begin{align}
		\int_{\partial D} du \sqrt{g} \, K \, &= \, 2\pi \chi(D) \, + \, A_D \, , \\
		\int_{\partial T} du \sqrt{g} \, K \, &= \, 2\pi \chi(T) \, + \, A_T \, , \\
		\int_{\partial D_\alpha} du \sqrt{g} \, K \, &= \, (2\pi - \alpha) \chi(D_\alpha) \, + \, A_{D_\alpha} \, , 
	\end{align}
where the area of the manifolds are given by
	\begin{gather}
		A_M \, \equiv \, \int_M d^2x \sqrt{g} \, , \\[2pt]
		A_D \, = \, 2\pi (\cosh \rho_0 - 1) \, , \ \quad A_T \, = \, b \sinh \rho_0 \, , \ \quad A_{D_\a} \, = \, (2\pi - \alpha) (\cosh \rho_0 - 1) \, .
	\end{gather}
Combining all of them, we find the on-shell actions
	\begin{align}
		I_D \, &= \, - \, 2\pi q \, \chi(D) \, - \, q A_D \, + \, q L_D \, , \\[2pt]
		I_T \, &= \, - \, 2\pi q \, \chi(T) \, - \, q A_T \, + \, q L_T \, , \\[2pt]
		I_{D_\alpha} \, &= \, - \, (2\pi - \a) q \, \chi(D_\alpha) \, - \, q A_{D_\a} \, + \, q L_{D_\a} \, ,
	\end{align}
where 
	\begin{align}
		q \, \equiv \, \phi_b \, .
	\end{align}
Even though we used the same notation $q$ for all three cases above, we must be careful to note that the actual value of $q$ is different for all three cases in their respective canonical ensembles.
We have $q_D = \g_D \cosh \r_0$ and $q_{D_\a} = \g_{D_\a} \cosh \r_0$. In the canonical ensemble, from (\ref{beta}) we have the relation $\g_{D_\a} = \z \g_D$,
implying that the charges are related by $q_{D_\a} = \z q_D$. In the following, we denote $q \equiv q_D$ and $q_\a \equiv q_{D_\a}$. Neglecting the Euler characteristic term for now, the fact that we may write all three actions as a constrained length piece plus an area piece implies that the boundary dynamics of the JT gravity can be understood as a charged particle in the corresponding two-dimensional manifold, though their explicit Lagrangian is different.
Charged particles on the Poincare disk ($D$) was previously studied in \cite{Comtet:1984mm, Comtet:1986ki, Pioline:2005pf}.
The area term can be regarded as a constant magnetic field contribution. 
For example, for $A_D$ we can define the corresponding gauge field 
	\begin{align}
		\mathbf{A}_D \, = \, 
		\begin{pmatrix}
		a_\r \\ a_\vp 
		\end{pmatrix}
		\, = \, (\cosh \r - 1)
		\begin{pmatrix}
		0 \\ 1
		\end{pmatrix} \, .
	\end{align}
Because $dB = (\partial_\r a_\vp - \partial_\vp a_\r ) d\r \wedge d\vp = \sinh \r \, d\r \wedge d\vp$,
$dB$ has the {\it constant} $\times$ {\it volume} form and thus interpreted indeed as a constant magnetic field.
For $A_{D_\a}$, the gauge field is more subtle.  One reason that the area term can be thought of as arising 
from a constant magnetic field is that it can be related to the Ricci scalar which can be written in terms of the exterior derivative of
the spin connection of an orthonormal frame \cite{Kitaev:2018wpr},
	\begin{align}
		A_M \, = \, - \, \int_M d^2x \sqrt{g} \, \frac{R}{2} \, = \, - \, \int_M d\omega \, ,
	\end{align}
where
	\begin{align}
		g_{\mu\nu} \, = \, e^{a}_\mu \, e^b_\nu \, \delta_{ab} \, , \qquad de^a = - \, \omega^{a}_{\;\;b}\wedge e^b \, , \qquad \omega \, \equiv \, \omega^{a}_{\;\; a} \, .
	\end{align}
This provides a mapping between the geometric variables and the gauge field in the charged particle picture.
The puncture shows up as a delta function from the Ricci scalar by the equation of motion and hence should appear as a purely topological component in the 
corresponding field strength \cite{Mertens:2019tcm}. 

In other words, we must include an Aharonov-Bohm (AB) field,
$\mathbf{A}_{D_\a} = \mathbf{A}_D + \mathbf{A}^{(\a)}$ where
	\begin{align}
		\mathbf{A}^{(\a)} \, = \, - \, \frac{\a \z}{2\pi} \, \big(\cosh\r_0 -1 -\sinh\r_0 \big) d\vp \, .
	\label{A^alpha}
	\end{align}
The last term in the AB field comes from matching the $L_{D_\alpha}$ contribution to the on-shell action, as we now explain. In writing the on-shell action in this manner, we have chosen a different gauge than (\ref{conegauge1}) in which the metric on the cone is now written
	\begin{align}
	d\tilde{s}^2 \, = \, d\rho^2 \, + \, \z^2 \, \sinh^2\rho \, d\vp^2 \, , \qquad 0 \leq \vp < 2\pi.
	\label{conegauge2prime}
	\end{align}
This metric can be obtained by rescaling $\vp \to \z \vp$ from the coordinates in (\ref{conegauge1}).
Importantly, this rescaling brings the periodicity of $\vp$ from $2\pi - \a$ to $2\pi$.
Rewriting the on-shell action of the punctured disk given in the gauge (\ref{conegauge1}) in terms of the new coordinates (\ref{conegauge2prime}), we now have
	\begin{align}
		I_{D_\alpha} + (2\pi - \a)q_\a\, = \, -q_\a A_{D_\alpha} + q_\a L_{D_{\alpha}}
		\, = \, -q \tilde{A}_{D_{\alpha}} + q \tilde{L}_{D_\alpha}  - q \int \mathbf{A}^{(\alpha)} \, .
	\label{I_D_alpha}
	\end{align}
Here $L_{D_\alpha}$ and $A_{D_\alpha}$ are the area and lengths of the surface $\rho=\rho_0$ in the gauge (\ref{conegauge1}) and 
$\tilde{A}_{D_\alpha}=2\pi \z (\cosh \r_0 - 1)$ and $\tilde{L}_{D_\alpha}=2\pi \z \sinh \r_0$ are the area and lengths of the surface $\rho=\rho_0$ in the gauge (\ref{conegauge2prime}).

Solving the charged particle on the trumpet can be obtained from the problem on the cone by the analytic continuation
	\begin{align}
		\rho_{D_\alpha} \, = \, \rho_T \, + \, \frac{i \pi}{2} \, , \qquad \varphi_{D_\alpha} \, = \, i \varphi_T \, .
	\end{align}
and by relating the trumpet parameter $b$ to the conical deficit parameter $\alpha$ via $b =-i(2\pi-\alpha)$ \cite{Witten:2020wvy}. In particular we have
$\mathbf{A}_{T} = - \, \frac{(2\pi - ib)}{2\pi} \, ( \cosh \rho_0-i-\sinh \rho_0 ) d\vp_T$, so that we can write as
	\begin{align}
		I_T \, = \, - (2\pi - ib) q_T\chi(D) -iq_T A_D + iq_T L_D - iq_T \int \mathbf{A}_T  \, .
	\end{align}
As before, we drop the Euler characteristic term in the charged particle picture until the end, where it is used to give a finite result. Recognizing $q_\alpha=iq_T$, solving the quantum mechanical problem on the trumpet geometry is equivalent to solving the problem on the punctured disk up to a slightly modified AB field. In particular, as $\rho_0\to\infty$, $A_T\to i(2\pi-ib)/2\pi$ whereas $A_{D_\a} \to -\a\zeta/2\pi$. We will focus on solving the problem on the cone. It is a simple matter to make the appropriate analytic continuations to give the solution to the quantum mechanical problem on the trumpet.

In finding the wavefunctions and partition function on the cone, it will be useful to further rescale $\rho\to \rho/\z$,
	\begin{align}
	ds^2 \, = \, \z^{-2} \, d\rho^2 \, + \, \z^2 \, \sinh^2(\rho / \z ) d\vp^2 \, , \qquad 0 \leq \vp < 2\pi.
	\label{conegauge2}
	\end{align}
The rationale for doing so is that near the origin, $ds^2 \, \approx \, \z^{-2} \, d\rho^2 + \rho^2 \, d\vp^2$ which is the metric on the cone $x_1=\sqrt{(\z^{-2}-1)(x_2^2+x_3^2)}$ 
in flat three-space. In terms of the partition function, both gauges (\ref{conegauge1}) and (\ref{conegauge2}) give the same answer. A similar gauge
choice for quantum mechanics on a flat cone was made in \cite{Deser:1988qn} which can be compared to the result of \cite{Carslaw1, Carslaw2, Carslaw3}, detailed further in appendix \ref{app:coneappendix}. 
The partition functions are identical.

 Throughout this discussion, the Euler characteristic piece of the on-shell action has been left untouched as in \cite{Yang:2018gdb} since it only contributes an overall constant to the partition function. Nevertheless, it will be important in finding a finite result for the $q\to \infty$ limit. The Aharonov-Bohm effect on the Poincare disk 
was previously studied in \cite{Grosche:1998ff,Lisovyy:2007mj}. It is simple to extend those results to the theory on a hyperbolic cone.

\subsection{Quantum mechanical system}
\label{sec:qm}
Following the same discussion as in \cite{Yang:2018gdb,Kitaev:2018wpr}, the charged particle is described by the Lagrangian (density)
	\begin{align}
		\mathcal{L}_D \, &= \, \frac{1}{2} \, \big( \dot{\rho}^2 + \sinh^2\rho \, \dot{\varphi}^2 \big) \, + \, q \cosh \rho \dot{\varphi} \, , \\
		\mathcal{L}_T \, &= \, \frac{1}{2} \, \big( \dot{\rho}^2 + \cosh^2\rho \, \dot{\varphi}^2 \big) \, + \, q \sinh \rho \, \dot{\varphi} \, , 
	\end{align}
where the dot represents a derivative with respect to the boundary time $u$.
The first term comes from rewriting the length $L_M$ as an integral constrained to have fixed length
\footnote{The boundary curve with the length $L_M$ must be a self-avoiding loop and this is discussed in \cite{Stanford:2020qhm}.
In this paper, we neglect this issue and follow the naive discussion as in \cite{Yang:2018gdb,Kitaev:2018wpr}.
},
while the second comes from the coupling of the gauge field to the charged particle.
This Lagrangian can be regarded as describing a non-relativistic spinless particle with unit mass and electric charge $q$ under a unit magnetic field strength \cite{Kitaev:2017hnr}.
The canonical momenta for $\mathcal{L}_D$ are given by 
\footnote{The imaginary factor of $i$ for the gauge field contribution is due to the fact that we are using Euclidean time \cite{Kitaev:2018wpr}.}
	\begin{equation}
		p_{\rho} \, = \, \dot{\rho} \, , \qquad \quad p_{\varphi} \, = \, \sinh^2 \rho \, \dot{\varphi} \, + i q \, \cosh \rho \, ,
	\end{equation}
and for $\mathcal{L}_T$
	\begin{equation}
		p_{\rho} \, = \, \dot{\rho} \, , \qquad \quad p_{\varphi} \, = \, \cosh^2 \rho \, \dot{\varphi} \, + i q \, \sinh \rho \, .
	\end{equation}
Therefore, the (classical) Hamiltonians are written as
	\begin{align}
		\mathcal{H}_D \, &= \, \frac{1}{2} \, \left[ p_{\rho}^2 \, + \, \frac{1}{\sinh^2 \rho} \, \big( p_\varphi - b \cosh \rho \big)^2 \right] \, , \qquad (b \, = \, iq) \\
		\mathcal{H}_T \, &= \, \frac{1}{2} \, \left[ p_{\rho}^2 \, + \, \frac{1}{\cosh^2 \rho} \, \big( p_\varphi - b \sinh \rho \big)^2 \right] \, .
	\end{align}

In order to obtain quantum version of the Hamiltonians, it is useful to consider $SL(2, \mathbb{R})$ generators, which is studied for example in \cite{Kitaev:2017hnr} extensively.
The $SL(2, \mathbb{R})$ generators are given by
	\begin{equation}
		L_0 \, = \, p_\varphi \, , \qquad L_{\pm} \, = \, e^{\mp i \varphi} \left( \pm i p_\rho + \coth\rho \, p_\varphi - \, \frac{b}{\sinh\rho} \right) \, , 
	\end{equation}
for the disk ($D$) and 
	\begin{equation}
		L_0 \, = \, - i p_\varphi \, , \qquad L_{\pm} \, = \,  i e^{\pm \varphi} \left( \pm p_\rho - \tanh\rho \, p_\varphi - \frac{b}{\cosh\rho} \right) \, , 
	\end{equation}
for the trumpet ($T$).
They satisfy the $SL(2, \mathbb{R})$ commutations
	\begin{equation}
		[L_\pm \, , \, L_0] \, = \, \pm L_\pm \, , \qquad [L_+ \, , \, L_-] \, = \, 2 L_0 \, ,
	\end{equation}
and the quantum version of the Hamiltonian is expressed as
	\begin{equation}
		\hat{\mathcal{H}} \, = \, \frac{1}{2} \left( - L_0^2 \, + \, \frac{1}{2} L_+ L_- \, + \, \frac{1}{2} L_- L_+ \, + \, b^2 \right) \, .
	\label{hamiltonian}
	\end{equation}
Explicitly these are given by
	\begin{align}
		\hat{\mathcal{H}}_D \, &= \, \frac{1}{2} \, \left[ p_{\rho}^2 \, - \, i \coth\rho \, p_\rho \, + \, \frac{1}{\sinh^2 \rho} \, \big( p_\varphi - b \cosh \rho \big)^2 \right] \, , \\
		\hat{\mathcal{H}}_T \, &= \, \frac{1}{2} \, \left[ p_{\rho}^2 \, - \, i \tanh\rho \, p_\rho \, + \, \frac{1}{\cosh^2 \rho} \, \big( p_\varphi - b \sinh \rho \big)^2 \right] \, .
	\end{align}

We label the states by quantum numbers $j$ and $k$, so that $p_\varphi |j, k \rangle = k |j, k \rangle$ and $2\hat{\mathcal{H}} |j, k \rangle = \big( j(1-j) +b^2 \big) |j, k \rangle$.
Then the Schr\"{o}dinger equations are explicitly written as
	\begin{align}
		\left[ -\partial_{\rho}^2 \, - \, \coth \rho \, \partial_{\rho} \, + \, \frac{1}{\sinh^2 \rho} \, \big(k- b \cosh \rho \big)^2 \right] f^D_{j,k}(\rho, \varphi) \,
		&= \, \Big( j(1-j) + b^2 \Big) \, f^D_{j,k}(\rho, \varphi) \, , \label{Schro-eq-D} \\
		\left[ -\partial_{\rho}^2 \, - \, \tanh \rho \, \partial_{\rho} \, + \, \frac{1}{\cosh^2 \rho} \, \big(k- b \sinh \rho \big)^2 \right] f^T_{j,k}(\rho, \varphi) \,
		&= \, \Big( j(1-j) + b^2 \Big) \, f^T_{j,k}(\rho, \varphi) \, . \label{Schro-eq-T}
	\end{align}
Before we proceed, we can verify that all quantities shown above are related by the analytic continuation of the coordinates:
	\begin{equation}
		\rho_D \, = \, \rho_T \, + \, \frac{i \pi}{2} \, , \qquad \varphi_D \, = \, i \varphi_T \, .
	\end{equation}

\subsection{Charged particle on the Poincar\'{e} disk}
\label{sec:partitionfunctionD}
The wavefunctions on $D$ which obey (\ref{Schro-eq-D}) are discussed in \cite{Kitaev:2017hnr}. We summarize this here.
The total wavefunction is decomposed as $f^D_{j,k}(\rho, \varphi) = e^{i(k-b)\varphi} \, f^D_{j,k}(x)$, where $k = b + \mathbb{Z}$ \cite{Kitaev:2017hnr}.
After the change of variables
	\begin{equation}
		x \, = \, \tanh^2 \left( \tfrac{\rho}{2} \right) \, ,
	\end{equation}
the Schr\"{o}dinger equation (\ref{Schro-eq-D}) is written as
	\begin{equation}
		\left[ - (1-x)^2 (x \partial_x^2 + \partial_x) \, + \, \frac{1-x}{4x} \Big( (k-b)^2 - (k+b)^2 x \Big) \right] f^D_{j,k}(x) \, = \, j(1-j) \, f^D_{j,k}(x) \, .
	\label{Schro-eq-D-2}
	\end{equation}
The general solution is given by a linear combination
	\begin{equation}
		 f^D_{j,k}(x) \, = \, a^b_{j,k} \, A^b_{j,k}(x) \, + \, a^{-b}_{j,-k} \, A^{-b}_{j,-k}(x) \, ,
	\label{f^D}
	\end{equation}
with
	\begin{align}
		A^b_{j,k}(x) \, = \, x^{\frac{(b-k)}{2}} (1-x)^j \, \mathbf{F}(j-k, \, j+b; \, 1+b-k; \, x) \, ,
	\label{A^q}
	\end{align}
where $\mathbf{F}(a,b;c;x) = \Gamma(c)^{-1} {}_2F_1(a,b;c;x)$ is the regularized hypergeometric function.
The asymptotic to the center of the disk ($x\to 0$) behavior of the solution is given by 
	\begin{equation}
		A^{\pm b}_{j,\pm k}(x) \, = \, x^{\pm \left( \frac{b-k}{2} \right)} \left[ \frac{1}{\Gamma(1\pm b \mp k)} \, + \, \mathcal{O}(x) \right] \, .
	\end{equation}
Therefore, for $b>k$, $A^{b}_{j,k}$ is the regular solution while for $b<k$, $A^{-b}_{j,-k}$ is the regular solution.
The coefficients $a^{\pm b}_{j,\pm k}$ are fixed by the orthonormality condition in appendix \ref{app:orthonormality}, which are given by 
	\begin{equation}
		a^b_{j,k} \, = \, \frac{i}{2\sqrt{\pi}} \, \frac{\Gamma(1-j-k)\Gamma(1-j+b)}{\Gamma(1-2j)} \, .
	\end{equation}
In particular, we find
	\begin{equation}
		\big| a^b_{\frac{1}{2}+is,b} \big|^2 \, = \, \frac{s \sinh(2\pi s)}{\cos(2\pi b)+ \cosh(2\pi s)} \, .
	\end{equation}

Now we study the propagator \cite{Yang:2018gdb,Kitaev:2018wpr} which is defined by the diffusion equation 
	\begin{equation}
		\big( \partial_\t + \hat{\mathcal{H}} \big) G(x,\vp; x', \vp'; \t) \,= \, 0 \, .
	\end{equation}
By defining the resolvent in the form of Laplace transform as
	\begin{equation}
		G(x,\vp; x', \vp'; \t) \,= \, \int_{\frac{1}{2}(b^2+\frac{1}{4})}^{\inf} dE \, e^{-E \t} \, \Pi(x,\vp; x', \vp'; E) \, ,
	\label{Laplace transf}
	\end{equation}
the diffusion equation is reduce to the eigenvalue equation we studied above as
	\begin{equation}
		\big( \hat{\mathcal{H}} - E \big) \Pi(x,\vp; x', \vp'; E) \,= \, 0 \, .
	\end{equation}
Therefore, the resolvent is given by the summation over $k$:
	\begin{align}
		\Pi(x,\vp; x', \vp'; E) \, = \, \sum_{k=b+\mathbb{Z}} \, f_{j,k}^{D *}(\r, \vp) f_{j,k}^D(\r', \vp') \, ,
	\label{resolvent}
	\end{align}
where $E=\frac{1}{2}(s^2+b^2+\frac{1}{4})$.	
If we set $\r'=0$ ($x'=0$), then only the $k=b$ contribution is non-zero by (\ref{A^q}).
Therefore, we find 
	\begin{equation}
		\Pi(x,\vp; 0, \vp'; E) \, = \, \big|a^b_{j,b}\big|^2 \, e^{-i b (\vp - \vp')} (1-x)^j \, \mathbf{F}(j-b, \, j+b; \, 1; \, x) \, .
	\end{equation}
The expression for general $\r'$ can be found by the symmetry.
The partition function is now defined by
	\begin{equation}
		Z_D \, = \, G(x,\vp; x, \vp; \b) \, = \, A_D \int_0^{\inf} ds \, e^{-\b \frac{s^2}{2}} \, \r(s) \, .
	\end{equation}
Using this expression, we find 
	\begin{equation}
		\r(s) \, = \, \big| a^b_{\frac{1}{2}+is,b} \big|^2 \, = \, \frac{s \sinh(2\pi s)}{\cos(2\pi b)+ \cosh(2\pi s)} \, .
	\label{backgroundspectrum}
	\end{equation}
The Schwarzian limit is defined by setting $b=i q$ and taking $q \to \inf$ with $s$ fixed.
In this limit the spectral density becomes
	\begin{equation}
		\r(s) \, \approx \, 2s \sinh(2\pi s) \, .
	\end{equation}
The Schwarzian limit can also be phrased a the fixed temperature limit of the LHS of (\ref{betadef}) as $\r_0 \to \inf$.
Therefore, in the Schwarzian limit we have a relation
	\begin{equation}
		\sinh \r_0 \, \approx \, e^{\r_0} \, \approx \, q \, .
	\end{equation}

\subsection{Charged particle on a hyperbolic cone}

\subsubsection{Wavefunctions}
\label{sec:wavefunctioncone}
As discussed in \cite{Lisovyy:2007mj}, the effect of the magnetic Aharonov-Bohm field on the charged particle is to shift the angular momentum quantum numbers 
by $\xi$ where $\xi = -i\Phi/2\pi$ and $\P=\int \mathbf{A}^{(\a)}$ is the Aharonov-Bohm magnetic flux. 
In other words, starting with the $SL(2,\mathbb{R})$ invariant Hamiltonian (\ref{Schro-eq-D-2}), we take $k\to k+\xi$. 
Following our earlier discussion, for the punctured disk, $\xi = \frac{iq\a \z}{2\pi}(\cosh \rho_0 - 1-\sinh \rho_0)$. In the limit $q\to \infty$, we find $\xi = -iq\alpha\zeta/2\pi$.

This can also be seen from the BF description of the JT gravity. (We summarize relevant aspects of the BF theory in appendix \ref{app:monodromiesinBFtheory}).
In the gauge (\ref{conegauge1}), the variation of $\p$ leads to $R(x)+2=2\alpha \delta^2(x-x_1)$ and the defect gives a conical deficit $\vp \cong \vp+ 2\pi -\alpha$.
But of course the left hand side of the former equation is just $F$ from the BF theory \cite{Mertens:2019tcm,Mertens:2018fds,Blommaert:2018oro}.
This means that there is a delta-function source for the gauge field in the BF theory which can be written,
	\begin{align}
		F \, = \, \kappa \, p\lambda p^{-1} \, \delta^2(x-x_1) \, .
	\end{align}
Here, $\lambda = \vec{\lambda}\cdot \vec{H}$ where $\vec{H}$ are elements of the Cartan subalgebra in the highest weight of the representation
and $p$ is some constant element in the algebra. The constant $\kappa$ will be related to the conical deficit.
The delta-function source means that the gauge field is not everywhere flat. Then we write
	\begin{align}
		A_\rho \, = \, g^{-1}\partial_\rho g \, , \qquad \qquad
		A_\vp \, = \, g^{-1}\partial_\vp g \, + \, \frac{\kappa}{\pi} \, p \lambda p^{-1} \, ,
	\label{Adeficiteq}
	\end{align}
for $g \in \mathfrak{sl}(2,\mathbb{R})$. In this expression, $g(\vp+2\pi)=g(\vp)$.
By performing a global gauge transformation, $g \to gp^{-1}$, the BF action becomes
	\begin{align}
		I_{\rm BF} \, = \, -\int d\vp \, \Tr(A_\vp)^2 \, \to \, - \int d\vp \, \Tr(g^{-1}dg + \Lambda)^2 \, ,
	\label{monodromyBF}
	\end{align}
where $\Lambda = \frac{\kappa\lambda}{2\pi}$.
Finally, we can redefine $\tilde{g} = gU(\vp)$ for $U(\vp) = \exp(\frac{\kappa}{2\pi}p\lambda p^{-1}\vp)$ which makes $A_\vp = \tilde{g}^{-1}\partial_\vp \tilde{g}$ flat.
With this definition, we have a monodromy $\tilde{g}(\vp+2\pi) = \tilde{g}(\vp)U(2\pi)$. Hence, we should characterize conical deficits by monodromies.
This also explains why the solution (\ref{Adeficiteq}) is correct for $A$. The presence of the deficit implies that
	\begin{align}
		\int_M F \, = \, \oint_{\partial M} A_\vp \, = \, \kappa \, p \lambda p^{-1} \, .
	\end{align}
which is seen to be solved by the solution (\ref{Adeficiteq}).
In particular for the AB gauge field (\ref{A^alpha}), using this relation we find that $\kappa \, p \lambda p^{-1} = -\a \z (\cosh \r_0 - 1 - \sinh \r_0)$ and $U(\vp)=\exp(i \x \vp)$ where
	\begin{align}
		\x \, = \, - \frac{i q}{2\pi } \, \kappa \, p \lambda p^{-1} \, = \, \frac{i q \a \z}{2\pi} \, \big( \cosh \r_0 - 1 - \sinh \r_0 \big) \, ,
	\end{align}
where we also included the charge $q$ from (\ref{I_D_alpha}) in $\x$.
This implies that for the gauge (\ref{conegauge2}), where the element is obtained by the action of $U(\vp)$,
the effect of the AB field is implemented by simply shifting the quantum number of the angular coordinate $\vp$ by $\xi$.

In fact, this shift is expected from the monodromy condition on the hyperbolic disk \cite{Mertens:2019tcm}, which can be thought of as the group manifold $PSL(2,\mathbb{R})$. Starting from the universal cover $\widetilde{\frak{G}} = \widetilde{\mathrm{SL}}(2,\mathbb{R})$ discussed in \cite{Kitaev:2017hnr}, a choice of cross-section of the principle bundle $\widetilde{\frak{G}}\to H^2$ is equivalent to a local frame on the Poincar\'e disk. Representing an element $g\in \widetilde{\frak{G}}$ as $g = e^{\vp \Lambda_0} e^{\rho\Lambda_1} e^{-\vartheta \Lambda_0}$ with representation
\footnote{Here, $\Lambda_0 = iL_0$, $\Lambda_1 =( L_- - L_+)/2$, $\Lambda_2 = i(L_-+L_+)/2$.}
	 \begin{align}
		 \Lambda_0 \, = \, \frac{1}{2}\left(\begin{array}{cc} i&0\\0&-i\end{array}\right) \, , \qquad
		 \Lambda_1 \, = \, \frac{1}{2}\left(\begin{array}{cc}0&1\\1&0\end{array}\right) \, , \qquad
		 \Lambda_2 \, = \, \frac{1}{2}\left(\begin{array}{cc}0&i\\-i&0\end{array}\right) \, ,
	 \end{align}
a choice of cross-section is equivalent to a choice of $\vartheta$ for the orthogonal basis vectors $v_1$ and $v_2$ on $H^2$:
	\begin{align}
		\begin{pmatrix}
		v_1^\rho \\ v_1^\vp 
		\end{pmatrix}
		\, = \,
		\begin{pmatrix}
		\cos \vartheta \\ -\frac{\sin\vartheta}{\sinh \rho}
		\end{pmatrix}
		\, , \, \qquad
		\begin{pmatrix}
		v_2^\rho \\ v_2^\vp 
		\end{pmatrix}
		\, = \,
		\begin{pmatrix}
		\sin \vartheta \\ \frac{\cos\vartheta}{\sinh \rho}
		\end{pmatrix} \, .
	\end{align}
 Two particular choices for $z=e^{i\vp}\tanh(\rho/2)$ are the tilde gauge $\tilde{g}(\rho,\vp) = e^{\vp \Lambda_0} e^{\rho\Lambda_1} $ with $\vartheta=0$ and the disk gauge $\mathring{g}(\rho,\vp)= e^{\vp \Lambda_0} e^{\rho\Lambda_1}e^{-\vp\Lambda_0}$ with $\vartheta = \vp$. 
It is easily verified that $\tilde{g}(\rho,\vp+2\pi) = -\tilde{g}(\rho,\vp)$ and $\mathring{g}(\rho,\vp+2\pi) = \mathring{g}(\rho,\vp)$. Choosing $\vartheta = (\kappa-1)\vp$ we find the monodromy $g(\rho,\vp+2\pi) = g(\rho,\vp)e^{-2\pi\kappa\Lambda_0}$. The different monodromies between the disk and tilde gauges have $\kappa = 1$. This is a reflection of the fact that the true symmetry group of the Poincar\'e disk is $PSL(2,\mathbb{R}) = SL(2,\mathbb{R})/\mathbb{Z}_2$, discussed further in appendix~\ref{app:monodromiesinBFtheory}.

The group perspecitive is useful since it allows for construction of spinors. In particular, eigenfunctions of the Hamiltonian in section \ref{sec:partitionfunctionD} can be thought of as $\nu$-spinors with the identification $\nu = -iq$. From \cite{Kitaev:2018wpr}, $\nu$-spinors are functions on $H^2$ which are characterized by the property that the generator $\Lambda_0$ has
the right group action of multiplication by $-i\nu$.
In other words, two $\nu$-spinors are related by $\Psi(g_2)=e^{-i\nu\theta}\Psi(g_1)$ for two different cross-sections related by $ \theta= \vartheta_1-\vartheta_2$.
Given the wavefunctions on the disk in section \ref{sec:partitionfunctionD} by $f_{j,k}^D(\rho,\vp) = e^{ik\vp}f_{j,k}^D(u)$ with $\nu = -iq$,
the wavefunction corresponding to the orthonormal frame with $\vartheta = (\kappa-1)\vp$ is $f_{j,k,\kappa}^{D} = e^{i(k+iq[\kappa-1])\vp}f_{j,k}(u)$.
In section~\ref{sec:wavefunctioncone} and appendix~\ref{app:monodromiesinBFtheory}, we show that for the conical deficit parameter $\alpha$, 
we should identify $\kappa = 1-\alpha/2\pi = \zeta$, see also \cite{Witten:2020wvy}. If we take $q\to q_\alpha$, we see that the angular momentum quantum numbers get shifted 
$k\to k-iq\zeta \alpha/2\pi$. In the limit $q\to \infty$, we see exactly the shift of \cite{Lisovyy:2007mj}, as expected.

The monodromy condition motivates the change in angular momentum quantum numbers, but it is not the whole story for the punctured disk, 
since it only captures the AB flux. Following \cite{Kitaev:2017hnr}, the representation $g = e^{\vp\Lambda_0}e^{\rho \Lambda_1}e^{-\vartheta \Lambda_0}$ leads to the metric on the Poincar\'e disk when extremized over $d\vartheta$. Since our monodromy cross-section uses the same parametrization before fixing $\vartheta$, we would find the same metric up to potential angular identifications, $\vp \simeq \vp+(2\pi-\alpha)$, as pointed out in (\ref{conegauge1}). We have stated that it is more convenient to use (\ref{conegauge2}) which has $\vp \simeq \vp + 2\pi$, where we rescaled $\rho\to\rho/\z$ and $\vp\to \z\vp$.
Then we can proceed as \cite{Lisovyy:2007mj}.\footnote{To match notation, we must take $k\to l+b$ with $l\in \mathbb{Z}$.
}

Starting from the eigenvalue equation (\ref{Schro-eq-D-2}), we rescale $\rho\to \rho/\zeta$, $\vp\to \zeta\vp$, $q\to q/\zeta$ and include the AB field via $k\to k+\xi$. Defining $x=\tanh^2(\rho/2\zeta)$, we have
	\begin{align}
	\label{Schro-eq-Da-2}
	&\left[ - (1-x)^2 (x \partial_x^2 + \partial_x) \, + \, \frac{1-x}{4\zeta^2x} \Big( (k+\xi-b)^2 - (k+\xi+b)^2 x \Big) \right] f^{D_\alpha}_{j,k}(x) \\
	&\hspace{280pt}=  j(1-j) \, f^{D_\alpha}_{j,k}(x) \, , \nn
	\end{align}
where $k = b+\mathbb{Z}$ as before.
We now note that if $k\to \zeta k$, $\xi \to \zeta\xi$ and $q\to \zeta q$, we have the exact same equation as (\ref{Schro-eq-D-2}), up to the shift of $k\to k+\xi$.
This is not surprising since this transformation takes (\ref{conegauge2}) to (\ref{conegauge1}), though of course we have a different definition for $x$. 

Now, we ask what the consequences of the conical deficit are for the continuous spectrum. We first consider the algebra of $\widetilde{SL}(2,\mathbb{R})$ on disk eigenvectors. Starting with an eigenvector on the disk, $|l,j\rangle$ characterized by $L_0|l,j\rangle = \frac{1}{2}[L_+,L_-]|l,j\rangle= -l|l,j\rangle$ and $2\hat{\mathcal{H}}_D|l,j\rangle = j(1-j)+b^2$, we can write
	\begin{align}
	\langle l,j | L_- L_+ |l,j\rangle \, &= \,(l+j)(l+1-j) \, , \nn\\
	\langle l,j | L_+L_- |l,j\rangle \, &= \, (l-j)(l-1+j) \, .
	\end{align}
Since $L_-^\dagger = L_+$, these matrix elements must be positive. The continuous principle series that we discussed in section \ref{sec:partitionfunctionD} have $j = \frac{1}{2}+is$ which leads to positive matrix elements for all $l \in \mathbb{Z}$.

For the cone, Under $\vp\to\zeta\vp$ and $q \to \frac{q}{\zeta}$, the same eigenvectors have $L_0|l,j\rangle = -\frac{l}{\zeta}|l,j\rangle$ and we can write
	\begin{align}
		\langle l,j | L_- L_+ |l,j\rangle \, &= \, \frac{l}{\zeta} \left(\frac{l}{\zeta}+1\right) + j(1-j) \, = \, \frac{(l+j_\zeta)(l+\zeta-j_\zeta)}{\zeta^2} \, , \nn\\
		\langle l,j | L_+L_- |l,j\rangle \, &= \, \frac{l}{\zeta} \left(\frac{l}{\zeta}-1\right) + j(1-j) \, = \, \frac{(l-j_\zeta)(l-\zeta+j_\zeta)}{\zeta^2} \, .
	\end{align}
Positivity is ensured if $j_\zeta = \zeta(\frac{1}{2}+is)$, but this means that $j=\frac{1}{2}+is$ does not change. This long-winded explanation shows that the effects of the conical geometry can be undone via a simple rescaling of the quantum numbers. Hence we may use the results already derived in section \ref{sec:partitionfunctionD} for contributions from the background field plus results from the literature \cite{Lisovyy:2007mj} for the Aharonov-Bohm filed contribution. The algebra also implies that the energy of the continuous spectrum on the cone is nearly unchanged, $E = \frac{1}{2}(s^2 + \frac{1}{4} - \frac{q^2}{\zeta^2})$. Following \cite{Yang:2018gdb, Kitaev:2018wpr}, the energy in the Schwarzian limit is $E_{\rm Sch} = E+\frac{q^2}{2\zeta}-\frac{1}{8} = s^2/2$.

Working on the punctured disk with $\vp \cong \vp +2\pi$, i.e. in the gauge (\ref{conegauge2}), the angular momentum component of the wavefunctions do not change. However, the radial wavefunctions do. Comparing the two Schr\"{o}dinger equations (\ref{Schro-eq-D-2}) and (\ref{Schro-eq-Da-2}), it is easily seen that the wavefunctions are given by
	\begin{align}
	f^{D_\alpha}_{j_\zeta,k}(x,\vp,\xi) \, = \, e^{i\frac{k-b}{\zeta}\vp}
	\left[a^{b}_{j_\zeta,k+\xi}A_{\frac{j_\zeta}{\zeta},\frac{k+\xi}{\zeta}}^{b/\zeta} \, + \, a^{-b}_{j_\zeta,k+\xi}A_{\frac{j_\zeta}{\zeta},-\frac{k+\xi}{\zeta}}^{-b/\zeta}(x) \right] \, ,
	\label{conewavefunctions}
	\end{align}
where $b= iq\zeta$, $k\in b + \zeta\mathbb{Z}$, 
	\begin{align}
	\xi \, = \, -i\frac{q\alpha\zeta}{2\pi}\left[\cosh(\frac{\rho_0}{\zeta}) - 1 -\sinh(\frac{\rho_0}{\zeta})\right] \, , \ \;\;
	j_\zeta \, = \, \zeta(\frac{1}{2}+is) \, , \ \;\;
	x \, = \,\tanh^2(\frac{\rho}{2\zeta}) \, ,
	\label{coneparameters}
	\end{align}
and
	\begin{align}
	A^{b/\zeta}_{\frac{j_\zeta}{\zeta},\frac{k+\xi}{\zeta}} \, = \, x^{\frac{b-k-\xi}{2\zeta}}(1-x)^\frac{j_\zeta}{\zeta} \,
	\mathbf{F}\left( \frac{j_\zeta-k-\xi}{\zeta},\frac{j_\zeta+b}{\zeta};1+\frac{b-k-\xi}{\zeta};x \right) \, .
	\end{align}
As an aside, we note that, under the rescaling $k\to \zeta k$, $\xi \to \zeta \xi$ and $b\to \zeta b$, we can relate
	\begin{equation}
	f^{D_\alpha}_{j,k} \, = \, e^{-i\xi\vp} \, f^{D}_{j,k+\xi} \, .
	\end{equation}
In other words, the regular solution on the punctured disk is related to a solution on the unpunctured disk with monodromy $f_{j,k+\xi}^{D}(x,\vp+2\pi) = e^{2\pi i\xi}f_{j,k+\xi}(x,\vp)$. This is a modification of periodic boundary conditions which is not regular at the origin, but is neverthless allowed by quantum mechanics through the theory of self-adjoint extensions \cite{ssolvable}. If $\xi = 1/2$, these would be Neveu-Schwarz boundary conditions, though here $\xi$ is imaginary and can be irrational. This monodromy justifies the matching we made in section~\ref{sec:partitionfunctioncone}.

\subsubsection{Propagator}
The propagator can be found by exactly following the prescription in \cite{Lisovyy:2007mj} with the appropriate identifications of parameters (\ref{coneparameters}).
This is because we are using a gauge in which $\vp\cong \vp+2\pi$ and the radial coordinate $x$ hides the conical geometry. One of our interests in this work is to calculate R\'enyi and entanglement entropies and for this, we must introduce a new conical deficit, which we label by $\theta$, that impliments the replica trick \cite{renyi1961, Calabrese:2004eu, Nishioka:2009un, Rangamani:2016dms, Lewkowycz:2013nqa}.
In what follows, it will be important to distinguish the conical manifold arising from the deformation of JT gravity, which has $\theta = 1$, from the conical manifold arising from the replicas, which has general $\theta$.

We start by describing the propagator for $D$ and then modify it to accommodate the AB field and conical deficit from the replicas. The propagator for $D$ can be written in a number of ways, including as a contour integral \cite{Lisovyy:2007mj}, a sum over the wavefunctions in a particular representation $s$ \cite{Kitaev:2017hnr, Kitaev:2018wpr}, and in closed form in terms of a hypergeometric function \cite{Comtet:1986ki}. In analogy with the flat cone analysis in appendix \ref{app:coneappendix}, we choose to represent this as a contour integral. A similar discussion of contour integrals is given by the map relating $\nu$-spinor representations to ``$\mu$-twisted $\lambda$-form" representations of the $\widetilde{\text{Diff}}_+(S^1)$ subgroup of $\widetilde{SL}(2,\mathbb{R})$ that is discussed in \cite{Kitaev:2017hnr}, see eq. (109). This map is equally well interpreted as a boundary-bulk propagator mapping a specific choice of boundary $\vp(u)$ to a bulk point.

For $\th=1$ case, the resolvent can be written in terms of a contour integral \cite{Lisovyy:2007mj}
(see also \cite{Kitaev:2017hnr} and appendix \ref{app:theta=1}) as
	\begin{align}
		\Pi^{(1)}(z;z') \, = \, \frac{1}{2 i}\int_{C_0} dw_1 \int_{C_+'\cup C_-'} dw_2 \, \Psi_{-}^{-b}(x,\vp,w_1)\Psi_{+}^b(x',\vp',w_2) \frac{e^{w_1}}{e^{w_1}-e^{w_2}} \, ,
	\label{kernelnoABfield2}
	\end{align}
where $\Psi_{\pm}$ are so-called horocyclic waves
	\begin{align}
		\Psi^{b}_{\pm}(x, \vp ,w) = \frac{1}{2\pi}\frac{(1-|z|^2)^{j_{\pm}}}{(1+ze^{-w})^{j_{\pm}-b}\,(1+\bar{z}e^{w})^{j_{\pm}+b}} \, .
	\end{align}
with $j_{\pm} = \frac{1}{2}\pm(j-\frac{1}{2})$ and $z = \sqrt{x} \, e^{i\vp}$.
The contour $C_0$ is running from $w = i(\vp + \pi) + \ln x$ to $w = i(\vp + \pi) - \ln x$ along the real axis and the contours $C_{\pm}$ are shown in figure~\ref{fig:hyperbolicdiskcontours}.
Defining $K(z;z')$ to be the part of the resolvent that arises from the background field, rather than the AB field, we can write this as
	\begin{align}
		\Pi^{(1)}_{\xi=0} \equiv K^{(1)}(z;z') \, = \, \pi \int_{C_0} dw \, \Psi_{-}^{-b}(x,\vp,w)\Psi_{+}^{b}(x',\vp',w) \, ,
	\end{align}
Finally the full heat kernel, or propagator, is then given by the Laplace transform (\ref{Laplace transf}). 

The wavefunctions in eq.~(\ref{conewavefunctions}) can also be written in terms of a contour integral over the horocyclic waves \cite{Lisovyy:2007mj}, discussed further in appendix \ref{app:theta=1}. From this perspective, following \cite{Kitaev:2017hnr}, the wavefunction with parameter $k$ is related to a specific Fourier component of the boundary time reparametrization, $e^{-k w_i}$.\footnote{There is no $i$ because we are in Euclidean time. This Fourier component when quantized becomes a boundary graviton.} The contour integral is then the bulk-boundary map of this component and the horocyclic wave is the bulk-boundary propagator. The propagator we derive here can be understood as a boundary two-point function written as the convolution of two bulk-boundary propagators. This same perspective was discussed in \cite{Yang:2018gdb, Kitaev:2018wpr}.

\begin{figure}[t]
\begin{center}
\begin{tikzpicture}[>=stealth]
	\draw[dotted, thick] (0,-1.5) -- (0,4);
	\draw[dotted, thick] (-4,0) -- (4,0);
	\draw[thick] (-.1,-1.0) -- (.1,-1.0);
	\draw[thick] (-.1,-.2) -- (.1, -.2);
	\draw[thick] (-.1,.6) -- (.1,.6);
	\draw[thick] (-.1,1.4) -- (.1,1.4);
	\draw[thick] (-.1,2.2) -- (.1,2.2);
	\draw[thick] (-.1,3.0) -- (.1,3.0);
	\draw[thick] (-.1,3.8) -- (.1,3.8);
	\draw[thick, dashed] (-4,3.0) -- (-2,3.0);
	\filldraw [black] (-2,3) circle (2pt);
	\draw[thick, dashed] (-4,1.4) -- (-2,1.4);
	\filldraw [black] (-2,1.4) circle (2pt);
	\draw[thick, dashed] (-4,-.2) -- (-2,-.2);
	\filldraw [black] (-2,-.2) circle (2pt);
	\draw[thick, dashed] (2,3.0) -- (4,3.0);
	\filldraw [black] (2,-.2) circle (2pt);
	\draw[thick, dashed] (2,1.4) -- (4,1.4);
	\filldraw [black] (2,1.4) circle (2pt);
	\draw[thick, dashed] (2,-.2) -- (4,-.2);
	\filldraw [black] (2,3) circle (2pt);
	\draw[very thick,dash dot, gray,->] (-4,1) -- (-2.5,1);
	\draw[very thick,dash dot, gray] (-2.5,1) -- (-1,1);
	\draw[very thick,dash dot, gray,->] (-1,1) -- (-1,1.8);
	\draw[very thick,dash dot, gray] (-1,1.8) -- (-1,2.6);
	\draw[very thick,dash dot, gray] (-4,2.6) -- (-2.5,2.6);
	\draw[very thick,dash dot, gray,<-] (-2.5,2.6) -- (-1,2.6);
	\draw (-4,2.5) node [fill=white!20]{\large $C_{+}$};
	\draw[very thick,dash dot, gray] (2.5,1) -- (4,1);
	\draw[very thick,dash dot, gray,->] (1,1) -- (2.5,1);
	\draw[very thick,dash dot, gray] (1,1) -- (1,1.8);
	\draw[very thick,dash dot, gray,<-] (1,1.8) -- (1,2.6);
	\draw[very thick,dash dot, gray,<-] (2.5,2.6) -- (4,2.6);
	\draw[very thick,dash dot, gray] (1,2.6) -- (2.5,2.6);
	\draw[thick] (1,-.1)--(1,.1);
	\draw (1,-.5) node {\large $\lambda_r$};
	\draw [thick, <->] (3.8,0) -- (3.8,1);
	\draw (4.2,.5) node {\large $\lambda_i$};
	\draw (4,2.5) node [fill=white!20]{\large $C_{-}$};
	\draw[ultra thick,->] (-2,1.4) -- (0,1.4);
	\draw[ultra thick] (0,1.4) -- (2,1.4);
	\draw[thick] (-2,-.1)--(-2,.1);
	\draw[thick] (2,-.1)--(2,.1);
	\draw (-2,.5) node {\large $\frac{1}{2}\ln x$};
	\draw (2,.5) node {\large $-\frac{1}{2}\ln x$};
	\draw (0,2.0) node [fill=white!20]{\large $C_0$};
	\draw (-1,-1) node [fill=white!20]{\large $\varphi - 2\pi$};
	\draw (-1,3.8) node [fill=white!20]{\large $\varphi + 4\pi$};
	\draw (-.4,.6) node [fill=white!20]{\large $\varphi $};
	\draw (-3.8,3.8) node [fill=white!20, draw=black]{\Large $w$};
\end{tikzpicture}
\caption{Contours on the complex $w$-plane used to calculate the propagator. The contours are defined for $z = \sqrt{x}e^{i\vp}$ and $0<x<1$.}
\label{fig:hyperbolicdiskcontours}
\end{center}
\end{figure}

Following the flat space procedure, one way to implement the effect of the conical deficit is to take the $\th=1$ expression (\ref{kernelnoABfield2}) and rescale the weight function,
	\begin{align}
	\Pi^{(\theta)}(z;z') \, = \,\frac{1}{2 i}\int_{C_0} dw_1 \int_{C_+'\cup C_-'} dw_2 \, \Psi_{-}^{-b}(x,\vp,w_1)\Psi_{+}^b(x',\vp',w_2) \frac{e^{w_1/\theta}}{e^{w_1/\theta}-e^{w_2/\theta}} \, .
	\label{kernelnoABfield-2}
	\end{align}
Here, the contours remain the same as in the disk geometry. Next, define $w_2 = w_2'+i\vp'$. This removes the explicit $\vp'$ dependence in $\Psi_{+}^{b}$. Furthermore, it becomes evident that the integral is periodic in $\vp'\to \vp' + 2\pi\theta$. We could similarly remove the $\vp$ dependence in $\Psi^{-b}_-$ by redefining $w_1 = w_1'+i\vp$. Now, we run the argument in reverse, converting the integral into a sum. The important point is that as $w_2' \to +\infty$, i.e. for $C_-'$, we write
	\begin{equation}
		\frac{e^{w_1/\theta}}{e^{w_1/\theta}-e^{w_2/\theta}} \, = \, -\sum_{\ell=1}^\infty e^{\ell w_1/\theta}e^{-\ell w_2/\theta} \, ,
	\label{lsum1}
	\end{equation}
whereas for $w_2' \to -\infty$, i.e. for $C_+'$, we write
	\begin{equation}
		\frac{e^{w_1/\theta}}{e^{w_1/\theta}-e^{w_2/\theta}} \, = \, \sum_{\ell=0}^\infty e^{- \ell w_1/\theta}e^{\ell w_2/\theta} \, .
	\label{lsum2}
	\end{equation}
Not surprisingly, the effect of the cone is to take $\ell\to \ell/\theta$.
This is the exact rescaling of the Schr\"{o}dinger equation (\ref{Schro-eq-Da-2}) and hence confirms the wavefunctions in (\ref{conewavefunctions}) with $\xi = 0$.
Returning to the double contour representation (\ref{kernelnoABfield-2}), we can deform the contour $C_+'\cup C_-'$ into two lines parallel to the real axis separated by $2\pi i$ plus contours that circle any poles that fall in between these two lines, like we did for the disk. The difference here is that, in addition to the pole at $e^{w_2}=e^{w_1}$ we can have a pole at $e^{w_2+2\pi i \theta} = e^{w_1}$. Furthermore, the two parallel lines no longer provide equal and opposite contributions. These are non-perturbative contributions to the propagator. As discussed in appendix \ref{app:coneappendix}, these arise from paths which intersect the puncture of the disk and we can discard them if we don't allow such paths in our path integral.

Finally, to include an AB field in the above discussion, we shift $\ell$ by the AB flux $\xi$ and sum over $\ell \in \mathbb{Z}$ in eqs. (\ref{lsum1}) and (\ref{lsum2}). The result is a simple multiplication of the integrand
	\begin{align}
	 \frac{e^{w_1/\theta}}{e^{w_1/\theta}-e^{w_2/\theta}} \, \to \, e^{\frac{\xi}{\theta}(w_1-w_2)} \, \frac{e^{w_1/\theta}}{e^{w_1/\theta}-e^{w_2/\theta}} \, .
	\end{align}

Starting with $\theta = 1$, the result from \cite{Lisovyy:2007mj} is
	\begin{align}
	\Pi^{(1)}_\xi(x,\vp;x',\vp') \, = \, \Delta^{(1)}(x,\vp;x',\vp') \, + \, e^{-i\xi(\vp-\vp'+2\pi \sigma)}K^{(1)}(x,\vp;x',\vp') \, ,
	\end{align}
where $\sigma = 1$ for $-2\pi<\vp-\vp'<-\pi$, $\sigma=0$ for $-\pi<\vp-\vp'<\pi$, and $\sigma=1$ for $\pi<\vp-\vp'<2\pi$. The function $\Delta^{(1)}$ is the contribution to the kernel arising from the AB field. It is given by
	\begin{align}
	\Delta^{(1)}(z,z') \, = \, \frac{1-e^{-2\pi i\xi}}{2i}e^{-i\xi(\vp-\vp')}\int_{C_0}dw_1\int_{\text{Im} w_2=\vp'} dw_2 \, 
	\Psi_{-}^{-b}(z,w_1)\Psi_{+}^{b}(z',w_2)\frac{e^{(1+\xi)(w_1-w_2)}}{e^{w_1-w_2}-1} \, .
	\label{ABfieldkernel}
	\end{align}
Interestingly, the separation of $\Pi^{(1)}$ into $\Delta^{(1)}$ and $K^{(1)}$ arises from the contribution to the kernel from the AB field and background field, respectively. In the contour integral, the latter arises from the pole contribution at $e^{w_1}=e^{w_2}$ when deforming $C_+'\cup C_-'$. The extra contribution from this deformation are two lines parallel to the real axis that can be deformed into a single line Im$w_2 = \vp'$ at the cost of a phase. If $\xi \el \mathbb{Z}$, then these two contributions cancel, as can be seen from the factor in front of eq. (\ref{ABfieldkernel}). This reflects an enhanced symmetry so that the stabilizer group is not $U(1)_\theta$ but $SL^n(2,\mathbb{R})$ discussed for instance in \cite{Mertens:2019tcm}.

Next, to include the conical deficit from the replica, we rescale the weight function in the contour integral as before. We can no longer write the deformed contour $C_+'\cup C_-'$ as a single line. The kernel is 
	\begin{align}
	\Delta^{(\theta)}(z,z') \, = \, \frac{1}{2i}e^{-i\xi(\vp-\vp')}\int_{C_0}dw_1\int_{C_1} dw_2 \,
	\Psi_{-}^{-b}(z,w_1)\Psi_{+}^{b}(z',w_2)\frac{e^{(1+\xi)(\frac{w_1}{\theta}-\frac{w_2}{\theta})}}{e^{\frac{w_1}{\theta}-\frac{w_2}{\theta}}-1} \, .
	\end{align}
Here $C_1$ is the contour $(\text{Im}w_2 = \vp')_\rightarrow\cup (\text{Im}w_2 = \vp' +2\pi)_\leftarrow$. Aside from the weight function, no changes occur to the kernel. In particular, $\Psi^{\pm b}_{\mp}$ have definite transformation properties under $SL(2,\mathbb{R})$. Then, we can follow the steps in appendix A of \cite{Lisovyy:2007mj}: interchanging the order of integration, defining $\tilde{w}_2=w_2-w_1$, then $\tilde{w}_1=w_1+\tilde{w}_2$, we may rewrite the kernel as
	\begin{align}
	\Delta^{(\theta)}(z,z') \, = \, \frac{1}{2i}e^{-i\xi(\vp-\vp')}\int_{\tilde{C}_1}d\tilde{w}_2\frac{e^{-(1+\xi)\frac{\tilde{w}_2}{\theta}}}{e^{-\frac{\tilde{w}_2}{\theta}}-1}
	\int_{\tilde{C}_0} d\tilde{w}_1\, \Psi_{-}^{-b}(z,\tilde{w}_1-\tilde{w}_2)\Psi_{+}^{b}(z',\tilde{w}_1) \, .
	\end{align}
Since $C_0$ was the contour defined at $w_1 = i\vp+i\pi$, $\tilde{C}_1$ is the contour $C_1$ shifted down by this amount. The contour $\tilde{C}_0$ is defined by shifting the contour $C_0$ by $\tilde{w}_2$. The important point is that since the contours and the functions $\Psi^{\pm b}_{\mp}$ do not get altered by the conical deficit, the integral is the same as performed in \cite{Lisovyy:2007mj}. Hence we may write
	\begin{equation}
	\Delta^{(\theta)}(z,z') \, = \, \frac{1}{2\pi i} e^{-i\xi(\vp-\vp')} \int_{-\infty}^\infty du 
	\left[\frac{e^{\frac{1+\xi}{\theta}(u - i(\vp'-\vp+\pi))}}{e^{(u-i(\vp'-\vp+\pi))/\theta}-1}-\frac{e^{\frac{1+\xi}{\theta}(u - i(\vp'-\vp-\pi))}}{e^{(u-i(\vp'-\vp-\pi))/\theta}-1} \right] \mathcal{F} \, , 
	\end{equation}
where
	\begin{equation}
	\mathcal{F} \, = \, \left(\frac{1+\sqrt{xx'}e^{-u}}{1+\sqrt{xx'}e^{u}}\right)^b \lambda\left(\frac{x+x'+2\sqrt{xx'}\cosh u}{1+xx'+2\sqrt{xx'}\cosh u}\right) \, ,
	\end{equation}
and 
	\begin{equation}
	\lambda(y) \, = \, \frac{1}{4\pi} \Gamma(j+b)\Gamma(j-b)(1-y)^j \, \mathbf{F}(j+b,j-b,2j,1-y) \, .
	\end{equation}
To find the density of states, we first set $z= z'$, giving
	\begin{align}
	\Delta^{(\theta)}(x) \, &= \, \frac{1}{2\pi i} \int_{-\infty}^\infty du
	\left[\frac{e^{\frac{1+\xi}{\theta}(u - i\pi)}}{e^{(u-i\pi)/\theta}-1}-\frac{e^{\frac{1+\xi}{\theta}(u +i\pi))}}{e^{(u+i\pi)/\theta}-1}\right] \nn\\
	&\hspace{140pt} \times \left(\frac{1+xe^{-u}}{1+xe^{u}}\right)^b \lambda\left(\frac{2x(1+\cosh u)}{1+x^2+2x\cosh u}\right) \, .
	\end{align}
Again, outside of the term in square brackets, there are no changes compared to \cite{Lisovyy:2007mj}. Finally, we can integrate this over the punctured disk giving,
	\begin{align}
	&\left(\frac{2\pi i}{\theta}\right)\text{Tr}\Delta^{(\theta)} \, \equiv \, \left(\frac{2\pi i}{\theta}\right)\int_{D_\alpha} dxd\vp\sqrt{-g} \Delta^{(\alpha)}(x) \nn\\
	&=\frac{1}{2j-1}\int_{0}^{1}dv\int_{-\infty}^{\infty}du\left[\frac{e^{\frac{1+\xi}{\theta}(u - i\pi)}}{e^{(u-i\pi)/\theta}-1}-\frac{e^{\frac{1+\xi}{\theta}(u +i\pi))}}{e^{(u+i\pi)/\theta}-1}\right] \frac{v^{j-1-b}}{(1+ve^{-u})(1+e^u)}\nn \\
	&-\frac{1}{2j-1}\int_0^1 dv \int_{-\infty}^{\infty} du \left[\frac{e^{-(u-i\pi)\frac{\xi}{\theta}}}{e^{(u-i\pi)/\theta}-1}-\frac{e^{-(u+i\pi)\frac{\xi}{\theta}}}{e^{(u+i\pi)/\theta}-1}\right]\frac{v^{j-1+b}}{(1+ve^{-u})(1+e^u)}.
	\label{traceintegral1}
	\end{align}
Inspecting the last two lines of this expression, we note that the pole structure is the same and that the two lines are related by $\xi \to -1-\xi$.
The $u$ integrals are done by residue theorem, which we evaluate in appendix \ref{app:u-integral}, and the result is given by
	\begin{align}
	\text{Tr}\Delta^{(\theta)} = \frac{\theta}{2j-1}\int_0^1 dv \Big(v^{j-1-b}\left[I_1(v)-I_2(v)\right] - v^{j-1+b}\left[I_3(v)-I_4(v)\right] \Big) \, ,
	\label{kernelTrace}
	\end{align}
where 
	\begin{align}
		I_1(v) - I_2(v) \, &= \, \frac{2\theta - (1-v)(\theta+1+2\xi)}{2(1-v)^2}+\frac{v^{\frac{1+\xi}{\theta}}}{(v^{\frac{1}{\theta}}-1)(1-v)} \, , \\
		I_3(v)-I_4(v) \, &= \, \frac{2\theta - (1-v)(\theta - 1 -2\xi)}{2(1-v)^2}+ \frac{v^{-\frac{\xi}{\theta}}}{(v^{\frac{1}{\theta}}-1)(1-v)} \, .
	\end{align}
Note that under $b\to -b, \xi\to -1-\xi$ this expression changes by a sign. Hence, it vanishes for $\xi = -1/2$ and $b=0$. To demonstrate that these integrals converge, we note that
	\begin{align}
	\lim_{v\to 1} \big( I_1-I_2 \big) \, = \, \lim_{v\to 1} \big( I_3-I_4 \big) \, = \, \frac{\theta^2 - 6\xi(1+\xi) -1}{12\theta} \, .
	\end{align}
Eq. (\ref{kernelTrace}) is an exact expression for the trace of the propagator for any $\theta$. However, for $\theta = 1/p$ for $p\in\mathbb{Z}$, eq. (\ref{kernelTrace}) can be computed in closed form, for instance the $\theta=1$ result of \cite{Lisovyy:2007mj}. The result is a sum over digamma functions. As an example, for $p=2$, 
	\begin{align}
	&(2j-1)\text{Tr}\Delta^{(\frac{1}{2})} \nn\\
	&\hspace{10pt}= \, \biggl\{\frac{1}{2}\left[\psi(\frac{j+b}{2}-\xi)+\psi(\frac{j-b}{2}+\xi)-\psi(\frac{1+j+b}{2}-\xi)-\psi(\frac{1+j-b}{2}+\xi)\right]\nn\\
	&\hspace{40pt}+\psi(j+b-2\xi)-\psi(j+b)+\psi(j-b)-\psi(j-b+2\xi)\nn\\
	&\hspace{40pt}+2(2\xi-b+j)\left[\psi(j-b)-\psi(j-b+2\xi)\right]\nn\\
	&\hspace{140pt}+2(2\xi-b+1-j)\left[\psi(j+b-2\xi)-\psi(j+b)\right] \biggr\}.
	\end{align}
In this expression, the last two lines are identical to \cite{Lisovyy:2007mj} with the substitution $\xi\to \xi/\theta$ and an overall factor of $1/\theta$. For any $\theta$ there will be such terms and they will dominate in the Schwarzian limit.
To obtain the density of states, we use
	\begin{equation}
	\rho^{(\theta)} \, = \, \frac{1}{\pi}\text{Im }\text{Tr}\Delta^{(\theta)} \, .
	\end{equation}
In particular, for the continuous principle series, since $j=is+1/2$, we look at the real part of the right hand side of  (\ref{kernelTrace}). Using the identity
	\begin{equation}
	\Im\left[\psi\left(\frac{1}{2}+i t \right) \right] \, = \, \frac{\pi}{2}\tanh(\pi t) \, ,
	\end{equation}
and identifying parameters in eq. (\ref{coneparameters}), we find that for $\theta=1$
	\begin{equation}
	\lim_{q\to\infty} \rho^{(1)} \, = \, 2qe^{-(2\pi-\alpha) q_\alpha}\cosh(2\pi s) \, ,
	\end{equation}
reproducing the earlier result.

\subsubsection{Partition function for $D(\a)$}
\label{sec:partitionfunctioncone}

The spectral density for the continuous mode in the Poincar\'e disk  with Aharonov-Bohm flux $\Phi = 2\pi\xi$ and magnetic field $b$ has been calculated by \cite{Lisovyy:2007mj} as
	\begin{align}
		\r_{s}^{(\xi)}(E, b) \, &\propto \, \frac{1}{s} \bigg[ \frac{s \sinh2\pi s + (\frac{1}{2}-b+\x)\sin2\pi (b-\x)}{\cosh2\pi s + \cos2\pi (b-\x)} \nn\\
		&\qquad\qquad\qquad\qquad\qquad \ - \, \frac{s \sinh2\pi s + (\frac{1}{2}-b+\x)\sin2\pi b}{\cosh2\pi s + \cos2\pi b} \bigg] \, ,
		\label{Lisovyydensityofstates}
	\end{align}
where $E = \frac{1}{2}(s^2+ \frac{1}{4}+b^2)$.
Then, the exact partition function is given by 
	\begin{equation}
		Z(\beta) \, = \, e^{S_0} e^{(2\pi - \a)q_\a} \int_{\frac{1}{2}( \frac{1}{4}+b^2)}^\infty dE \, e^{-\frac{\beta E}{\z^2}} \, \r_{s}^{(\xi)}(E, iq_\a) \, ,
	\end{equation}
where we included the background contributions from the Euler characteristic and ground state entropy as in \cite{Yang:2018gdb}.
The extra factor $\z^{-2}$ in the Boltzmann factor is explained as follows.
If we want the punctured disk partition function to arise as the leading correction to the JT partition function as in eq. (\ref{expanded-partition-func}), we must match $\beta$ between the disk and the punctured disk using eq. (\ref{beta}). In the coordinates we have been using, this means we must rescale $\beta_{D_\alpha} =  \beta_D/\zeta^2$.

The Schwarzian limit is taken by setting $b=iq_\alpha$, $\xi = \frac{iq_\alpha\alpha}{2\pi}$ and sending $q_\alpha \to \infty$ with $s$ fixed.
We find for the punctured disk,
	\begin{equation}
		\lim_{q_\alpha\to\infty}\r_{s,\alpha}^{(\x)}(E, iq) \, = \, \frac{2 q_\alpha \zeta}{s} \, e^{-(2\pi-\alpha) q_\alpha} \, \cosh2\pi s \, .
		\label{punctureddiskdensityofstates}
	\end{equation}
Including the background contributions from the Euler characteristic and ground state entropy, this leads to the gravitational density of states
	\begin{equation}
	\rho_{\a}(s) \, = \, e^{S_0}e^{(2\pi-\alpha)q_\alpha}\r_{\l,\a}^{(\x)}(\lambda) \, = \, 2q_\alpha\zeta e^{S_0} \, \frac{\cosh2\pi s}{s} \, .
	\end{equation}
As a final step, to bring the Boltzmann factor to the standard form, taking $s \to \zeta s$, we find the partition function in the Schwarzian limit is given by
	\begin{equation}
		Z(\b) \, = 4q_\alpha\zeta^2\, \int_0^{\infty} ds \, e^{- \frac{\b s^2}{2}} \cosh[(2\pi-\alpha)s]. 
	\label{punctureddiskpartitionfunction}
	\end{equation}
matching the results obtained via the other methods up to an overall constant.

We note that for the trumpet, the analytic continuation gives
	\begin{equation}
	Z(\beta) \, \propto \, \int_0^\infty ds \, e^{-\frac{\beta s^2}{2}}\cos[bs] \, ,
	\end{equation}
in agreement with \cite{Mertens:2019tcm,Saad:2019lba,Witten:2020wvy}.

Before completing the discussion of the partition function, there are three important points that we must address. The first point is that the partition function of a charged particle on the punctured disk contains in addition to the contribution arising from the AB field, a contribution arising from the background magnetic field. This has the same spectrum as (\ref{backgroundspectrum}), subject to the appropriate rescaling of the representation parameter $s$. However, in the large $q$ limit, this spectrum is subleading by $O(q^{-1})$ compared to the spectrum arising from the effect of the AB field.  Hence, it may be safely ignored in the large $q$ limit. Taking this a step further, the fact that the $\alpha\to 0$ limit does not smoothly connect the density of states of $D_\alpha$ to $D$ is a consequence of the fact that the AB field contribution is parametrically larger than the background field contribution which does connect smoothly. Had the AB field been an independent parameter, we could take $\xi\to 0$ in (\ref{Lisovyydensityofstates}) and see that $\rho_s^{(\xi)}\to 0$ and the two would smoothly connect.

 The second point is that while allowing for the AB contribution to dominate, the overall factor of $q$ in (\ref{punctureddiskdensityofstates}) seems at first a little troublesome. However, it can be absorbed into the definition of $\epsilon$ in (\ref{expanded-partition-func}) which must be sufficiently small to view the contribution of the punctured disk as a correction to to the leading order gravity answer arising from the disk. Finally, the third point is that the Schwarzian result neglects non-perturbative contributions that could arise from paths that intersect the puncture on the disk. As we discuss in appendix \ref{app:coneappendix}, it is reasonable to disregard these contributions if we remove this point from our manifold or similarly throw out any paths that intersect this point. Among other rationales for removing such paths, any geodesic intersecting this point is incomplete and new physics must be put in by hand to continue paths further.

\subsection{R\'enyi and von Neumann entropies in the Schwarzian limit}
Finally in this section, we study R\'enyi and von Neumann entropies for the JT gravity on the punctured disk ($D_\a$).
The $n$-th R\'enyi entropy $S_n$ is computed from the $n$-th R\'enyi partition function via
	\begin{equation}
		S_n \, \equiv \, \frac{1}{1-n}\log Z_n \, ,
	\end{equation}
where $Z_n$ is defined as the propagator on an $n$-sheeted replicated geometry with $\theta = n$. These entropies are determined for positive $n\in \mathbb Z$, but they can be analytically continued to real $n$ and in the limit $n\to 1$, they give the von Neumann or entanglement entropy:
	\begin{equation}
		S \, \equiv \, - \, \partial_n \left( \frac{\, Z_n \, }{Z_1^n} \right) \biggr|_{n=1} \, .
		\label{vonNeumannentropy}
	\end{equation} 
The entanglement entropy has been computed in pure JT gravity in \cite{Lin:2018xkj,Jafferis:2019wkd} and captures the contribution from the constant background field.
There, the trivial topology of the disk allowed for $Z_n(\b) = Z_1(n \b)$.
This is related to the fact that $\Pi^{(\theta)} = K^{(1)}$ in (\ref{kernelnoABfield-2}) since it arises from the pole contribution $w_1=w_2$ and doesn't see the branch cuts.

With the AB field, $Z_n$ is not as simple because of the non-trivial topology. We write
	\begin{equation}
	Z_n \, = \,  C_n \frac{e^{(2\pi-\alpha)q_\alpha}}{\pi}\int_0^\infty ds\, s\, e^{-n\beta_\alpha s^2} \text{Im} \left[\text{Tr}\, \Pi^{(n)}\right].
	\label{Zn}
	\end{equation}
We can obtain the topological contributions to $Z_n$ from the trace of the resolvent (\ref{kernelTrace}) via
	\begin{equation}
		Z^{(\rm AB)}_n \, = \,C_n \, \frac{e^{(2\pi-\alpha)q_\alpha}}{\pi}\int_0^\infty ds\, s\, e^{-n\beta_\alpha s^2} \, \text{Im}\left[\text{Tr}\Delta^{(n)}\right] \, .
	\label{ZABn}
	\end{equation}
To simplify notation, we will write
	\begin{equation}
		\rho^{(\alpha)}_{n} \, = \, \frac{e^{(2\pi-\alpha)q_\alpha}}{\pi} \, \text{Im}\left[\text{Tr}\Delta^{(n)}\right] \, .
	\end{equation}
As discussed earlier $\beta_\alpha = \beta/\zeta^2$.
$C_n$ is a numerical constant that comes from appropriately matching the $\epsilon$ expansion in (\ref{expanded-partition-func}). From now on we will set $C_n = 1$.
The background field contribution is given by the part of eq.~(\ref{Zn}) coming from $K^{(n)}=\Pi^{(1)}$.
However, with an AB field, in the Schwarzian limit limit $q\to \infty$, this is subleading to the topological contribution.
Now we evaluate eq. (\ref{vonNeumannentropy}). It will be useful to look at two separate contributions: : $Z^{(\rm AB)}_n= Z^{(\rm AB1)}_n + Z^{(\rm AB2)}_n$.
The first is similar to the R\'enyi partition function of the disk JT gravity
	\begin{align}
	Z^{({\rm AB}1)}_n \, &= \, \int_0^\infty ds\, s\, e^{-n\beta_\alpha s^2}\rho^{(\alpha)}_1\nn\\
	&= \, Z_1^n \int_0^\infty ds\, s\, \Big( \rho^{(\alpha)}_1 \Big)^{1-n} \Big( Z_1^{-1}\rho^{(\alpha)}_1 e^{-\beta_\alpha s^2} \Big)^n \, ,
	\end{align}
leading to the its contribution to the entanglement entropy
	\begin{equation}
		S_1 \, = \, \frac{1}{Z_1}\int_0^\infty ds\, s\, e^{-\beta_\alpha s^2} \, \rho^{(\alpha)}_1 \, \Big(\log Z_1 + \beta_\alpha s^2 \Big) \, .
	\end{equation}
As in \cite{Lin:2018xkj}, this can be written as
	\begin{equation}
		S_1 \, = \, \int_0^\infty ds\, s\, p_s \Big[-\log p_s + \log(\text{dim} R) \Big] \, ,
	\end{equation}
where
	\begin{equation}
		p_s \, = \, \frac{\text{dim} R}{Z_1} \, e^{-\beta_\alpha s^2} \, , \qquad \text{dim} R \, = \, \rho^{(\alpha)}_1 \, .
	\end{equation}
The other accounts for the non-trivial contribution from the conical defect as
	\begin{equation}
		Z^{({\rm AB}2)}_n \, = \, \int_0^\infty ds\,s\, e^{-n\beta_\alpha s^2} \Big( \rho^{(\alpha)}_n-\rho^{(\alpha)}_1\Big) \, .
	\end{equation}
The entanglement entropy from this contribution is
	\begin{equation}
		S_2 \, = \frac{1}{Z_1}\int_0^\infty ds\, s\, e^{-\beta_\alpha s^2} \Big( -\partial_n \rho^{(\alpha)}_n \Big)_{n=1} \, .
	\end{equation}
Using the trace of the resolvent (\ref{kernelTrace}), we start with
	\begin{align}
		\partial_n \big( I_1-I_2 \big) \Bigr|_{n=1} \, &= \, \frac{1-v^2 +2v^{1+\xi}\left[1+(1-v)\xi\right]\log v}{2(1-v)^3} \, , \nn\\
		\partial_n \big( I_3-I_4 \big) \Bigr|_{n=1} \, &= \, \frac{1-v^2 -2v^{-\xi}\left[-v + (1-\nu)\xi\right]\log v}{2(1-v)^3} \, .
	\end{align}
The integrals in the resolvent (\ref{kernelTrace}) can be found in closed form and can again be written in terms of digamma functions.
Here we just write the Schwarzian limit, $q\to\infty$,
	\begin{equation}
	\Big[-\partial_n \rho^{(\alpha)}_n \Big]_{n=1} \, = \, q_\alpha\frac{1+\zeta}{s}\cosh(2\pi s) \, = \, \frac{1+\zeta}{2} \, \rho_1^{(\alpha)} \, .
	\end{equation}
The extra contribution is hence proportional to $\dim R$.\footnote{Here, we must perform the same rescaling of $s$ as in eq.~(\ref{punctureddiskpartitionfunction}).}

Combining both results, we find
	\begin{equation}
	S \, = \, S_1 \, + \, S_2 \, = \, \frac{1+\zeta}{2} \, + \, \int_0^\infty ds\, s\, p_s \big[-\log p_s + \log(\text{dim}R) \big] \, .
	\label{totalentropy}
	\end{equation}
Recall that we cannot take $\zeta\to 1$ to obtain the disk result since we must also rescale $q_\alpha$. Nevertheless, the result looks very similar to that of \cite{Lin:2018xkj}. In particular, the $\log(\text{dim} R)$ term gives rise to the semi-classical entanglement entropy of the Hartle-Hawking state. Evaluating $S_1$ contribution in the semi-classical limit,
	\begin{equation}
	S_{HH} \, = \, \log \, \cosh\big[(2\pi-\alpha) s\big]_{s = \frac{2\pi-\alpha}{\beta}} \, = \, \frac{(2\pi-\alpha)^2}{\beta} \, = \, (1-\beta\partial_\beta)\log Z_{D_\alpha}(\beta) \, .
	\end{equation}
Finally, the constant contribution from $S_2$ does not depend on $s$ and is therefore a global IR contribution. It can therefore be interepreted as a topological entanglement entropy \cite{Kitaev:2005dm}. A surprising feature of this topological entropy is that at low temperatures, it dominates the classical contribution to the entropy. It would be interesting to understand this deeper.

\section{$SL(2,\mathbb{Z})$ Black Holes}
\label{sec:sl(2,z) bh}
In \cite{Maxfield:2020ale}, it was shown that the $SL(2,\mathbb{Z})$ black holes of pure AdS$_3$ gravity are reduced to the JT gravity
with a defect in the S-wave dimensional reduction in the near-extremal limit.
Therefore in this section, we will further investigate the near-extremal limit of the $SL(2,\mathbb{Z})$ black holes.

Because of the absence of bulk propagating degrees of freedom in three-dimensional pure gravity with negative cosmological constant, 
all classical geometries are locally isometric to empty AdS$_3$.
In particular with torus boundary, this allows us to completely classify all classical solutions \cite{Maloney:2007ud}.
These solutions are characterized by the $SL(2,\mathbb{Z})$ mapping group of the boundary torus; hence, called the `$SL(2,\mathbb{Z})$ black holes'.
To be explicit, we take complex coordinates $(z,\bar{z})$ to label the boundary torus 
	\begin{equation}
		z \, = \, \frac{\vp + i t_E}{2\pi} \, , \qquad z \, \approx \, z+1 \, \approx \, z + \t \, , \qquad \t \, = \, \frac{\th + i \b}{2\pi} \, ,
	\end{equation}
where $\b$ is the inverse temperature and $\th$ is the chemical potential for the angular momentum.
Then, the modular group of the boundary torus acts on the modular parameter $\t$ as
	\begin{equation}
		\t \, \mapsto \, \t' \, = \, \frac{a \t + b}{c \t + d} \, , \qquad \quad \g \, \equiv \,
		\begin{pmatrix}
			a & b \\
			c & d
		\end{pmatrix}
		\, \in \, PSL(2,\mathbb{Z})/\mathbb{Z} \, ,
	\end{equation}
where we required $ad-bc=1$

The $SL(2,\mathbb{Z})$ black hole solutions \cite{Maloney:2007ud} are given by 
	\begin{equation}
		\ell_3^{-2} ds_3^2 \, = \, - f(r) dt^2 \, + \, \frac{dr^2}{f(r)} \, + \, r^2 \left( d\vp - \frac{r_+ r_-}{r^2} dt \right)^2 \, , 
	\label{metrix-lorentzian}
	\end{equation}
where $\ell_3$ is the AdS$_3$ radius, $r_{\pm}$ are the outer and inner horizon locations, with 
	\begin{equation}
		f(r) \, = \, \frac{(r^2 - r_+^2)(r^2 - r_-^2)}{r^2} \, .
	\end{equation}
The Euclidean time is obtained by Wick rotation: $t \to i t_E$ which leads to 
	\begin{equation}
		\ell_3^{-2} ds_3^2 \, = \, f(r) dt_E^2 \, + \, \frac{dr^2}{f(r)} \, + \, r^2 \left( d\vp - \frac{i\,  r_+ r_-}{r^2} dt_E \right)^2 \, , 
	\end{equation}
where 
	\begin{equation}
		(t_E, \vp) \, \simeq \, (t_E, \vp + 2\pi) \, \simeq \, (t_E+\b, \vp + \th) \, ,
	\end{equation}
with
	\begin{equation}
		\b \, = \, \frac{2\pi}{c} \frac{r_+}{r_+^2 - r_-^2} \, , \qquad \th \, = \, - 2\pi \frac{d}{c} \, + \, \frac{2\pi i}{c} \frac{r_-}{r_+^2 - r_-^2} \, .
	\label{SL2zbetathetadef}
	\end{equation}

We note that by defining a near-horizon coordinate as $r-r_+ = \frac{r_+^2 - r_-^2}{2r_+} \r^2$, the near-horizon geometry is given by
	\begin{equation}
		\ell_3^{-2} ds_2^2 \, = \, d\r^2 + \frac{\r^2}{c^2} \left( \frac{2\pi}{\b} \, dt_E \right)^2 \, ,
	\end{equation}
where we omitted the third direction.
Since the period of $t_E$ is $\b$, for $c>1$ the geometry close to $\r=0$ has a conical defect with opening angle $2\pi/c$ \cite{Maxfield:2020ale}.

\subsection{Lyapunov exponent}
Before we move on to the dimensional reduction of the $SL(2,\mathbb{Z})$ black holes, we consider out-of-time-order correlators (OTOC) for this background.
The OTOC characterizes quantum chaos \cite{Shenker:2013pqa} and its growing rate in time is represented by the Lyapunov exponent $\l_L$.
This exponent has a sharp upper bound \cite{Maldacena:2015waa}
	\begin{equation}
		\l_L \, \le \, \frac{2\pi}{\b} \, .
	\label{lambda_L bound}
	\end{equation}
In this section, we show that the Lyapunov exponent of the $SL(2,\mathbb{Z})$ black hole is lower than the upper bound (\ref{lambda_L bound}), being instead given by $\l_L=2\pi /(\b c)$.

First, since we will be interested in the near-horizon geometry, it is more convenient to work with co-rotating coordinates defined by
	\begin{equation}
		\p \, \equiv \, \vp \, - \, \O \, t \, , \qquad \quad \O \, = \, \frac{r_-}{r_+} \, .
	\end{equation}
where $\O$ is the angular velocity of the outer horizon.
For these coordinates, the metric is written as
	\begin{equation}
		\ell_3^{-2} ds_3^2 \, = \, - f(r) dt^2 \, + \, \frac{dr^2}{f(r)} \, + \, r^2 \Big( N^{\p}(r) dt + d\p \Big)^2 \, , 
	\end{equation}
where
	\begin{equation}
		N^{\p}(r) \, = \, \frac{r_-}{r_+} \frac{r^2 - r_+^2}{r^2} \, . 
	\end{equation}
Next for the right exterior region, we introduce Kruskal coordinates by
	\begin{equation}
		U \, = \, - \, e^{-\k (t- r_*)} \, , \qquad V \, = \, e^{\k (t+ r_*)} , \qquad \k \, = \, \frac{r_+^2 - r_-^2}{r_+} \, ,
	\end{equation}
and the tortoise coordinate $r_*$ is given by \cite{Jahnke:2019gxr}
	\begin{equation}
		r_* \, = \, \frac{1}{2\k} \, \log\left( \frac{\sqrt{r^2 - r_-^2} - \sqrt{r_+^2 - r_-^2}}{\sqrt{r^2 - r_-^2} + \sqrt{r_+^2 - r_-^2}} \right) \, .
	\end{equation}
In the Kruskal coordinates, the metric takes the form 
	\begin{equation}
		\ell_3^{-2} ds_3^2 \, = \, \frac{- 4dU dV - 4 r_- (U dV - V dU) d\p + [(1-UV)^2 r_+^2 + 4UV r_-^2] d\p^2}{(1+UV)^2} \, .
	\end{equation}
To study the butterfly effect, we consider releasing a particle from the boundary of AdS at a time $t$ in the past.
Then for late times (i.e. $t \gg \k$) the energy density of this particle in Kruskal coordinates is exponentially boosted and the boost factor is given by
	\begin{equation}
		E \, \sim \, E_0 \, e^{\k t} \, ,
	\end{equation}
where $E_0$ is the initial asymptotic energy of the particle.
This exponential boost gives the Lyapunov exponent as
	\begin{equation}
		\l_L \, = \, \k \, = \, \frac{2\pi}{c \b} \, .
	\label{lambda_L}
	\end{equation}

We can also explicitly compute the out-of-time-order correlators and identity the Lyapunov exponent.
The OTOC of the BTZ black hole background was explicitly computed by using eikonal approximation in \cite{Shenker:2014cwa} and 
later, the eikonal approximation method was also applied for the rotating BTZ black holes in \cite{Jahnke:2019gxr}.
Our derivation of the OTOC for the $SL(2,\mathbb{Z})$ black holes is almost identical to that of \cite{Jahnke:2019gxr}, so we will just summarize main points.
In order to study the out-of-time-order correlators (OTOC)
	\begin{equation}
		F(t, \vec{x}) \, = \, \big\la V(0) W(t, \vec{x}) V(0) W(t, \vec{x}) \big\ra \, ,
	\end{equation}
where the expectation value is evaluated on the thermofield double states, we consider all the operators acting on the right side of the geometry.
Two-sided configurations can be obtained by analytical continuation of times from this correlator \cite{Shenker:2014cwa}.
The operators $V$ and $W$ introduce particles in the bulk, which are called the $V$-particle and the $W$-particle, respectively.
We focus on the $W$-particle traveling along the $U=0$ at the angle $\p_W$.
For large momentum along the $V$ direction, $p^V$, the energy-momentum tensor of the particle takes the form
	\begin{equation}
		T_{UU} \, = \, \frac{A_0}{r_+} \, p^V \d(U) \d(\p - \p_W) \, .
	\end{equation}
Then, the resulting metric can be written in the form
	\begin{equation}
		\ell_3^{-2} ds_3^2 \, = \, - 2 A(UV) dU dV \, + \, r^2(U V) d\p^2 \, + \, h_{UU} dU^2 \, .
	\end{equation}
where $A_0 \equiv A(0)$ and 
	\begin{equation}
		h_{UU} \, = \, 16 \pi G_N \, r_+ A_0 \, p^V \d(U) \, h(\p - \p_W) \, .
	\end{equation}
The angular profile $h(\p)$ can be fixed by the linearized Einstein's equation:
	\begin{equation}
		h''(\p) \, - \, 2 r_- \, h'(\p) \, - \, (r_+^2 - r_-^2) h(\p) \, = \, \d(\p) \, ,
	\end{equation}
and the general solution of the above equation has the form
	\begin{equation}
		h(\p) \, = \, c_1 \, e^{(r_+ + r_-) \p} \, + \, c_2 \, e^{-(r_+-r_-) \p} \, ,
	\end{equation}
where $c_1$ and $c_2$ are constants.

In the limit $\D_W \gg \D_V \gg 1$, where $\D_{W, V}$ are the dimensions of the $W, V$ operators, 
the elastic eikonal gravity approximation leads to the OTOC \cite{Jahnke:2019gxr}
	\begin{align}
		\frac{\big\la V(i \e_1) W(t+i \e_2) V(i \e_3) W(t+i\e_4) \big\ra}{\big\la V(i \e_1) V(i \e_3) \big\ra\big\la W(i \e_2) W(i\e_4) \big\ra} 
		\, &= \, \left( 1 + \frac{16 \pi i G_N \D_W}{\e_{13} \e_{24}^*} \, e^{\k t} \, h(\p_{2}) \right)^{\D_V} \nn\\
		\, &\approx \, 1 \, + \, \frac{16 \pi i G_N \D_V \D_W }{\e_{13} \e_{24}^*} \, e^{\k t} \, h(\p_{2}) \, .
	\end{align}
We can now identify the Lyapunov exponent: $\l_L=2\pi /(\b c)$ as in (\ref{lambda_L})
We note that the Lyapunov exponent is lower than the upper bound (\ref{lambda_L bound}) by the factor $c^{-1}$.

\subsection{Partition functions}
\label{sec:partition functions}
In this section, we study the partition function of the $SL(2,\mathbb{Z})$ black hole and its behavior in the near-extremal limit.
The partition function of the $SL(2,\mathbb{Z})$ black hole is given by 
	\begin{equation}
		Z_{c,d}(\t) \, = \, \c_{\mathds{1}}(\t') \c_{\mathds{1}}(- \bar{\t}') \, , 
	\end{equation}
where 
	\begin{equation}
		\t' \, \equiv \, \g \t \, = \, \frac{a \t + b}{c \t + d} \, , \qquad \g \, = \,
		\begin{pmatrix}
			a & b \\
			c & d
		\end{pmatrix}
	\end{equation}
and 
	\begin{equation}
		\c_{\mathds{1}}(\t) \, = \, \frac{(1-q)q^{-k+\frac{1}{24}}}{\h(\t)} \, , \qquad \h(\t) \, = \, q^{\frac{1}{24}} \prod_{n=1}^{\inf} (1-q^n) \, , \qquad q \, = \, e^{2\pi i \t} \, .
	\label{chi1}
	\end{equation}
Since $({\rm Im}{\t})^{1/4} \, \h(\t)$ is modular invariant, it is sometimes convenient to express \cite{Maloney:2007ud}
	\begin{equation}
		\c_{\mathds{1}}(\t') \, = \, \frac{1}{({\rm Im}{\t})^{\frac{1}{4}} \h(\t)} \left[ ({\rm Im}{\t'})^{\frac{1}{4}} \, (1-q')q'^{-k+\frac{1}{24}} \right] \, ,
	\label{chi2}
	\end{equation}
where $q' = e^{2\pi i \t'}$.

\subsubsection{BTZ Black Hole ($c=1$, $d=0$)}
We first summarize the BTZ case, which was discussed in \cite{Ghosh:2019rcj}.
The BTZ black hole is given by 
	\begin{equation}
		\g \, = \,
		\begin{pmatrix}
			0 & -1 \\
			1 & 0
		\end{pmatrix}
		\, , \qquad \t' \, = \, - \frac{1}{\t} \, .
	\end{equation}
The near-extremal limit is obtained by setting 
	\begin{equation}
		\t \, = \, \frac{i \b_L}{2\pi} \, , \qquad \bar{\t} \, = \, - \, \frac{i \b_R}{2\pi} \, ,
	\label{beta's}
	\end{equation}
and then taking the limit
	\begin{equation}
		\b_L , k \, \gg \, 1 \, \gg \, \b_R \, .
	\label{ext-limit}
	\end{equation}
For the holomorphic part, it is more convenient to use (\ref{chi2}) to get
	\begin{equation}
		\c_{\mathds{1}}(\t') \, \simeq \, 2\pi \left( \frac{2\pi}{\b_L} \right)^{\frac{3}{2}} \, \exp\left[ \frac{\b_L}{24}+\frac{4\pi^2 k}{\b_L} \right] \, ,
	\end{equation}
while for the anti-holomorphic part, it is more convenient to use the original expression (\ref{chi1}) to find
	\begin{equation}
		\c_{\mathds{1}}(-\bar{\t}') \, = \, e^{\frac{4\pi^2 k}{\b_R}} \prod_{n=2}^\inf \left( 1 - e^{-\frac{4\pi^2 n^2}{\b_R}} \right)^{-1} \nn\\
		\, \simeq \, e^{\frac{4\pi^2 k}{\b_R}} \, .
	\end{equation}
Therefore, we obtain the near-extremal limit of the BTZ black hole partition function
	\begin{equation}
		Z_{1,0}(\b_L, \b_R) \, \simeq \, 2\pi \left( \frac{2\pi}{\b_L} \right)^{\frac{3}{2}} \, \exp\left[ \frac{\b_L}{24}+\frac{4\pi^2 k}{\b_L} +\frac{4\pi^2 k}{\b_R} \right] \, .
	\end{equation}
As explained in \cite{Ghosh:2019rcj}, the left-moving sector of the above partition function agrees with the disk JT gravity partition function $Z_D$ (\ref{Z})
by identification of the temperature $1/\b = 2k/\b_L$ together with extracting the extremal contribution $e^{\b_L/24}$.

\subsubsection{$SL(2,\mathbb{Z})$ Black Holes ($c>1$)}
Next we consider generic $SL(2,\mathbb{Z})$ black holes ($c>1$), which are parametrized by 
	\begin{equation}
		\g \, = \,
		\begin{pmatrix}
			a & b \\
			c & d
		\end{pmatrix}
		\, , \qquad \t' \, = \, \frac{a \t + b}{c \t + d} \, .
	\end{equation}
We again use (\ref{beta's}) and take the near-extremal limit (\ref{ext-limit}).
In this case, in the near-extremal limit we have 
	\begin{align}
		\t' \, &= \, \frac{a}{c} \, + \, \frac{2\pi i}{c^2 \b_L} \, + \, \mathcal{O}(\b_L^{-2}) \, , \\
		- \bar{\t}' \, &= \, - \frac{b}{d} \, + \, \frac{i \b_R}{2\pi d^2} \, + \, \mathcal{O}(\b_R^2) \, . 
	\label{nearextremaltau}
	\end{align}
Using these expansions, we find
	\begin{align}
		\c_{\mathds{1}}(\t') \, \simeq \, \left( \frac{2\pi}{c \b_L} \right)^{\frac{1}{2}} \,
		\exp\left[ \frac{\b_L}{24} + \frac{4\pi^2k}{c^2\b_L} \right] \left( 1 - e^{-\frac{2\pi i a}{c}} \right) e^{-\frac{2\pi i a k}{c}}\, ,
	\end{align}
and $\c_\mathds{1}(-\bar{\t}') \simeq \c_\mathds{1}(-b/d)$.
Now for this case ($c>1$) the left-moving sector agrees with the punctured disk JT gravity partition function $Z_{D_\a}$ (\ref{Z})
with the same identification of the temperature and the extremal contribution as in the BTZ case above.
For this case, the deficit angle is identified by $2\pi /c = 2\pi -\a$.

\subsection{Spectral densities}
\label{sec:spectral densities}
Next, we study the spectral density of the $SL(2,\mathbb{Z})$ black hole and its behavior in the near-extremal limit.
In \cite{Benjamin:2020mfz}, a Virasoro character
	\begin{equation}
		\c_h(\t) \, = \, \frac{q^{h-k+\frac{1}{24}}}{\eta(\t)} \, ,
	\end{equation}
was expressed in terms of the modular crossing kernel $\mbK^{(\g)}$ as
	\begin{equation}
		\c_h(\t') \, = \, \int_{k - \frac{1}{24}}^{\inf} dh' \, \mbK^{(\g)}_{h',h} \, \c_{h'}(\t) \, ,
	\end{equation}
where
	\begin{equation}
		\mbK^{(\g)}_{h',h} \, = \, \e(\g) \sqrt{\frac{2}{c}} \, e^{\frac{2\pi i}{c} \big( a (h-k+\frac{1}{24}) + d (h'-k+\frac{1}{24})\big)} \, 
		\frac{\cos\Big(\frac{4\pi}{c} \sqrt{(h-k+\frac{1}{24})(h'-k+\frac{1}{24})}\Big)}{\sqrt{(h'-k+\frac{1}{24})}} \, ,
	\end{equation}
with some $h$, $h'$-independent phase factor $\e(\g)$.

In the previous section, we have seen that the partition function of the Schwarzian theory is obtained by the near-extremal limit of the holomorphic sector.
Therefore, we focus on the holomorphic sector.
First we note that the untransformed Virasoro character leads to the Boltzmann factor:
	\begin{equation}
		\c_h(\t) \, = \, \c_h\left( \frac{i \b_L}{2\pi} \right) \, \simeq \, e^{-\b_L (h-k)} \, .
	\end{equation}
Since the vacuum character is obtained by subtracting the null state: $\c_{\mathds{1}}(\t') = \c_{0}(\t') - \c_{1}(\t')$, now we find
	\begin{align}
		\c_{\mathds{1}}(\t') \, &= \, \int_{k-\frac{1}{24}}^{\inf} dh \, \c_h(\t) \Big( \mbK^{(\g)}_{h,0} - \mbK^{(\g)}_{h,1} \Big) \nn\\
		&\simeq \, \int_0^{\inf} dE \, e^{-\b_L E} \Big( \mbK^{(\g)}_{E+k,0} - \mbK^{(\g)}_{E+k,1} \Big) \nn\\
		&= \, \int_0^{\inf} ds \, e^{-\b_L s^2/2} \r^{(\g)}(s) \, ,
	\label{chi_1}
	\end{align}
where we used $E=h-k=s^2/2$ and the spectral density is given by
	\begin{equation}
		\r^{(\g)}(s) \, = \, s \Big( \mbK^{(\g)}_{\frac{s^2}{2}+k,0} - \mbK^{(\g)}_{\frac{s^2}{2}+k,1} \Big) \, .
	\end{equation}

\subsubsection{BTZ Black Hole ($c=1$, $d=0$)}
First we consider the near-extremal limit of the BTZ black hole, which has already been discussed in \cite{Ghosh:2019rcj}.
In order to take the near-extremal limit, it is convenient to introduce a new set of parameters following from the Liouville CFT description,
	\begin{equation}
		P^2 \, = \, h - k + \frac{1}{24} \, , \qquad k \, = \, \frac{1}{24} + \frac{Q^2}{4} \, , \qquad Q \, = \, u^{-1} + u \, .
	\label{Q}
	\end{equation}
By writing $\mbK^{(\g)}_{P',P} \equiv \mbK^{(\g)}_{h',h}$, we find 
	\begin{equation}
		\mbK^{(c=1,d=0)}_{P',P} \, = \,  \frac{\sqrt{2} \, \e(\g)}{P'} \, \cos\Big( 4\pi P P' \Big) \, .
	\end{equation}
Since $h=0$ ($h=1$) corresponds to $P=\frac{i}{2}(u^{-1}+u)$ ($P=\frac{i}{2}(u^{-1}-u)$), the spectral density is
	\begin{equation}
		\r^{(c=1,d=0)}(P') \, = \, 4 \, \e(\g) \, \sinh(2\pi P' u) \sinh(2\pi P' / u) \, .
	\end{equation}
The near-extremal limit is given by taking 
	\begin{equation}
		u \, \to \, 0 \, , \qquad P' \, \to \, 0 \, , \qquad {\rm with} \quad P'/u \, = \, s \, : \, {\rm fixed} \, .
	\label{ext-limit(2)}
	\end{equation}
In this limit, the spectral density becomes
	\begin{equation}
		\r^{(c=1,d=0)}(s) \, \propto \, s \sinh(2\pi s) \, .
	\end{equation}
This agrees with the spectral density of the disk JT gravity $\r_D$ (\ref{rho}).

\subsubsection{$SL(2,\mathbb{Z})$ Black Holes ($c>1$)}
Next we consider the general $SL(2,\mathbb{Z})$ black holes.
For this case, we have
	\begin{equation}
		\mbK^{(\g)}_{P',P} \, = \, \e(\g) \sqrt{\frac{2}{c}} \frac{1}{P'} \, e^{\frac{2\pi i}{c} (aP^2 + d P'^2)}\, \cos\Big( 4\pi P P' / c \Big) \, .
	\end{equation}
Subtraction of the null state is given by 
	\begin{align}
		&\quad \ \mbK^{(\g)}_{P',P=\frac{i}{2}(u^{-1}+u)} \, - \, \mbK^{(\g)}_{P',P=\frac{i}{2}(u^{-1}-u)} \\
		&= \, \e(\g) \sqrt{\frac{2}{c}} \frac{1}{P'} \, e^{\frac{2\pi id}{c} P'^2}
		\bigg[ 2 \left( e^{-\frac{i\pi a}{2c} (u^{-2}+u^2)} + \cos\left( \tfrac{\pi a}{c} \right) \right) \sinh\left(\tfrac{2\pi P' u}{c} \right) \sinh\left(\tfrac{2\pi P'}{c u} \right) \nn\\
		&\hspace{200pt} - 2i \sin\left( \tfrac{\pi a}{c} \right) \cosh\left(\tfrac{2\pi P' u}{c} \right) \cosh\left(\tfrac{2\pi P'}{c u} \right) \bigg] \, . \nn
	\end{align}
In the near-extremal limit (\ref{ext-limit(2)}), the factor $e^{-\frac{i\pi a}{2c} u^{-2}}$ rotates rapidly; therefore we set it to zero, 
while $e^{\frac{2\pi id}{c} P'^2} \to 1$ and $e^{-\frac{i\pi a}{2c} u^2} \to 1$.
Since $\sinh(2\pi P' u) \to  \mathcal{O}(P' u)$ and $\cosh(2\pi P' u) \to 1$, the leading contribution is given by the $\cosh(2\pi P' u)$ term.
This leads to the near-extremal limit of the spectral density 
	\begin{equation}
		\r^{(\g)}(s) \, \propto \, \cosh\left( \frac{2\pi s}{c}\right) \, .
	\label{rho^gamma}
	\end{equation}
This agrees with the Schwarzian theory with defect (\ref{rho}), where $\z = 1/c$.
We note that for the BTZ black hole (for which $a=0$), this leading $\cosh(2\pi P' u)$ does not exist.
As an aside, returning to eq.~(\ref{nearextremaltau}), we see that the temperatures of the $c\neq 1$ black holes are related to the $c=1$ black holes via $\beta^{c}_L = c^2\beta_L^{c=1}$. This is the same matching we used in eq.~(\ref{punctureddiskpartitionfunction}). Furthermore, there is an overall factor of $1/P'$ which is analogous to the overall factor of $q$ in eq.~(\ref{punctureddiskpartitionfunction}).

\subsubsection{Topological entanglement entropy}
In \cite{McGough:2013gka}, the Bekenstein-Hawking entropy of a BTZ black hole was related to a topological entanglement entropy via the relation
	\begin{equation}
		S_{BH} \, = \, \lim_{P_+\to\infty, \; P_-\to \infty}\log\left(S_0^{P_+}S_0^{P_-}\right) \, = \, \frac{2\pi r_+}{4} = S_{BH,0}+\frac{2\pi^2}{\beta} \, ,
	\label{BekensteinHawkingEntropy}
	\end{equation}
where
	\begin{equation}
		S_0^{P_\pm} \, = \, \mbK^{(c=1,\; d= 0)}_{P_\pm,P=\frac{i}{2}(u^{-1}+u)} \, - \, \mbK^{(c=1,\;d=0)}_{P_\pm,P=\frac{i}{2}(u^{-1}-u)} \, .
	\end{equation}
Our results are relevant to the near-extremal BTZ case, in which case eq.~(\ref{BekensteinHawkingEntropy}) comes from appropriately taking the classical limit of our Schwarzian result: first, take the Schwarzian limit
	\begin{equation}
		S_0^{P_\pm} \, = \, s_\pm \sinh(2\pi s_\pm) \, ,
	\end{equation}
then look at states with $s_\pm\gg 1$ with $s_+ + s_- = r_+/4$. Following the same derivation, for the $SL(2,\mathbb{Z})$ black holes, we find
 	\begin{equation}
 		S_{BH} \, = \, \frac{2\pi (s_++s_-)}{c} = \frac{2\pi r_+}{4} \, = \, S_{BH,0}+\frac{2\pi^2}{c^2\beta_L} \, .
 	\end{equation}
Here, we used that $r_+ = 4(s_++s_-)/c$ which follows from eq.~(\ref{SL2zbetathetadef}) and $\beta_L^{c} = c^2\beta_L^{c=1}$. The result is that at the semi-classical order, the $SL(2,Z)$ black holes have the same Bekenstein-Hawking entropy as the BTZ black hole. They will differ at subleading order. The difference from the extremal entropy, $S_{BH,0}$ matches the results we found from the charged particle picture, see also \cite{Engelsoy:2016xyb, Lin:2018xkj}. It would be interesting if we could also see the topological entropy contribution of eq.~(\ref{totalentropy}) from this perspective.

\section{Schwarzian Correlation Functions}
\label{sec:schwarzian}
In section \ref{sec:charged}, we used the charged particle picture to study the dynamics of JT gravity. As the boundary cutoff $\rho_0\to\infty$, the gravitational degree of freedom of the JT gravity is reduced to the Schwarzian theory (\ref{S[f]}) which has a residual gauge symmetry that depends on the spacetime manifold. This theory can be studied directly in the Schwarzian limit.
For the Poincare disk ($D$) the dynamics is reduced to the Schwarzian theory on the Diff($S^1$)/$SL(2,\mathbb{R})$ Virasoro coadjoint orbit \cite{Stanford:2017thb},
while the dynamics of the punctured disk ($D_\a$) is reduced to the Schwarzian theory on the Diff($S^1$)/$U(1)$ orbit \cite{Mertens:2019tcm}.
In this section, we study the latter theory perturbatively in the Schwarzian coupling.
The discussion of this section is mostly parallel to the study of the Diff($S^1$)/$SL(2,\mathbb{R})$ orbit case \cite{Maldacena:2016upp},
but for completeness we exhibit all discussions explicitly here.

From (\ref{S[f]}) making the identification $\vp(u) = 2\pi \z f(u)/\b$, where $f(u)$ is a monotonically increasing function of $u$ with $f(u+\b) = f(u) + \b$, 
the Schwarzian theory on the Diff($S^1$)/$U(1)$ orbit is described by the action
	\begin{equation}
		S_{\z}[f] \, = \, - \, C \int_0^\b du \, \left\{ \tan\left( \frac{\pi \z f(u)}{\b} \right) \, , \, u \right\} \, , 
	\label{S_theta}
	\end{equation}
where $C$ is the Schwarzian coupling constant with dimension of length and the Schwarzian derivative is defined by
	\begin{equation}
		 \left\{ F(u) \, , \, u \right\} \, = \, \frac{F'''(u)}{F'(u)} \, - \, \frac{3}{2} \left( \frac{F''(u)}{F'(u)} \right)^2 \, .
	\end{equation}
For large $C$ the saddle-point solution is given by $f(u)=u$.
Therefore, taking an infinitesimal fluctuation around this saddle by $f(u) = u + \ve(u)$, we obtain the quadratic action of $\ve(u)$ as
	\begin{align}
		S_{\z}[\ve] \, &= \, \frac{C}{2} \int_0^\b du \left[ \ve'(u)^2 - \left( \frac{2\pi \z}{\b} \right)^2 \ve'(u)^2\right] \nn\\
		&= \, \pi C \left( \frac{2\pi}{\b} \right)^3 \, \sum_{n \in \mathbb{Z}} \, n^2 (n^2 - \z^2) \, \ve_n \ve_{-n} \, ,
	\end{align}
where we used the Fourier mode expansion $\ve(\t) = \sum_{n \in \mathbb{Z}} e^{\frac{2\pi i}{\b} n u} \, \ve_n$.
Therefore, the bare two-point function of the Schwarzian mode is given by 
	\begin{equation}
		\big\la \ve_n \ve_{-n} \big\ra_\z \, = \, \frac{1}{2\pi C} \, \frac{1}{n^2(n^2 - \z^2)} \, .
	\end{equation}
Here and henceforth, we set $\b = 2\pi$ to simplify the expressions.
We note that the $n=0$ mode is the zero mode, while for $\zeta\neq 1$, $n = \pm 1$ are no longer zero modes, unlike the Diff($S^1$)/$SL(2,R)$ orbit case \cite{Stanford:2017thb}.
Fourier transforming back with excluding the zero mode, we obtain
	\begin{equation}
		\big\la \ve(u) \ve(0) \big\ra_\z \, = \, \frac{1}{2\pi \z^2 C} \, \bigg[ - \frac{1}{2} \, (|u| - \pi)^2 + \frac{\pi^2}{6} + \z^{-2}
		- \frac{\pi}{\z} \, \cot( \pi \z ) \cos( \z u ) - \frac{\pi}{\z} \, \sin| \z u | \bigg] \, .
	\label{<ee>}
	\end{equation}
We show a derivation of this expression in Appendix \ref{app:2pt}.

\subsection{Schwarzian contribution to the four-point functions}
Now we suppose to couple the Schwarzian theory to a conformal operator $\mathcal{O}$ with conformal dimension $\D$.
We would like to study the Schwarzian mode contribution in the leading order of the Schwarzian coupling $C$ for the four-point functions of $\mathcal{O}$ \cite{Maldacena:2016upp}.
To this end, we start from the disconnected contribution of the four-point function in the $s$-channel
$\la \mathcal{O}(u_1) \mathcal{O}(u_2) \mathcal{O}(u_3) \mathcal{O}(u_4) \ra = |u_{12}|^{-2\D} |u_{34}|^{-2\D}$.
The two-point function $\la \mathcal{O}(u_1) \mathcal{O}(u_2) \ra = |u_{12}|^{-2\D}$ transforms under the reparametrization $u \to \tan(\z f(u)/2)$ as
	\begin{equation}
		\big\la \mathcal{O}(u_1) \mathcal{O}(u_2) \big\ra \, = \, \frac{\big| f'(u_1) f'(u_2) \big|^{\D}}{\big| \frac{2}{\z} \, \sin\frac{\z ( f(u_1)-f(u_2))}{2} \big|^{2\D}} \, .
	\end{equation}
Furthermore considering a small fluctuation $f(u) = u + \ve(u)$, up to the first order of $\ve(u)$, this can be written as
	\begin{equation}
		\big\la \mathcal{O}(u_1) \mathcal{O}(u_2) \big\ra \, = \, \frac{\D \mathcal{B}_{\z}(u_1, u_2)}{\big| \frac{2}{\z} \, \sin\frac{\z u_{12}}{2} \big|^{2\D}} \, ,
	\end{equation}
where
	\begin{equation}
		\mathcal{B}_{\z}(u_1, u_2) \, = \, \ve'(u_1) \, + \, \ve'(u_2) \, - \, \z \left( \frac{\ve(u_1) - \ve(u_2)}{\tan \frac{\z u_{12}}{2}} \right) \, .
	\end{equation}
We note that when $\z=1$, $\mathcal{B}_{\z}$ is invariant under the $SL(2,\mathbb{R})$ generated by $\d \ve(u)= \{ e^{-i u}, \, 1, \, e^{i u} \}$.
On the other hand, when $\z<1$, $\mathcal{B}_{\z}$ is only $U(1)$ invariant corresponding to a constant translation of $\ve(u)$.

Now we consider the four-point function.
At leading order in the Schwarzian coupling, the Schwarzian contribution is given by the exchange diagram of the Schwarzian mode as
	\begin{equation}
		\frac{\la \mathcal{O}(u_1) \mathcal{O}(u_2) \mathcal{O}(u_3) \mathcal{O}(u_4) \ra}
		{\la \mathcal{O}(u_1) \mathcal{O}(u_2) \ra \la \mathcal{O}(u_3) \mathcal{O}(u_4) \ra}
		\, = \, \D^2 \big\la \mathcal{B}_{\z}(u_1, u_2) \mathcal{B}_{\z}(u_3, u_4) \big\ra \, ,
	\end{equation}
where the expectation value is evaluated by the Schwarzian mode propagator (\ref{<ee>}).
For the time-ordered case ($u_1 > u_2 > u_3 > u_4$), the four-point function is factorized as 
	\begin{equation}
		\frac{\la \mathcal{O}(u_1) \mathcal{O}(u_2) \mathcal{O}(u_3) \mathcal{O}(u_4) \ra}
		{\la \mathcal{O}(u_1) \mathcal{O}(u_2) \ra \la \mathcal{O}(u_3) \mathcal{O}(u_4) \ra}
		\, = \, \frac{\D^2}{2\pi C} \left[ \left( - \frac{2}{\z} + \frac{u_{12}}{\tan\frac{\z u_{12}}{2}} \right) \left( - \frac{2}{\z} + \frac{u_{34}}{\tan\frac{\z u_{34}}{2}} \right) \right] \, .
	\end{equation}
On the other hand, for the out-of-time-ordered case ($u_1 > u_3 > u_2 > u_4$), the four-point function is given by
	\begin{align}
		&\quad \frac{\la \mathcal{O}(u_1) \mathcal{O}(u_2) \mathcal{O}(u_3) \mathcal{O}(u_4) \ra}
		{\la \mathcal{O}(u_1) \mathcal{O}(u_2) \ra \la \mathcal{O}(u_3) \mathcal{O}(u_4) \ra} \\
		\, &= \, \frac{\D^2}{2\pi C} \bigg[ \left( - \frac{2}{\z} + \frac{u_{12}}{\tan\frac{\z u_{12}}{2}} \right) \left( - \frac{2}{\z} + \frac{u_{34}}{\tan\frac{\z u_{34}}{2}} \right) \nn\\
		&\hspace{40pt} \quad + \frac{2\pi \big( \sin\frac{\z (u_1 - u_2 - u_3 + u_4)}{2c} - \sin \frac{\z(u_1 + u_2 - u_3 - u_4)}{2} \big)}{\z \, \sin\frac{\z u_{12}}{2} \sin\frac{\z u_{34}}{2}} 
		\, + \, \frac{2\pi \, u_{23}}{\tan\frac{\z u_{12}}{2} \tan\frac{\z u_{34}}{2}} \bigg] \, . \nn
	\end{align}
We note that now the four-point function also depends of the separation between the two pairs of the $s$-channel from the second $\sin$ term as well as the last term.
In order to see the exponential growth, we continue to Lorentzian time $u \to - i t$ and into the chaos region, the OTOC has the form
	\begin{equation}
		\frac{\big\la \mathcal{O}(\e_1) \mathcal{O}(\e_2-it) \mathcal{O}(\e_3) \mathcal{O}(\e_4-it) \big\ra}
		{\big\la \mathcal{O}(\e_1) \mathcal{O}(\e_3) \big\ra\big\la \mathcal{O}(\e_2) \mathcal{O}(\e_4) \big\ra} 
		\, \sim \, \frac{\D^2 \b}{C \z} \, e^{\frac{2\pi \z t}{\b}} \, , \qquad \quad \frac{\b}{2\pi \z} \, \ll \, t \, \ll \, \frac{\b}{2\pi \z} \log\frac{C\z}{\b} \, ,
	\end{equation}
where we retrieved the explicit $\b$ dependence.
From this we identity the Lyapunov exponent as $\l_L = 2\pi \z /\b$,
which agrees with the result we found for the $SL(2,\mathbb{Z})$ black holes (\ref{lambda_L}) by identifying $\z = c^{-1}$.

\subsection{One-loop correction to the two-point function}
Next we consider one-loop correction for the two-point function.
Expanding the reparametrized two-point function
	\begin{equation}
		\frac{1}{|u_{12}|^{2\D}} \, \to \, \frac{(1+\ve_1')^\D(1+\ve_2')^\D}{\big| \frac{2}{\z} \, \sin\frac{\z(u_{12} + \ve_1 - \ve_2)}{2} \big|^{2\D}} \, ,
	\end{equation}
up to the quadratic order in $\ve$, we obtain
	\begin{align}
		\frac{\la \mathcal{O}(u_1) \mathcal{O}(u_2) \ra_{\rm one\mbox{-}loop}}{\la \mathcal{O}(u_1) \mathcal{O}(u_2) \ra_{\rm tree}}
		\, &= \, \D \left\la \frac{\z^2(\ve_1 - \ve_2)^2}{4\sin^2\frac{\z u}{2}} - \frac{1}{2} (\ve_1'{}^2 + \ve_2'{}^2) \right\ra
		\, + \, \frac{\D^2}{2} \left\la \left( \ve_1' + \ve_2' - \frac{\z(\ve_1 - \ve_2)}{\tan\frac{\z u}{2}} \right)^2 \right\ra \nn\\
		&= \, \frac{1}{2\pi \z^2 C} \bigg[  \D \z^2 \left( \frac{u^2 - 2\pi u + 2b (1- \cos \z u)+ \frac{2\pi}{\z} \sin\z u}{4 \sin^2\frac{u}{2c}} - \z^{-2} - \frac{b}{\z} \right) \nn\\
		&\hspace{46pt} \, + \, \D^2 \left( - \frac{2}{\z} + \frac{u}{\tan\frac{\z u}{2}} \right) \left( - \frac{2}{\z} + \frac{(u-2\pi \z^2)}{\tan\frac{\z u}{2}} \right) \bigg] \, ,
	\end{align}
where $u= u_1 - u_2$ and $b = -(\pi/\z) \cot(\pi \z)$. For the second equality we used the propagator (\ref{<ee>}).

\section{Conclusion}
\label{sec:conclusion}
A major motivation of this work was to understand the theory of Jackiw-Teitelboim gravity on arbitrary genus Riemann surfaces both in the Schwarzian limit and beyond.
As a starting point, one must understand this theory on the punctured hyperbolic disk and the hyperbolic trumpet.
In this paper, we have provided a complimentary approach to other studies of this problem by solving the related problem of the quantum mechanics of a charged particle on a hyperbolic cone with a constant background magnetic field plus an Aharonov-Bohm field.
This allowed us to find the exact density of states of the theory with confirming previously derived results in the Schwarzian limit.
Solving this problem allows us to go further, for instance the computation of the bulk-boundary propagator and the entanglement entropy of the Hartle-Hawking state.
We then confirmed our results in a complimentary picture of dimensionally reduced $SL(2,\mathbb{Z})$ black holes of the 3D pure gravity.
In this context, we derived the Lyapunov exponent characterizing quantum chaos and showed that it is decreased in comparison to the theory on a hyperbolic disk by a factor proportional to the conical deficit.
We then reproduced this result directly in the 1D boundary Schwarzian theory.

Our work raises several interesting questions. The first is the nature of the Lorentzian continuation of Hartle-Hawking state. In particular, the Lorentzian continuation of the Hartle-Hawking state on the disk is a smooth geometry, the maximally extended AdS$_2$ black hole \cite{Maldacena:2016upp, Maldacena:2017axo, Kitaev:2018wpr}. The Lorentzian continuation of the conical singularity at the origin of the punctured disk likely causes the horizons to become singular, see for instance \cite{Kyono:2017jtc}. It is well-known that smooth horizons require special correlations in the Hartle-Hawking state which can be seen from the entanglement entropy \cite{VanRaamsdonk:2010pw}. We may be able to gain some insight into this question as the $\eta\to\infty$ limit of the scalar potential in \cite{Kyono:2017jtc} is also exponential in the scalar field, $U(\phi) \sim \eta e^{\eta\phi}$. By appropriately taking the limit $T\to0$ and $\eta\to \infty$, the entropy has a residual contribution $S\sim \eta^{-1}$. This is in addition to any potential extremal entropy contribution arising from dimensional reduction of an extremal black hole. 
Another open question is the role of these states in the factorization problem of the Lorentzian setup with two boundaries, where the thermofield double state constructed by a tensor-product of the two boundary CFT states appears to be contradiction with the existence of gauge constraints in the bulk \cite{Harlow:2018tqv}.
Computation of the entanglement entropies between the two boundaries in this setup requires a particular `defect operator' \cite{Jafferis:2019wkd, Kitaev:2018wpr}
which factories the Hilbert space. It is interesting to understand this defect operator from our charged particle picture.

Given the role of JT gravity in recent work on the black hole information problem \cite{Penington:2019npb, Almheiri:2019psf,Almheiri:2019hni, Penington:2019kki, Almheiri:2019qdq}, it would be interesting to understand the impact of the topological entropy contribution. In particular, finite entropy zero temperature states are reminiscent of remnants \cite{Susskind:1995da}. Nevertheless, from a three-dimensional perspective, these states seem to resolve some problems of unitarity in pure three-dimensional gravity \cite{Maxfield:2020ale}. Clearly, a better understanding is necessary. Furthermore, our method of calculating the entanglement entropy does not depend on assumptions about saddle points and it would be interesting to compare our results to purely gravitational calculation \cite{Akers:2020pmf}.

One would also like to understand matter fields coupled to the JT gravity with defects.
As a first step, we computed two and four-point correlation functions of boundary conformal operators coupled to the gravitational dynamics of JT gravity.
This was done by two methods: dimensionally reducing $SL(2,\mathbb{Z})$ black holes of the pure 3D gravity and computing the exchanged diagram of the boundary Schwarzian modes.
In particular, the out-of-time-ordered correlation function (OTOC) showed that the Lyupanov exponent is decreased as $\lambda_L = (2\pi - \alpha)/\beta$ for conical deficit $\alpha$.
Exactly solvable quantum systems, with the Lyupanov exponent interpolating between the maximal chaos bound and away from it, are highly rare.
Apart from the JT gravity with defects we studied here, the only known example which has this feature is the large $q$ limit of the SYK model \cite{Streicher:2019wek, Choi:2019bmd},
where the Lyupanov exponent is given by $\lambda_L = 2\pi v/\beta$ with an RG parameter $0\le v \le 1$  \cite{Maldacena:2016hyu}.
It is interesting to investigate a relation between these two models, but we also believe that our computation itself will be valuable for the study of non-maximal chaotic systems.

A final question is how to extend our results to higher genus Riemann surfaces. In particular, in the charged particle picture, the solutions for the wavefunctions and propagator are regular for $\rho\to\infty$, but we do not need to take this limit.\footnote{There are some issues with including self-intersecting paths which we ignore for now, see \cite{Stanford:2020qhm}.} This means that the non-geodesic boundary of the hyperbolic trumpet need not be at the boundary of AdS. One can ask whether the three-holed sphere can be constructed from stitching together such trumpet solutions along their non-geodesic boundaries. In addition to providing a complementary route to the density of states, understanding higher genus solutions would allow for a deeper understanding of quantum gravity through quantum informatic perspectives including probes beyond the horizon and entanglement inequalities \cite{Hayden:2011ag, Hartman:2013qma, Bao:2015bfa}.

\acknowledgments
This work is supported by the European Research Council (ERC) under the European Union's Horizon 2020 research and innovation programme (grant agreement No758759 and No818066).

\appendix
\section{Three routes to the heat kernel on a flat cone}
\label{app:coneappendix}
There is a very long history for solving the Schrodinger equation on a cone embedded in flat spacetime because the Euclidean problem is equivalent to solving for the kernel of the heat diffusion equation, a classic problem of the late nineteenth century. Solving the problem on the cone is related to many other problems that interested nineteenth century physicists including diffraction of waves by wedges and it is within this context that Sommerfeld and Carslaw provided the first solutions for these problems \cite{Carslaw1, Carslaw2, Carslaw3}. In the spirit of a self-contained discussion and to motivate the approach to hyperbolic cones, we reproduce their arguments here, following the reference \cite{Carslaw2}. 

\begin{figure}[t]
\centering
\vspace{-10pt}
\begin{tikzpicture}[scale=.8,>=stealth]
	\draw (-.7,-.9) node [fill=white!20]{\large $\varphi'$};
	\filldraw[fill=gray!20,draw=gray!20] (-3,-3) rectangle +(2,6);
	\filldraw[fill=gray!20,draw=gray!20] (1,-3) rectangle +(2,6);
	\draw[dotted, thick] (-3,0) -- (3,0);
	\draw[dotted, thick](-2,-3) -- (-2,3);
	\draw[dotted, thick](2,-3) -- (2,3);
	\draw[thick](-.75,-.15) -- (-.75,.15);
	\draw[thick](0,-.15) -- (0,.15);
	\draw[thick, <->] (-1,2) -- (1,2);
	\draw[thick, ->] (-.75,.3) arc (-270:90:.3cm);
	\draw (-2,-1) node {\large $\varphi - \pi$};
	\draw (2,-1) node {\large $\varphi + \pi$};
	\draw (0,-1) node [fill=white!20]{\large $\varphi$};
	\draw (0,2) node [fill=white!20]{\large $\pi$};
	\draw (-2.6,2.6) node [fill=white!20, draw=black]{\large $z$};
	\draw (7.3,-.5) node [fill=white!20]{\large $\varphi'$};
	\filldraw[fill=gray!20,draw=gray!20] (5,-3) rectangle +(2,6);
	\filldraw[fill=gray!20,draw=gray!20] (9,-3) rectangle +(2,6);
	\draw[dotted, thick] (5,0) -- (11,0);
	\draw[dash dot,thick](6,-3) -- (6,0);
	\draw[dash dot,thick,<-](6,0) -- (6,3);
	\draw[dash dot,thick,->](10,-3) -- (10,0);
	\draw[dash dot,thick](10,0) -- (10,3);
	\draw[thick](7.25,-.15) -- (7.25,.15);
	\draw[thick](8,-.15) -- (8,.15);
	\draw[thick] (6.1,3) arc (-180:0:1.9cm);
	\draw[thick] (9.9,-3) arc (0:180:1.9cm);
	\draw[thick, ->] (7.8,-1.1) -- (8.2,-1.1);
	\draw[thick, <-] (7.8,1.1) -- (8.2,1.1);
	\draw[thick] (6,-3) -- (6.1,-3);
	\draw[thick] (9.9,-3) -- (10,-3);
	\draw[thick] (6,3) -- (6.1,3);
	\draw[thick] (9.9,3) -- (10,3);
	\draw (8,-.6) node [fill=white!20]{\large $\varphi$};
	\draw[thick,->] (6,3.1) -- (6,3.7);
	\draw[thick,->] (10,3.1) -- (10,3.7);
	\draw[thick,->] (6,-3.1) -- (6,-3.7);
	\draw[thick,->] (10,-3.1) -- (10,-3.7);
	\draw (10,4) node [fill=white!20]{\large $+i\infty$};
	\draw (6,4) node [fill=white!20]{\large $+i\infty$};
	\draw (10,-4) node [fill=white!20]{\large $-i\infty$};
	\draw (6,-4) node [fill=white!20]{\large $-i\infty$};
	\draw (5.4,2.6) node [fill=white!20, draw=black]{\large $z$};
\end{tikzpicture}
\vspace{10pt}
\caption{The contour C circling a single pole at $\vp = \vp'$ (left) can be deformed into the contour at right. The gray areas are regions for which $\cos(z-\vp)$ has a negative real part where the contour can be deformed to $\pm i\infty$. The dot-dashed lines at $z = \vp\pm \pi$ provide equal and opposite contributions which cancel and be neglected when $\zeta$ is a rational number.}
\label{fig:contourflatcone}
\end{figure}
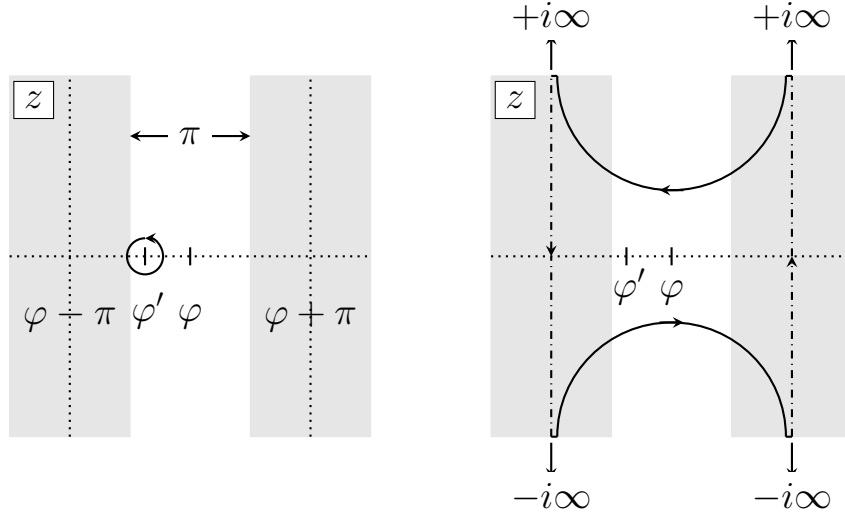

We start with the well-known representation of the non-relativistic propagator, Wick-rotated to Euclidean time as $t\to -i\tau$, \cite{Shankar}
	\begin{equation}
	G(r,\vp;r',\vp';t) \, = \, \frac{1}{4\pi \k t} \, e^{- \frac{|\vec{r}-\vec{r}'|^2}{4\k\t}} \, ,
	\label{heatkernel}
	\end{equation}
which is a solution to the diffusion equation
	\begin{equation}
	\left(-\frac{\partial}{\partial \tau} +\kappa \nabla^2\right)G(r,\vp;r',\vp';\tau) \, = \, \delta(\tau)\delta(\vp-\vp')\frac{\delta(r-r')}{\sqrt{rr'}} \, .
	\label{diffusioneq}
	\end{equation}
Without loss of generality, we may consider $|\vp-\vp'|<\pi$. Consider the contour integral 
	\begin{equation}
	\frac{1}{8\pi^2\k \tau} \, e^{-\frac{(r^2+r'^2)}{4\k \tau}} \oint_C dz \, e^{\frac{rr'\cos(z - \vp)}{2k\tau}} \, \frac{e^{iz}}{e^{iz}-e^{i\vp'}} \, ,
	\label{contour1}
	\end{equation}
along the contour $C$ depicted on the left of figure \ref{fig:contourflatcone} circling only the pole at $e^{iz}=e^{i\vp'}$. Clearly this integral reproduces $G$. The contour can be deformed into the contour on the right which is made up of two curves. Some salient features of this contour integral are that it is explicitly periodic as $\vp \to \vp+2\pi$ and that as $\tau\to 0$, the integrand vanishes unless $r=r'$ and $\vp = \vp'$ and it vanishes for $r\to\infty$. Now consider the expression
	\begin{equation}
	\frac{1}{8\pi^2 \kappa \zeta \tau} \, e^{-\frac{(r^2+r'^2)}{4\kappa\tau}}
	\oint_C dz \, e^{\frac{rr'\cos(z-\vp)}{2\kappa\tau}} \, \frac{e^{i\frac{z}{\zeta}}}{e^{i\frac{z}{\zeta}}-e^{i\frac{\vp'}{\zeta}}} \, ,
	\end{equation}
over the same contour $C$ as in (\ref{contour1}). For $\zeta = 1$, this is clearly the same integral as (\ref{contour1}). Consider letting $z-\vp = z'$. The integral becomes
	\begin{equation}
	\frac{1}{8\pi^2 \kappa \zeta \tau} \, e^{-\frac{(r^2+r'^2)}{4\kappa\tau}}
	\oint_C dz' \, e^{\frac{rr'\cos z'}{2\kappa\tau}} \, \frac{e^{i\frac{z'+\vp}{\zeta}}}{e^{i\frac{z'+\vp}{\zeta}}-e^{i\frac{\vp'}{\zeta}}} \, .
	\label{periodicdeformation}
	\end{equation}
From this expression, it is clear that the solution is periodic as $\vp\to \vp+2\pi\zeta$ and $\vp'\to\vp'+2\pi\zeta$. For $\zeta = (2\pi-\alpha)/2\pi$, this describes the flat space limit of the cone discussed in the main text. Furthermore, it can be checked that this solves the same diffusion equation as before (\ref{diffusioneq}), with the appropriate $t\to 0$ limit to reproduce the delta function on the right hand side. Hence, this represents the solution on a cone with metric
	\begin{equation}
	ds^2 \, = \, dr^2 \, + \, r^2\vp^2 \, , \qquad \vp \, \cong \, \vp+2\pi-\alpha \, .
	\end{equation}
We can represent the function inside the integrand as an infinite sum
	\begin{equation} 
	2\frac{e^{i\frac{z'+\vp}{\zeta}}}{e^{i\frac{z'+\vp}{\zeta}}-e^{i\frac{\vp'}{\zeta}}}
	\, = \, -\sum_{n=1}^\infty e^{+ i n (\vp-\vp')/\zeta} \, e^{+ i n z'/\zeta} \, + \, \sum_{n=0}^\infty e^{- i n (\vp-\vp')/\zeta} \, e^{- i n z'/\zeta} \, ,
	\end{equation}
where the first term comes from the upper part of the contour and the second comes from the lower, chosen by convergence of $e^{\pm i n z'/\zeta}$ as $z'\to \pm i\infty$. Taking these considerations into account, this contour integral can be written
	\begin{equation}
	G_\zeta(r,\vp;r',\vp';t') = \frac{e^{-(r^2+r'^2)/4\kappa\tau}}{4\pi\kappa\zeta \tau} \sum_{n=-\infty}^\infty e^{-i|n|\pi/2\zeta}e^{in(\vp-\vp')/\zeta}J_{|n|/\zeta}\left(\frac{irr'}{2\tau}\right).
	\label{Carslawsresult}
	\end{equation}
We can arrive at this expression from other routes. First, we can take the $\zeta\to\infty$ limit which represents the multi-sheeted Riemann surface originally used by Sommerfeld to solve the diffusion equation \cite{Sommerfeld}. The solution without a conical deficit can be obtained by summing over images with the correct periodicity \cite{Dowker:1977zj}. For instance, to reproduce the result of Carslaw, we simply take
	\begin{equation}
	G_{\zeta}(r,\vp;r',\vp',\tau) = \sum_{n=-\infty}^\infty G_{\infty}(r,\vp+2\pi n \zeta; r',\phi'; \tau)
	\end{equation}
Poisson resummation of this expression gives (\ref{Carslawsresult}). As emphasized in \cite{Dowker:1977zj}, the inclusion of an AB field can be incorporated by multiplying each term in the sum by $e^{2\pi i n \delta}$  where $\delta$ is the electromagnetic flux through the origin. After Poisson resummation, this phase just shifts the angular momentum quantum numbers by $\delta$:
	\begin{equation}
	G_\zeta(r,\vp;r',\vp';t') = \frac{e^{-(r^2+r'^2)/4\kappa\tau}}{4\pi\kappa\zeta \tau} \sum_{n=-\infty}^\infty e^{-i|n+\delta|\pi/2\zeta}e^{i(n+\delta)(\vp-\vp')/\zeta}J_{|n+\delta|/\zeta}\left(\frac{irr'}{2\tau}\right).
	\label{FlatconeAB}
	\end{equation}
The inclusion of such phases has an interesting connection to the the theory of self-adjoint extensions \cite{ssolvable} which arise due to the fact that the Hamiltonian on the cone is not self-adjoint, which is also the case in hyperbolic space as emphasized in \cite{Yang:2018gdb, Lisovyy:2007mj}. The inclusion of phases is the unique self-adjoint extension with regular wavefunctions at the origin of the cone.

A third route \cite{Deser:1988qn} to the heat kernel is to consider rescaling $r\to r/\zeta, \vp\to \zeta \vp$ giving a metric
	\begin{equation}
	ds^2 = \frac{dr^2}{\zeta^2} + r^2d\vp^2,\quad \vp \, \cong \, \vp+2\pi
	\label{flatconegauge2}
	\end{equation}
describing a cone embedded in flat three-dimensional space with $x_3 = \sqrt{(\zeta^{-2}-1)(x_1^2+x_2^2)}$. This change of coordinates allows for a direct mapping onto the solutions in flat space since in this gauge, the angular momentum quantum numbers remain integers. The mapping is surprisingly simple: an eigenvector, $f(r,\vp) = f(r)e^{-il\vp}$, of the conical Laplacian with eigenvalue $-\lambda^2$ and with angular momentum quantum number $l$ solves the equation
	\begin{equation}
	\left(r\partial_r r\partial_r -\frac{l^2}{\zeta^2} + \lambda^2r^2\right)f(r,\vp) = 0
	\end{equation}
Letting $\rho = r/\zeta$, this is exactly the eigenvalue equation in flat space with the substitutions $l\to l/\zeta$ and $\lambda\to\lambda/\zeta$. Hence, we must also shift $\kappa \to \kappa/\zeta^{-2}$ in eq. (\ref{diffusioneq}). This is exactly what we find in eq. (\ref{Carslawsresult}) reminding ourselves that the coordinate $\vp$ is rescaled with respect to the gauge in eq. (\ref{flatconegauge2}).

We therefore find three complimentary routes to the same expression (\ref{Carslawsresult}). By far, the simplest was via the mapping using the gauge in eq. (\ref{conegauge2}), though this of course required expressing the flat space heat kernel, eq. (\ref{heatkernel}), as a sum over Bessel functions which is essentially the route taken by the other two approaches. Nevertheless, given an expression for the flat space heat kernel in terms of a sum over angular momentum, we can use gauge transformations of the metric to find an expression on the cone that is equivalent to rescaling the quantum numbers. This is the same procedure that we perform in hyperbolic space in the main text.

In fact, our derivation was too fast. In deforming the contour in figure \ref{fig:contourflatcone}, we performed two steps \cite{Dowker:1977zj}. Starting with the contour on the left of this figure, we added two vertical lines at $z=\vp\pm\pi$. When $\zeta=1/p$ for integer $p$ these contours cancel each other, a fact easily seen in eq. (\ref{periodicdeformation}). Then we deformed the contour circling the pole at $z=\vp'$ into two curves which meet the vertical lines at infinity. However, for generic $\zeta$, these two contributions do not cancel. In addition, for $\zeta\neq 1$, there can be extra contributions from poles at $z=\vp' \pm k(2\pi\zeta)$, for integer $k$, which satisfy $\vp-\pi < z < \vp+\pi$, which will appear when $\pi>|\vp-\vp'|>\pi\zeta$. For $\zeta>1$ no such pole contributes. Hence, the contributions of these poles and the vertical contours are non-perturbative contributions to the heat kernel arising from the apex of the cone in addition to eq. (\ref{Carslawsresult}). If we remove the apex of the cone, then the sum in eq. (\ref{Carslawsresult}) represents a sum of classical paths on the multiply connected space $\mathcal{M}_{\infty}/Z_\infty(\zeta)$ where $\mathcal{M}_\infty$ is a simply connected space (Sommerfeld's Riemann surface where $G_\infty$ is defined) and $Z_\infty(\zeta)$ is the infinite cyclic group with period $2\pi\zeta$. In the hyperbolic case, we will similarly remove the singular point at $\rho=0$, discarding the non-perturbative terms.

\section{Orthonormality condition}
\label{app:orthonormality}
In this appendix, we consider the orthonormality condition for the wavefunctions on the disk (\ref{f^D})
	\begin{align}
		\int_0^1 \frac{2dx}{(1-x)^2} \, f^{D*}_{j,k}(x) f^D_{j',k}(x) \, = \, \delta(s-s') \, ,
	\label{orthonormality}
	\end{align}
for the principal series ($j=1/2+is$, $s\in\mathbb{R}$, $s>0$).
Since the wavefunction is regular at the center of the disk ($x\to 0$), the singular contribution must come from the $x \to 1$ limit.
Let us now suppose $b>k$ so that $f^D_{j,k}(u) = a^b_{j,k} \, A^b_{j,k}(u)$.
The $x \to 1$ asymptotic behavior of this solution is
	\begin{align}
		 f^D_{j,k}(x) \, = \, a^b_{j,k} \, \left[ \, \frac{\Gamma(1-2j)}{\Gamma(1-j-k)\Gamma(1-j+b)} \, (1-x)^j \, + \, (j \to 1-j) \, \right] \, .
	\end{align}
Then, we decompose the orthonormality integral (\ref{orthonormality}) into two intervals: $0 \le x \le 1- e^{-1/\epsilon}$ and $1- e^{-1/\epsilon} \le x \le 1$.
The delta function in (\ref{orthonormality}) comes from the $\epsilon \to 0$ limit of the second interval.
Using the representation 
	\begin{equation}
		 \delta (y) \, = \, \lim_{\epsilon \to 0} \frac{\sin(y/\epsilon)}{\pi y} \, ,
	\end{equation}
indeed we find 
	\begin{equation}
		\lim_{\epsilon \to 0} \, \int_{1-e^{-1/\epsilon}}^1 \frac{2dx}{(1-x)^2} \, f^{D*}_{j,k}(x) f^D_{j',k}(x)
		\, = \, - 4\pi \big| a^b_{j,k} \big|^2 \left| \frac{\Gamma(1-2j)}{\Gamma(1-j-k)\Gamma(1-j+b)} \right|^2 \, \delta(s-s') \, .
	\end{equation}
Requiring the normalization (\ref{orthonormality}) fixes the coefficients as
	\begin{equation}
		a^b_{j,k} \, = \, \frac{i}{2\sqrt{\pi}} \, \frac{\Gamma(1-j-k)\Gamma(1-j+b)}{\Gamma(1-2j)} \, .
	\end{equation}

\section{Contour integral representation of the $\th=1$ propagator}
\label{app:theta=1}
In this appendix we summarize the contour integral representation of the $\th=1$ propagator without the effect of the AB field (i.e. $\xi =0$) \cite{Lisovyy:2007mj}.
Start with so-called horocyclic waves for the representation $j=\frac{1}{2}+is$ \cite{De_Micheli_2006}, 
	\begin{equation}
		\Psi^{b}_{\pm}(z,w) \, = \, \frac{1}{2\pi}\frac{(1-|z|^2)^{j_{\pm}}}{(1+ze^{-w})^{j_{\pm}-b}\,(1+\bar{z}e^{w})^{j_{\pm}+b}} \, , 
	\end{equation}
where $j_{\pm} = \frac{1}{2}\pm(j-\frac{1}{2})$, $z = \sqrt{x}e^{i\vp}$, and $w$ is an arbitrary complex parameter. It can be verified that the horocyclic waves are eigenfunctions of eq. (\ref{Schro-eq-D-2}) with an infinite number of branch points at $w = \pm \frac{1}{2}\ln x + i(\vp + 2\pi\mathbb{Z}+\pi)$. Furthermore, $L_0$ acts on $\Psi$ as $L_0\Psi = -\partial_w \Psi$. We will relate the wavefunctions and propagator to a series of contour integrals over the contours shown in figure \ref{fig:hyperbolicdiskcontours}. Consider the following integral
	\begin{equation}
		I_+^l \, = \, \int_{C_+} dw \, e^{lw} \, \Psi^{b}_{+}(z,w) \, ,
	\end{equation}
along some contour $C_+$. For $C_+$ described by $w = i(\lambda+\pi)$ with $\lambda\in[0,2\pi)$ and $l \in \mathbb{Z}$, this gives exactly eq. (109) of \cite{Kitaev:2017hnr} for $m=\nu + l$. We can add two rays to the contour, one running along the real axis from $w=-\infty$ to the origin, $w=0$, and the other running parallel to the real axis from $w=2\pi i$ to $w=-\infty +2\pi i$. Because $l$ is an integer, these two sections of the contour cancel each other and the integral is unchanged. In fact, we could have considered two other rays which are a reflection of $C_+$ about the imaginary axis. Changing the direction of this contour to give a counterclockwise orientation, we get the same result up to a minus sign. This contour is labelled $C_-$ in figure \ref{fig:hyperbolicdiskcontours}. In fact, we can equally shift $C_+$ by some amount $-\lambda_r$ along the real axis and some amount $\lambda_i$ along the imaginary axis while shifting $C_-$ some amount $\lambda_r$ along the real axis and some amount $\lambda_i$ along the imaginary axis. So long as $|\vp-\lambda_i|<\pi$ and $0\leq \lambda_r < - \ln x$, the integrals are unchanged. One can check explicitly that, for $l = m+b$,
	\begin{equation}
		I_+^{m+b}(x,\vp) \, = \, e^{i\pi(m+b+\frac{1}{2})}\frac{\Gamma(j+m)}{\Gamma(j-b)}e^{i(m+b)\vp} A^{b}_{j,-m}(x) \, ,
	\end{equation}
which exactly matches the $\nu$-spinor representation of the eigenfunctions up to an overall phase (eqs. (105) and (107) of \cite{Kitaev:2017hnr}).

Now, consider the integral
	\begin{equation}
	I_0^l \, = \, \int_{C_0} dw \, e^{lw} \, \Psi^b_{-}(z,w) \, ,
	\end{equation}
for the contour $C_0$ in figure \ref{fig:hyperbolicdiskcontours} running from $w = i(\vp + \pi) + \ln x$ to $w = i(\vp + \pi) - \ln x$. One can check by explicit calculation that, for $l = m+b$,
	\begin{equation}
	I_0^{m+b} \, = \, e^{i(m+b)\vp} \, x^{(m+b)/2} \, (1-x)^j \, \mathbf{F}(j+m,j+b,2j;1-x) \, .
	\end{equation}
The radial, $x$-dependent, part of this function can be expressed as a linear combination of the radial functions $A^{b}_{j,m}$ and $A^{-b}_{j,-m}$ in eq. (\ref{A^q}) (see eq. (59) of \cite{Kitaev:2017hnr}).

The heat kernel of $\theta = 1$ for $x>x'$ can be written in terms of a summation over the angular momentum $l$ as in (\ref{resolvent})
	\begin{align}
	2 i \Pi^{(1)}(x,\vp;x',\vp') &= \sum_{l\leq 0} \int_{C_0} dw_1\int_{C'_+}dw_2\Psi_{-}^{-b}(x,\vp,w_1)\Psi_{+}^{b}(x',\vp',w_2)e^{l(w_1-w_2)}\nn\\
	& -\sum_{l> 0} \int_{C_0} dw_1\int_{C'_-} dw_2\Psi_{-}^{-b}(x,\vp,w_1)\Psi_{+}^{b}(x',\vp',w_2)e^{l(w_1-w_2)}.
	\label{kernelnoABfield}
	\end{align}
Here, the contour $C_0$ is defined with respect to the un-primed coordinates while $C'_+$ and $C'_-$ are defined with respect to the primed coordinates.
We bring the summation into the integrand and perform the sum over $l$ depending on whether Re$(w_1-w_2)<0$ as for $\theta_2 \in C_-'$ or Re$(w_1-w_2)>0$ as for $\theta_2\in C_+'$.
The result is a contour integral over $C_+'\cup C_-'$,
	\begin{equation}
	\Pi^{(1)}(x,\vp;x',\vp') = \frac{1}{2 i}\int_{C_0} dw_1 \int_{C_+'\cup C_-'} dw_2 \, \Psi_{-}^{-b}(x,\vp,w_1)\Psi_{+}^{b}(x',\vp',w_2) \frac{e^{w_1}}{e^{w_1}-e^{w_2}} \, .
	\end{equation}
Here, we see the role of periodicity in $\vp$ and $\vp'$. Consider $\vp \to \vp+2\pi$. The function $\Psi_{-}$ is unchanged. Furthermore, the only contour that depends on un-primed variables is $C_0$ which shifts by $2\pi$ vertically with the only effect of having $e^{w_1}\to e^{w_1+2\pi i}= e^{w_1}$. Hence this function is $2\pi$-periodic in $\vp$. For $\vp' \to \vp'+2\pi$, the function $\Psi_{+}$ is unchanged. The branch cuts all shift vertically by $2\pi$. This is equivalent to shifting the contour $\int_{C_+'\cup C_-}$ downward vertically by $2\pi$. For each component parallel to the real axis, this means $e^{w_2}\to e^{w_2-2\pi i} = e^{w_2}$. The components parallel to the imaginary axis change their starting and end points, but otherwise the integral is unchanged. Hence, it is $2\pi$-periodic in $\vp$ and $\vp'$. 

Finally, the contour  $C_+'\cup C_-'$ can be deformed so that the parts parallel to the imaginary axis cancel each other, picking up the pole at $e^{w_2}=e^{w_1}$. The parts parallel to the real axis also cancel each other and we have
	\begin{equation}
	\Pi^{(1)}(x,\vp;x',\vp') \, = \, \pi \int_{C_0} dw \, \Psi_{-}^{-b}(x,\vp,w)\Psi_{+}^{b}(x',\vp',w) \, .
	\end{equation}
Essentially, we have performed Carslaw's integral transform in reverse. This integral can be computed in closed form in which case it gives the result in \cite{Kitaev:2017hnr, Yang:2018gdb, Kitaev:2018wpr}.

\section{$u$ contour integral}
\label{app:u-integral}
Here we perform the $u$ contour integral for the resolvent (\ref{traceintegral1}) using the residue theorem. The first integral,
	\begin{equation}
		I_1(v) \, = \, \frac{1}{2\pi i}\int_{-\infty}^{\infty}du\frac{e^{\frac{1+\xi}{\theta}(u - i\pi)}}{e^{(u-i\pi)/\theta}-1} \frac{1}{(1+ve^{-u})(1+e^u)} \, ,
	\end{equation}
has poles at $u = i \pi + k(2\theta\pi i)$, $u = i \pi + k(2\pi i)$, and $u = \ln v + i \pi + k(2\pi i)$.
In the first and second set of poles $k=0$ is the same pole and is hence second order. If $\theta = n/p$ for co-prime $n,p\in \mathbb{Z}$, then poles with $k=mp$ from the first set overlap with the $k=mn$ pole from the second set and we would need to include an infinite number of second order pole contributions. We will ignore this for now. The result of integration is
	\begin{align}
	2I_1(v) &=\frac{2\theta - (1-v)(1+\theta+2\xi)}{2(1-v)^2}+\frac{v^{\frac{1+\xi}{\theta}}}{(v^{\frac{1}{\theta}}-1)(1-v)}\\
	&+\sum_{k\neq 0}\text{sgn}(k)\biggl(\frac{e^{i\frac{1+\xi}{\theta}(2\pi k)}}{(e^{i2\pi k/\theta}-1)(1-v)}+ \frac{v^{\frac{1+\xi}{\theta}}e^{i\frac{1+\xi}{\theta}(2\pi k)}}{(v^{\frac{1}{\theta}}e^{i\frac{2\pi k}{\theta}}-1)(1-v)}+\frac{e^{i2\pi k \xi}}{(1-ve^{-i2\pi k \theta})(1-e^{i2\pi k\theta})}\biggr).\nn
	\end{align}
The second integral
	\begin{equation}
		I_2(v) \, = \, \frac{1}{2\pi i}\int_{-\infty}^{\infty}du\frac{e^{\frac{1+\xi}{\theta}(u + i\pi)}}{e^{(u+i\pi)/\theta}-1} \frac{1}{(1+ve^{-u})(1+e^u)} \, ,
	\end{equation}
has poles at $u =-i \pi + k(2\theta\pi i)$, $u = \ln v - i \pi + k(2\pi i)$, and  $u = - i \pi +k(2\pi i)$. Again, the $k=0$ pole is a second-order pole. Again, if we had $\theta = n/p$ for co-prime $n,p$ then there would be extra contributions from second-order poles. The result is
	\begin{align}
	2I_2(v) &=-\frac{2\theta - (1-v)(1+\theta+2\xi)}{2(1-v)^2}-\frac{v^{\frac{1+\xi}{\theta}}}{(v^{\frac{1}{\theta}}-1)(1-v)}\\
	&\sum_{k\neq 0}\text{sgn}(k)\biggl(\frac{e^{i\frac{1+\xi}{\theta}(2\pi k)}}{(e^{i2\pi k/\theta}-1)(1-v)}+ \frac{v^{\frac{1+\xi}{\theta}}e^{i\frac{1+\xi}{\theta}(2\pi k)}}{(v^{\frac{1}{\theta}}e^{i\frac{2\pi k}{\theta}}-1)(1-v)}+\frac{e^{i2\pi k \xi}}{(1-ve^{-i2\pi k \theta})(1-e^{i2\pi k\theta})}\biggr).\nn
	\end{align}
These two integrals combine to give
	\begin{equation}
		I_1(v) - I_2(v) \, = \, \frac{2\theta - (1-v)(\theta+1+2\xi)}{2(1-v)^2}+\frac{v^{\frac{1+\xi}{\theta}}}{(v^{\frac{1}{\theta}}-1)(1-v)} \, .
	\label{I1I2result}
	\end{equation}
The remaining integrals can be evaluated from eq. (\ref{I1I2result}) under $\xi\to -1-\xi$. Defining
	\begin{equation}
		I_3(v)-I_4(v) \, = \, \frac{2\theta - (1-v)(\theta - 1 -2\xi)}{2(1-v)^2}+ \frac{v^{-\frac{\xi}{\theta}}}{(v^{\frac{1}{\theta}}-1)(1-v)} \, ,
	\label{I3I4result}
	\end{equation}
we have
	\begin{equation}
		\text{Tr}\Delta^{(\theta)} \, = \, \frac{\theta}{2j-1}\int_0^1 dv\left\{v^{j-1+\nu}\left[I_1(v)-I_2(v)\right] - v^{j-1-\nu}\left[I_3(v)-I_4(v)\right]\right\} \, .
		\label{Resolventintegraltheta}
	\end{equation}
For $\theta = 1$, we can solve this in closed form giving eq.~(\ref{Lisovyydensityofstates}). 

In the Schwarzian limit, $\nu  = -iq_\alpha$, $\xi = iq_\alpha \alpha/2\pi$, we can also write this in closed form. To do so, move the integral slightly off the real axis, $v\to v +i\epsilon$. As per usual, the sign of $\epsilon$ depends on convergence which can be determined from $v^{\pm i q} \sim e^{\mp \epsilon q}$. Now, compute the integral via a contour integral. Because $q\to\infty$, the ray $v \in [0,1)$ can be extended to $v\to \pm\infty$ at no cost. We can also add a semi-circular contour at infinity. The important point is that the first terms in eq.~(\ref{I1I2result}) and eq.~(\ref{I3I4result}) do not contribute at leading order in $q$. The second terms have poles at $v = e^{2\pi i k}$ and $e^{2\pi i k\theta}$. By moving the integral slightly off the axis, the $k=0$ poles do not contribute. It is a simple matter to calculate the result via residues. The result will be a sum over poles. Importantly, as $q\to\infty$ only one pole will dominate. The result is, for the cases relevant to this work with $\theta\geq1$:
	\begin{equation}
		\lim_{q\to\infty}\text{Tr}\Delta^{(\theta=1)} \, = \, \frac{2\pi i q_\alpha\zeta}{s}e^{-(2\pi-\alpha) q_{\alpha}}\cosh(2\pi s) \, ,
	\end{equation}
and	
	\begin{align}
	\lim_{q\to\infty}\text{Tr}\Delta^{(\theta>1)} &= -\frac{2\pi\theta}{s}e^{-(2\pi-\frac{\alpha}{\theta}) q_\alpha}\left[\frac{e^{-\frac{2\pi i}{\theta}}e^{2\pi s}}{e^{-\frac{2\pi i}{\theta}}-1}-\frac{e^{-2\pi s}}{e^{\frac{2\pi i}{\theta}}-1}\right]\nn\\
	 &= \frac{2\pi \theta}{is}e^{-(2\pi-\frac{\alpha}{\theta})q_\alpha}e^{-\frac{\pi i}{\theta}}\left[\frac{\cosh(2\pi s)}{\sin(\pi/\theta)}\right]
	\end{align}
The limit $\theta\to 1$ of this expression must be taken very carefully since naively this diverges. Note also that $\alpha\to 0$ naively converges, but in fact this limit does not commute with $q_\alpha\to\infty$. Finally, for integer $\theta$, we can use this expression to derive the R\'eyni entropies.

\section{Schwarzian two-point function}
\label{app:2pt}
In this appendix, we compute the Fourier transform of the Schwarzian two-point function
	\begin{equation}
		\big\la \e(u) \e(0) \big\ra_\z \, = \, \sum_{n \ne 0} e^{i n u} \, \big\la \e_n \e_{-n} \big\ra_\z \, = \, \frac{1}{2\pi C} \sum_{n \ne 0} \frac{e^{i n u}}{n^2(n^2 - \z^2)} \, ,
	\end{equation}
where we excluded the zero mode ($n=0$) from the summation.
First, it is convenient to decompose it into the $\z$-independent and dependent parts as
	\begin{equation}
		\big\la \e(u) \e(0) \big\ra_\z \, = \, \frac{1}{2\pi \z^2 C} \sum_{n \ne 0} e^{i n u} \left[ \frac{1}{n^2 - \z^2} \, - \, \frac{1}{n^2} \right] \, .
	\end{equation}
The $\z$-independent part can be summed directly by using $\sum_{n=1}^{\infty} z^n/n^2 = {\rm Li}_2(z)$ as
	\begin{equation}
		\sum_{n \ne 0} e^{i n u} \frac{1}{n^2} \, = \, \frac{1}{2} \, (|u| - \pi)^2 \, - \, \frac{\pi^2}{6} \, .
	\end{equation}
For the $\z$-dependent part, it is more convenient to express the summation in terms of a contour integral as
	\begin{equation}
		\sum_{n \ne 0} e^{i n u} \frac{1}{n^2 - \z^2} \, = \, \frac{1}{2i} \oint_C \frac{dz}{\sin(\pi z)} \frac{e^{i(u - \pi)z}}{z^2 - \z^2} \, ,
	\end{equation}
where the contour $C$ is a collection of small circles centered at $z = n$ ($n \in \mathbb{Z}$ excluding $n=0$).
Deforming the contour, we can write the integral in terms of residues at $z = 0$ and $z = \pm \z$.
Evaluating these residues, we find
	\begin{equation}
		\sum_{n \ne 0} e^{i n u} \frac{1}{n^2 - \z^2} \, = \, \z^{-2} \, - \, \frac{\pi}{\z} \Big[ \cot(\pi \z) \cos(\z u) \, + \, \sin| \z u | \Big] \, .
	\end{equation}
Therefore, the Fourier transform is given by 
	\begin{equation}
		\big\la \ve(u) \ve(0) \big\ra_\z \, = \, \frac{1}{2\pi \z^2 C} \, \bigg[ - \frac{1}{2} \, (|u| - \pi)^2 + \frac{\pi^2}{6} + \z^{-2}
		- \frac{\pi}{\z} \, \cot( \pi \z ) \cos( \z u ) - \frac{\pi}{\z} \, \sin| \z u | \bigg] \, .
	\end{equation}

\section{Monodromies in the BF description of deformed JT gravity}
\label{app:monodromiesinBFtheory}

Here, we will be very explicit about the relation of 2 dimensional quantum gravity to the BF gauge theory. This is discussed in many places, including in \cite{Jackiw:1992bw, Saad:2019lba}. It is important to distinguish between Lorentzian and Euclidean theories. Here, we will discuss the Euclidean theory. Following Kitaev \cite{Kitaev:2017hnr}, we write the fundamental representation of $\mathfrak{sl}(2,\mathbb{R})$ as
\begin{align}
\Lambda_0 = \frac{1}{2}\left(\begin{array}{cc}i&0\\0&-i\end{array}\right)\quad
\Lambda_1 = \frac{1}{2}\left(\begin{array}{cc}0&1\\1&0\end{array}\right)\quad
\Lambda_2 = \frac{1}{2}\left(\begin{array}{cc}0&i\\-i&0\end{array}\right)\nonumber
\end{align}
with commutators
\begin{align}
\left[\Lambda_1,\Lambda_0\right] =-\Lambda_2, \quad \left[\Lambda_2,\Lambda_0\right] =\Lambda_1,\quad \left[\Lambda_1,\Lambda_2\right] = -\Lambda_0.
\end{align}
If we worked in Lorentzian signature, to match with the notation of \cite{Jackiw:1992bw}, we would identify $P_1 = \Lambda_2, P_2 = \Lambda_1, J = \Lambda_0$, forming the set $J_A = \{J,P_1,P_2\}$ so that
\begin{align}
\left[P_1,P_2\right] = J, \quad \left[P_1,J\right] = P_2, \quad \left[P_2,J\right] = -P_1.
\end{align}
In our case in Euclidean signature, we match \cite{Saad:2019lba} and set $J =i\Lambda_2, P_1 = i\Lambda_0, P_2 = \Lambda_1$ and $J_A = \{J,P_1,P_2\}$, i.e.
\begin{align}
J = \frac{1}{2}\left(\begin{array}{cc} 0&-1\\1&0\end{array}\right),\quad P_1 = \frac{1}{2}\left(\begin{array}{cc}-1&0\\0&1\end{array}\right),\quad P_2 = \frac{1}{2}\left(\begin{array}{cc}0&1\\1&0\end{array}\right).
\end{align}
This choice makes the generators all real. The fundamental commutation relations are
\begin{align}
\left[P_1,P_2\right] = J,\quad \left[P_1,J\right]=  P_2,\quad \left[P_2,J\right] = -P_1.
\end{align} 
Then we have a Killing-Cartan metric
\begin{align}
tr(J_AJ_B) =\frac{1}{2}\eta_{AB},\quad \eta_{AB} = \left(\begin{array}{ccc}-1&0&0\\0&1&0\\0&0&1\end{array}\right).
\end{align}
so that $\mathfrak{sl}(2,\mathbb{R})\simeq \mathfrak{so}(1,2)$. Now, define a gauge field
\begin{align}
A = \left(J\omega_\mu + P_1 e^1_\mu + P_2 e^2_\mu \right)dx^\mu = \frac{1}{2}\left(\begin{array}{cc}-e^1 & e^2-\omega\\ e^2+\omega & e^1\end{array}\right)
\end{align}
where $e^a_\mu$ is the zweibein defined via $g_{\mu\nu} = e^{a}_\mu e^b_\nu \delta_{ab}$  and $\omega_\mu$ is the trace of the spin-connection of the two-dimensional AdS metric. Notably $e^{a}_\mu$ is a $2\times 2$ matrix. In component form,
\begin{align}
\omega_{\mu\;\;\;b}^{\;\;a}=e_\nu^a e^\lambda_b \Gamma^\nu_{\mu\lambda} - e^{\lambda}_b \partial_\mu e_\lambda^a
\end{align}
which is defined to satisfy the tetrad postulate, $\nabla_\mu e_\nu^a = 0$. We note that this can be rewritten as
\begin{align}
de^a = -\omega^{a}_{\;\;b}\wedge e^b.
\end{align}
Here, $a,b$ denote the orthonormal basis, whereas $\mu,\nu$ refer to the AdS basis. In particular, $e_{\lambda}^a = (e^{-1})^\lambda_{a}$. Up and down position of indices are irrelevant since we are in Euclidean signature. Furthermore, $\omega_{ab} = -\omega_{ab}$, which since we are in two dimensions implies that $\omega_{\mu\, 12} = -\omega_{\mu\,21}=\omega_\mu$ and hence defining $\omega=\omega_\mu dx^\mu$.
$\\ \\$
The gauge field has an associated field strength
\begin{align}
F = dA + A\wedge A
\end{align}
can be written in component form as
\begin{align}
F = (\mathcal{D}e)^aP_a + \left(d\omega +\frac{1}{2} \epsilon_{ab}e^{a}e^b\right)J = \frac{1}{2}\left(\begin{array}{cc}-(\mathcal{D}e)^1& (\mathcal{D}e)^2 - (d\omega + e^1\wedge e^2)\\ (\mathcal{D}e)^2 + (d\omega + e^1\wedge e^2) & (\mathcal{D}e)^1\end{array}\right)
\end{align}
where $(\mathcal{D}e)^a =de^a +\omega^{a}_{\;\;b}\wedge e^b $. Notably, $(\mathcal{D}e)^a$ is identical to the torsion density which vanishes by definition of $\omega$. In deriving this, we used that $\omega\wedge \omega = 0$. Next, the curvature two-form is defined as
\begin{align}
R^{a}_{\;b} = d\omega^{a}_{\;b} + \omega^{a}_{\;c}\wedge \omega^{c}_{\;b} = \frac{1}{2}e^{a}_\mu e^{\nu}_b R^{\mu}_{\;\;\nu\rho\lambda}dx^\rho\wedge dx^\lambda
\end{align}
and hence has 2 greek and 2 latin indices. In two dimensions, $R_{\mu\nu\rho\lambda} = \frac{1}{2}R\left(g_{\mu\rho}g_{\nu\lambda}-g_{\nu\rho}g_{\mu\lambda}\right)$ so that we have
\begin{align}
R^{a}_{\; b} = d\omega^a_b = \frac{1}{2} R e^a \wedge e_b.
\end{align}
As before, since $\omega_{ab}$ is antisymmetric, so is $R_{ab}$, and we also have the $d\omega^{a}_{b} = \epsilon^{a}_{b}d\omega$, serving as a definition of $d\omega$. Then we consider
\begin{align}
e^1 \wedge e^2 = e^1_\mu e^2_\nu dx^\mu\wedge dx^\nu =  \sqrt{g}d^2x.
\end{align}
where the last equality is the definition of the volume form. Combining these, we have
\begin{align}
dw = \frac{R}{2}\sqrt{g}d^2x.
\end{align}
Up to now, we haven't specified the theory we care about. Now, we specifically look at JT gravity,
\begin{align}
I_1 = \int d^2 x\sqrt{g} \phi(R+2)
\end{align}
where, for now, we have dropped the boundary terms.The equations of motion are
\begin{align}
R &= -2\nonumber\\
\left(\nabla_\mu\nabla_\nu - g_{\mu\nu}\nabla^2 + g_{\mu\nu}\right)\phi&= 0
\end{align}
By taking traces, we can rewrite these equations as
\begin{align}
\left(\nabla_\mu\nabla_\nu - \frac{1}{2}g_{\mu\nu}\nabla^2\right)\phi &=0\nonumber\\
\left(\nabla^2 - 2\right)\phi &=0.
\end{align}
For reasons that will become clear later, if we identify $\phi^0 = \phi$, then 
\begin{align}
\int d^2x\sqrt{g}\phi(R+2) = 2\int \phi^0\left(d\omega + e^1\wedge e^2\right).
\end{align}
We must in addition enforce that the solution is torsion-free, which can be implemented via two Lagrange multipliers, $\phi^1$ and $\phi^2$, and we have an action
\begin{align}
I_1' = \int\sqrt{g}\phi(R+2) = 2\int \phi^0 \left(d\omega + \frac{1}{2}\epsilon_{ab}e^ae^b\right) + \phi_a(\mathcal{D}e)^a.
\end{align}
It is easily seen that this reproduces $R=-2$ upon variation of $\phi^A$. 
 $\\ \\$
If instead, we vary $e^a$ and $\omega$, we find
\begin{align}
d\phi_a +\omega \epsilon_a^{\;\;b}\phi_b + \phi_0\epsilon_{ab}\phi^a &=0\nonumber\\
d\phi_0 + \epsilon^{a}_{b}\phi_a e^b &=0.
\end{align}
If we take a spacetime covariant derivative of the second and use the first equation, we recover $\left(\nabla_\mu\nabla_\nu - g_{\mu\nu}\nabla^2 - g_{\mu\nu}\right)\phi = 0$. 
$\\ \\$
Finally, to rewrite this as the ``BF'' gauge theory, we identify certain components of the theory. First, we write
\begin{align}
F = f^aP_a + fJ
\end{align}
so that there are {\it three} field strengths, combined as $F^A = (f,f^a)$. Importantly, we see that $F\in \mathfrak{sl}(2,\mathbb{R})$ since the field strengths just multiply the generators. Furthermore, our equations of motion actually imply that
\begin{align}
F = 0
\end{align}
which makes sense given the introduction of Lagrange multipliers. Hence, we are working with pure gauge $A$, i.e. flat connections.  In other words, we may write 
\begin{align}
A = g^{-1}dg
\end{align}
for $g \in \mathfrak{sl}(2,\mathbb{R})$. In component form, recall (it had been a while since I did any non-abelian gauge theory, so forgive me for writing this)
\begin{align}
g^{-1}dg = g^{-1}\partial_\mu g dx^\mu \Rightarrow d(g^{-1}dg) = -(g^{-1}\partial_\mu g)(g^{-1} \partial_\nu g) dx^\mu \wedge dx^\nu = -A\wedge A
\end{align} 
demonstrating that $F =0$ for this A.
$\\ \\$
Now, the Lagrangian can be written
\begin{align}
\mathcal{L}_1' =2\sum_{A=0}^2\phi_A F^A.
\end{align}
Defining a $2\times2$ matrix $B$ as 
\begin{align}
B = -i\left(\begin{array}{cc}-\phi^1&\phi^2+\phi\\\phi^2-\phi&\phi^1\end{array}\right)
\end{align}
we may also write the Lagrangian as
\begin{align}
\mathcal{L}_1' = i \Tr(BF)
\end{align}
where we have rescaled the Lagrangian and pulled out an overall factor of $i$ since we are in Euclidean signature. This is just a convention. In other words, we are describing the so-called ``BF'' theory. Importantly, we note that we can write $B$ as
\begin{align}
\frac{i}{2}B = -\phi^0J + \phi^1 P_1 + \phi^2 P_2.
\end{align}
Purely imaginary $\phi$ lead to $B\in \mathfrak{sl}(2,\mathbb{R})$. Finally, if we vary $A$ in the Lagrangian, we derive a new equation of motion for $B$,
\begin{align}
\partial_\mu B + [A_\mu, B] = 0
\end{align}
which implies that
\begin{align}
B = g^{-1}\Phi g
\end{align}
for some constant (i.e. $d\Phi = 0$) element $\Phi\in \mathfrak{sl}(2,\mathbb{R})$. Here $g$ is the same group element used to define $A$. Now, note that under a group transformation, 
\begin{align}
A \to A' = g_1^{-1} A g_1 + g_1^{-1}d g_1
\end{align}
which of course, since $A$ is pure gauge, can be written as $A' = g_2^{-1}dg_2$ where $g_2 = gg_1$. Hence,
\begin{align}
B' = g_2^{-1}\Phi g_2
\end{align}
In order for this to agree with the group transformation for $B$, we must also have $g_1$ be a constant element. Hence, there is a residual global $\mathfrak{sl}(2,\mathbb{R})$ symmetry.

\subsection{BF theory on the disk}
Now, let's consider the theory on the disk with solution for the metric and dilaton is
\begin{align}
ds^2 = d\rho^2 + \sinh(\rho)^2d\vp^2, \quad \phi=  C\;\cosh\rho
\end{align}
The zweibeins are
\begin{align}
e^1 = d\rho,\quad e^2 = \sinh(\rho) d\vp
\end{align}
 We use
\begin{align}
de^1 = 0, \; de^2 = \cosh(\rho)d\rho\wedge d\vp = -\omega_{21}\wedge e^1 \Rightarrow \omega_{12} = -\cosh(\rho) d\vp
\end{align}
Since $\omega_{12} = \omega$, we can write
\begin{align}
A = \frac{1}{2}\left(\begin{array}{cc}-d\rho &\exp(\rho) d\vp\\ -\exp(-\rho)d\vp & d\rho\end{array}\right).
\end{align}
Now, we must solve the equations of motion for $B$. We start by writing
\begin{align}
B = -i \left(\begin{array}{cc} -\phi^1 & \phi^2 + C\cosh(\rho)\\ \phi^2 - C\cosh(\rho) &\phi^1\end{array}\right)
\end{align}
The equation of motion
\begin{align}
\partial_\mu B =- [A_\mu,B] 
\end{align}
with solution
\begin{align}
B = -i\left(\begin{array}{cc} 0& C\exp(\rho)\\ -C\exp(-\rho) & 0\end{array}\right) = - i CA_\vp.
\end{align}
Now, we would like to recover the Schwarzian theory from the BF theory. As in the second order formalism, we must introduce a boundary term. Notably, since $B\propto A_\vp$, we would like a boundary term that respects this. In other words, we would like on the boundary to have
\begin{align}
A_\vp \delta B - B\delta A_\vp |_{\partial M} = 0
\end{align}
which would imply $B\propto A_\vp$. A convenient choice of boundary is then
\begin{align}
I = -i\int_M \Tr(BF) + \frac{i}{2}\int_{\partial M} \Tr (B A).
\end{align}
where it is understood that $A\to A_\vp$ on the boundary. Now, we may integrate out the bulk degrees of freedom and arrive at
\begin{align}
I = \frac{1}{2l} \int_{\partial M} \Tr(A_\vp^2) d\vp
\end{align}
More generally, if we parametrize the boundary as some curve $\{\vp(w), \rho(w)\}$ and $\phi = \phi_b/\epsilon$ on this boundary, we have
\begin{align}
I = -\phi_b \int dw \; \text{Sch}(w)
\end{align}
where
\begin{align}
\text{Sch}(w) \equiv \text{Sch}\left(\tan\frac{\vp(w)}{2},w\right)
\end{align}
and the right hand side means to take the Schwarzian derivative of $\tan(\vp(w)/2)$ with respect to $w$. To arrive at this, we write that asymptotically (see \cite{Cotler:2019nbi}), 
\begin{align}
A = \frac{dr}{2}\left(\begin{array}{cc}-1&0\\0&1\end{array}\right) + \frac{dw}{2}\left(\begin{array}{cc}0&e^\rho\\-2\text{Sch}(w) e^{-\rho} & 0 \end{array}\right).
\end{align}
We note that the boundary has
\begin{align}
\phi_b = C \frac{e^{\rho_0}}{2}
\end{align}
and we require $e^{\rho_0}=\beta/\pi\epsilon$. Here, we have defined
\begin{align}
\beta = \lim_{\rho_0\to\infty} \frac{L}{\phi_b}
\end{align}
for the boundary circumference $L$, hence $C=2\pi/\beta$. Notably, we can represent 
\begin{align}
A_w = \left(\begin{array}{cc} 0 & -\text{Sch}(w)/2\\ 1&0\end{array}\right)
\end{align}
which can be arrived at from a global $sl(2,\mathbb{R})$ transformation. This matrix will be important for characterizing the monodromies/holonomies.

\subsection{Monodromies in the BF theory}
Following the discussion in section \ref{sec:wavefunctioncone}, now we care about flat connections which reproduce the representation of $A_\vp$ above on the boundary. In other words, we want
\begin{align}
A_w = \tilde{g}^{-1}\partial_w \tilde{g} = \left(\begin{array}{cc}0&-T(w)/2\\1&0\end{array}\right)
\end{align}
where $T(w) = \text{Sch}(w)$. As pointed out in \cite{Mertens:2019tcm}, these are easily classified by representing
\begin{align}
\tilde{g} = \left(\begin{array}{cc} A(w) & B(w)\\ C(w) & D(w)\end{array}\right)
\end{align}
with solutions to 
\begin{align}
A'' + \frac{A}{2}T &= 0,\quad B = A'\\
C'' + \frac{C}{2}T &=0,\quad D=C'
\end{align}
and $AC'- CA' =1$. Here, primes represent derivatives with respect to $w$. Then $\text{Sch}(\frac{A}{C},w) = T(w)$. Recalling that $\tilde{g}(w+2\pi) = \tilde{g}(w)U(2\pi)$ defines a monodromy matrix $U$, we may relate $U$ to solutions $F \equiv A/C$ via
\begin{align}
F(w+2\pi) \to \frac{a F(w) + b}{c F(w) + d},\quad U = \left(\begin{array}{cc} a & b\\ c& d\end{array}\right) \in \mathfrak{sl}(2,\mathbb{R})
\end{align}
which should be considered conjugate to $F$. Since $SL(2,\mathbb{R})$ is a gauge symmetry of the Schwarzian action, the boundary has its own conjugacy relation
\begin{align}
g \sim g\cdot S, \quad S \in SL(2,\mathbb{R})
\end{align}
which requires that
\begin{align}
U \sim S\cdot U \cdot S^{-1}.
\end{align}
This important fact leads to the modification of the density of states when we consider the path integral measure since the only gauge equivalent $S$ of the boundary which must be accounted for are those $S$ which commute with $U$.

Now, we will demonstrate that the monodromy matrix
\begin{align}
U = \left(\begin{array}{cc} \cos(\pi\zeta) & \sin(\pi\zeta)\\ -\sin(\pi\zeta) & \cos(\pi\zeta)\end{array}\right)\in U(1)_\zeta
\end{align}
leads to a conical deficit. First, note that this monodromy matrix means that
\begin{align}
F(w+2\pi) = \frac{F(w) + \tan{\pi\zeta}}{1-\tan(\pi\zeta)F(w)}
\label{Ftransformation}
\end{align}
For $\gamma = 1$, we have that $F(w+2\pi) = F(w)$ which is the case for the Poincar\'e disk. Hence, in this case, 
\begin{align}
F(w) = \tan\left(\frac{\vp(w)}{2}\right)
\end{align}
which is the natural parametrization for the Schwarzian of Euclidean AdS. This requires that the boundary time reparametrization respect $\vp(w+2\pi) = \vp+2\pi$. A simple modification
\begin{align}
F(w) = \tan\left(\frac{\zeta}{2}\vp(w)\right)
\end{align}
can easily be seen to satisfy eq.~(\ref{Ftransformation}) for generic $\zeta$ as long as we maintain $\vp(w+2\pi) = \vp +2\pi$. Next, we can choose a specific gauge to connect the bulk coordinates to boundary coordinates. A simple choice is to use conformal gauge
\begin{align}
ds^2 = \frac{\partial_+ X^+ \partial_- X^-}{(X^+-X^-)^2}dx^+dx^-
\end{align}
and let $X^{\pm} = F(x^\pm)$. Near the horizon ($r\to 0$), we find
\begin{align}
ds^2 \approx \frac{ r^2\zeta^2d\vp^2 + dr^2}{\zeta^2}
\end{align}
so that there is a conical deficit $2\pi\zeta$. Another way to see this is by choosing a gauge with the same conical deficit,
\begin{align}
ds^2 = d\rho^2 + \zeta^2\sinh^2\rho d\vp^2
\end{align}
and perform the same first order formalism matching we saw earlier. Here, $\vp \in [0,2\pi)$. In particular
\begin{align}
A_\vp = \frac{1}{2}\left(\begin{array}{cc} 0&\zeta\exp(\rho)\\ -\zeta\exp(-\rho)&0\end{array}\right)
\end{align}
so that there is a holonomy around the origin equal to the monodromy
\begin{align}
\mathcal{P}\exp\left(\int_0^{2\pi} A_\vp d\vp\right) = \left(\begin{array}{cc}\cos(\zeta\pi)&\sin(\zeta\pi)\\-\sin(\zeta\pi)&\cos(\zeta\pi)\end{array}\right)
\end{align}

Now, let's understand how this appears in the path integral. In parametrizing our boundary, we are free to choose any $\vp(w)$ that obeys $\vp(w+2\pi) = \vp(w)+2\pi$. In other words, the boundary is a gauge theory with gauge group $diff(S_1)$. At the same time, the bulk and boundary are invariant under global $PSL(2,\mathbb{R})$ transformations, at least when we work with no deficits. In other words, transformations
\begin{align}
\tan(\frac{f(w)}{2})\to \frac{a\tan(\frac{f(w)}{2}) + b}{c\tan(\frac{f(w)}{2})+d}
\end{align}
with $ad-bc =1$ and $\{a,b,c,d\}\sim-\{a,b,c,d\}$ leave the boundary metric invariant. Notably, the Schwarzian action is also invariant under this transformation. Hence, it is a gauge theory with gauge group $diff(S_1)/PSL(2,\mathbb{R})$. 

On the other hand, when we have a conical deficit, while generic fractional linear transforms are still a symmetry of the Schwarzian, it is only the diagonal $U(1)_\gamma$ which preserves the boundary. The best way to think about this is actually in terms of the connection $A = g^{-1}dg$. Starting with the disk, we can choose a connection parametrized in terms of the $SL(2,\mathbb{R})$ generators (this follows \cite{Cotler:2019nbi} and is different than the parametrization we use in the main text),
\begin{align}
g = \exp(-\vp J)\exp(\alpha_1 P_1)\exp(\Psi[P_2-J]) = \left(\begin{array}{cc}\cos(\frac{\vp}{2})&\sin(\frac{\vp}{2})\\-\sin(\frac{\vp}{2})&\cos(\frac{\vp}{2})\end{array}\right)\left(\begin{array}{cc}\Lambda&0\\0&\Lambda^{-1}\end{array}\right)\left(\begin{array}{cc}1&\Psi\\0&1\end{array}\right)
\end{align}
where $\Lambda = -\log(\alpha_1)>0$ and $\Psi \in \mathbb{R}$. It is easily checked that matching $A$ to its boundary values for $\Lambda, \Psi$ returns the Schwarzian action. Furthermore, $g$ and $hg$ for $h\in PSL(2,\mathbb{R})$ give the same flat connection. Notably, choosing 
\begin{align}
h = \left(\begin{array}{cc} d &-c\\-b&a\end{array}\right): \tan(\frac{\vp}{2}) \to \frac{a  \tan(\frac{\vp}{2})+b}{c \tan(\frac{\vp}{2})+d}
\end{align}
In other words, this gives a left coset construction. To introduce a conical deficit, we must introduce a monodromy, which is equivalent to letting
\begin{align}
\tilde{g} = \exp(w\lambda)g
\end{align}
for the same $g$ above. Here, $\lambda$ is some matrix parametrizing the monodromy. Given our earlier monodromy matrix, a good choice is $\lambda = -2\zeta J$. Then we have
\begin{align}
\tilde{g}=\left(\begin{array}{cc}\cos(\frac{\zeta w+\vp}{2})&\sin(\frac{\zeta w +\vp}{2})\\-\sin(\frac{\zeta w +\vp}{2})&\cos(\frac{\zeta w+\vp}{2})\end{array}\right)\left(\begin{array}{cc}\Lambda&0\\0&\Lambda^{-1}\end{array}\right)\left(\begin{array}{cc}1&\Psi\\0&1\end{array}\right)
\end{align}
Finally, redefining $\zeta w+ \vp = \zeta\tilde{\vp}$, if $\vp(w+2\pi) = \vp(w)$, now we have $\tilde{\vp}(w+2\pi) = \tilde{\vp}+2\pi\zeta$ describing a space with conical deficit. The important point is that the coset construction now only includes $h = \exp(\sigma J) \in U(1)$ with $\lambda\to h\lambda h^{-1}$ and $g\to h g$. Hence we must have a different measure in the path integral. We note that this discussion is identical to the BF discussion in \ref{monodromyBF}. This just gives an explicit parametrization. We are free to choose either a right or left coset construction.

Finally, we are ready to look at the path integral. The classical solutions are always of the form
\begin{align}
\tilde{\vp}(w) = w
\end{align}
which is easily seen to satisfy $\partial_w\{\tan(\zeta w/2),w\} = 0$. For $PSL(2,\mathbb{R})$, $\tilde{\vp}=\vp$, though this is not the case with conical deficits. The challenge is understanding the fluctuations. A generic element of $PSL(2,\mathbb{R})$ is specified by three numbers
\begin{align}
\lambda = \lambda^0 J + \lambda^1 P_1 + \lambda^2 P_2.
\end{align}
Hence, the quotient construction allows for the fixing of three modes of the fluctuation. On the other hand $U(1)$ is specified by only one number and we can fix only one mode. In terms of $\tilde{\vp}$, we write the mode expansion as
\begin{align}
\tilde{\vp(w)} = w + \sum_{n/\mathcal{H}} \epsilon_n e^{in w}
\end{align}
where $\epsilon_{-n} = \epsilon_n^{*}$ so that $\tilde{\vp}$ is real. For $PSL(2,\mathbb{R})$, we fix $\epsilon_{-1,0,1}=0$ whereas for $U(1)$ we can only fix $\epsilon_0 = 0$. This quotient suffices to eliminate the $vol(\mathcal{H})$ in the measure. On the other hand, we must include a measure for the modes $\epsilon_n$. The natural measure derived from the group is the Haar measure which is equivalent to the Pfaffian of a symplectic form
\begin{align}
\omega = C\int_0^{2\pi} \left(\frac{d\vp'\wedge d\vp''}{(\vp')^2}-d\vp\wedge d\vp'\right)dw
\end{align}
in terms of the naturally conjugate variable $de_n$ and $de_n^*$. In JT gravity, $C = 1/8\pi G$. Under a natural rescaling $w = 2\pi \tau/\beta$, we find that
\begin{align}
Z_{D_\alpha}(\beta) = \int \frac{[d\vp(\tau)]Pf(\omega)}{U(1)_\gamma}\exp\left(-\int_0^\beta d\tau H\right)= \frac{\exp(\frac{2\pi^2\zeta^2}{\beta})}{\sqrt{2\pi \beta}}
\end{align}
Compare this the partition function for the disk
\begin{align}
Z_D(\beta) = \int \frac{[d\vp(\tau)]Pf(\omega)}{PSL(2,\mathbb{R})}\exp\left(-\int_0^\beta d\tau H\right) =\frac{\exp(\frac{(2\pi^2)}{\beta})}{\beta\sqrt{2\pi\beta}}
\end{align}
The exponential piece differs only by the conical deficit. The overall factor differs by $\beta^{-1}$ which arises from the two extra zero modes in the conical deficit construction.


\bibliographystyle{JHEP}
\bibliography{Refs} 

\providecommand{\href}[2]{#2}\begingroup\raggedright\begin{thebibliography}{10}

\bibitem{Jackiw:1984je}
R.~Jackiw, \emph{{Lower Dimensional Gravity}},
  \href{https://doi.org/10.1016/0550-3213(85)90448-1}{\emph{Nucl. Phys. B}
  {\bfseries 252} (1985) 343}.

\bibitem{Teitelboim:1983ux}
C.~Teitelboim, \emph{{Gravitation and Hamiltonian Structure in Two Space-Time
  Dimensions}}, \href{https://doi.org/10.1016/0370-2693(83)90012-6}{\emph{Phys.
  Lett. B} {\bfseries 126} (1983) 41}.

\bibitem{Almheiri:2014cka}
A.~Almheiri and J.~Polchinski, \emph{{Models of AdS$_{2}$ backreaction and
  holography}}, \href{https://doi.org/10.1007/JHEP11(2015)014}{\emph{JHEP}
  {\bfseries 11} (2015) 014} [\href{https://arxiv.org/abs/1402.6334}{{\ttfamily
  1402.6334}}].

\bibitem{Maldacena:2016upp}
J.~Maldacena, D.~Stanford and Z.~Yang, \emph{{Conformal symmetry and its
  breaking in two dimensional Nearly Anti-de-Sitter space}},
  \href{https://doi.org/10.1093/ptep/ptw124}{\emph{PTEP} {\bfseries 2016}
  (2016) 12C104} [\href{https://arxiv.org/abs/1606.01857}{{\ttfamily
  1606.01857}}].

\bibitem{Jensen:2016pah}
K.~Jensen, \emph{{Chaos in AdS$_2$ Holography}},
  \href{https://doi.org/10.1103/PhysRevLett.117.111601}{\emph{Phys. Rev. Lett.}
  {\bfseries 117} (2016) 111601}
  [\href{https://arxiv.org/abs/1605.06098}{{\ttfamily 1605.06098}}].

\bibitem{Engelsoy:2016xyb}
J.~Engels\"oy, T.~G. Mertens and H.~Verlinde, \emph{{An investigation of
  AdS$_{2}$ backreaction and holography}},
  \href{https://doi.org/10.1007/JHEP07(2016)139}{\emph{JHEP} {\bfseries 07}
  (2016) 139} [\href{https://arxiv.org/abs/1606.03438}{{\ttfamily
  1606.03438}}].

\bibitem{Sachdev:1992fk}
S.~Sachdev and J.~Ye, \emph{{Gapless spin fluid ground state in a random,
  quantum Heisenberg magnet}},
  \href{https://doi.org/10.1103/PhysRevLett.70.3339}{\emph{Phys. Rev. Lett.}
  {\bfseries 70} (1993) 3339}
  [\href{https://arxiv.org/abs/cond-mat/9212030}{{\ttfamily
  cond-mat/9212030}}].

\bibitem{Polchinski:2016xgd}
J.~Polchinski and V.~Rosenhaus, \emph{{The Spectrum in the Sachdev-Ye-Kitaev
  Model}}, \href{https://doi.org/10.1007/JHEP04(2016)001}{\emph{JHEP}
  {\bfseries 04} (2016) 001}
  [\href{https://arxiv.org/abs/1601.06768}{{\ttfamily 1601.06768}}].

\bibitem{Maldacena:2016hyu}
J.~Maldacena and D.~Stanford, \emph{{Remarks on the Sachdev-Ye-Kitaev model}},
  \href{https://doi.org/10.1103/PhysRevD.94.106002}{\emph{Phys. Rev. D}
  {\bfseries 94} (2016) 106002}
  [\href{https://arxiv.org/abs/1604.07818}{{\ttfamily 1604.07818}}].

\bibitem{Jevicki:2016bwu}
A.~Jevicki, K.~Suzuki and J.~Yoon, \emph{{Bi-Local Holography in the SYK
  Model}}, \href{https://doi.org/10.1007/JHEP07(2016)007}{\emph{JHEP}
  {\bfseries 07} (2016) 007}
  [\href{https://arxiv.org/abs/1603.06246}{{\ttfamily 1603.06246}}].

\bibitem{Jevicki:2016ito}
A.~Jevicki and K.~Suzuki, \emph{{Bi-Local Holography in the SYK Model:
  Perturbations}}, \href{https://doi.org/10.1007/JHEP11(2016)046}{\emph{JHEP}
  {\bfseries 11} (2016) 046}
  [\href{https://arxiv.org/abs/1608.07567}{{\ttfamily 1608.07567}}].

\bibitem{Saad:2019lba}
P.~Saad, S.~H. Shenker and D.~Stanford, \emph{{JT gravity as a matrix
  integral}},  \href{https://arxiv.org/abs/1903.11115}{{\ttfamily 1903.11115}}.

\bibitem{Stanford:2019vob}
D.~Stanford and E.~Witten, \emph{{JT Gravity and the Ensembles of Random Matrix
  Theory}},  \href{https://arxiv.org/abs/1907.03363}{{\ttfamily 1907.03363}}.

\bibitem{Maxfield:2020ale}
H.~Maxfield and G.~J. Turiaci, \emph{{The path integral of 3D gravity near
  extremality; or, JT gravity with defects as a matrix integral}},
  \href{https://arxiv.org/abs/2006.11317}{{\ttfamily 2006.11317}}.

\bibitem{Witten:2020wvy}
E.~Witten, \emph{{Matrix Models and Deformations of JT Gravity}},
  \href{https://arxiv.org/abs/2006.13414}{{\ttfamily 2006.13414}}.

\bibitem{Bagrets:2016cdf}
D.~Bagrets, A.~Altland and A.~Kamenev,
  \emph{{Sachdev\textendash{}Ye\textendash{}Kitaev model as Liouville quantum
  mechanics}},
  \href{https://doi.org/10.1016/j.nuclphysb.2016.08.002}{\emph{Nucl. Phys. B}
  {\bfseries 911} (2016) 191}
  [\href{https://arxiv.org/abs/1607.00694}{{\ttfamily 1607.00694}}].

\bibitem{Stanford:2017thb}
D.~Stanford and E.~Witten, \emph{{Fermionic Localization of the Schwarzian
  Theory}}, \href{https://doi.org/10.1007/JHEP10(2017)008}{\emph{JHEP}
  {\bfseries 10} (2017) 008}
  [\href{https://arxiv.org/abs/1703.04612}{{\ttfamily 1703.04612}}].

\bibitem{Mertens:2017mtv}
T.~G. Mertens, G.~J. Turiaci and H.~L. Verlinde, \emph{{Solving the Schwarzian
  via the Conformal Bootstrap}},
  \href{https://doi.org/10.1007/JHEP08(2017)136}{\emph{JHEP} {\bfseries 08}
  (2017) 136} [\href{https://arxiv.org/abs/1705.08408}{{\ttfamily
  1705.08408}}].

\bibitem{Bardeen:1999px}
J.~M. Bardeen and G.~T. Horowitz, \emph{{The Extreme Kerr throat geometry: A
  Vacuum analog of AdS(2) x S**2}},
  \href{https://doi.org/10.1103/PhysRevD.60.104030}{\emph{Phys. Rev. D}
  {\bfseries 60} (1999) 104030}
  [\href{https://arxiv.org/abs/hep-th/9905099}{{\ttfamily hep-th/9905099}}].

\bibitem{NavarroSalas:1999up}
J.~Navarro-Salas and P.~Navarro, \emph{{AdS(2) / CFT(1) correspondence and near
  extremal black hole entropy}},
  \href{https://doi.org/10.1016/S0550-3213(00)00165-6}{\emph{Nucl. Phys. B}
  {\bfseries 579} (2000) 250}
  [\href{https://arxiv.org/abs/hep-th/9910076}{{\ttfamily hep-th/9910076}}].

\bibitem{Nayak:2018qej}
P.~Nayak, A.~Shukla, R.~M. Soni, S.~P. Trivedi and V.~Vishal, \emph{{On the
  Dynamics of Near-Extremal Black Holes}},
  \href{https://doi.org/10.1007/JHEP09(2018)048}{\emph{JHEP} {\bfseries 09}
  (2018) 048} [\href{https://arxiv.org/abs/1802.09547}{{\ttfamily
  1802.09547}}].

\bibitem{Sachdev:2019bjn}
S.~Sachdev, \emph{{Universal low temperature theory of charged black holes with
  AdS$_2$ horizons}}, \href{https://doi.org/10.1063/1.5092726}{\emph{J. Math.
  Phys.} {\bfseries 60} (2019) 052303}
  [\href{https://arxiv.org/abs/1902.04078}{{\ttfamily 1902.04078}}].

\bibitem{Ghosh:2019rcj}
A.~Ghosh, H.~Maxfield and G.~J. Turiaci, \emph{{A universal Schwarzian sector
  in two-dimensional conformal field theories}},
  \href{https://doi.org/10.1007/JHEP05(2020)104}{\emph{JHEP} {\bfseries 05}
  (2020) 104} [\href{https://arxiv.org/abs/1912.07654}{{\ttfamily
  1912.07654}}].

\bibitem{Penington:2019npb}
G.~Penington, \emph{{Entanglement Wedge Reconstruction and the Information
  Paradox}}, \href{https://doi.org/10.1007/JHEP09(2020)002}{\emph{JHEP}
  {\bfseries 09} (2020) 002}
  [\href{https://arxiv.org/abs/1905.08255}{{\ttfamily 1905.08255}}].

\bibitem{Almheiri:2019psf}
A.~Almheiri, N.~Engelhardt, D.~Marolf and H.~Maxfield, \emph{{The entropy of
  bulk quantum fields and the entanglement wedge of an evaporating black
  hole}}, \href{https://doi.org/10.1007/JHEP12(2019)063}{\emph{JHEP} {\bfseries
  12} (2019) 063} [\href{https://arxiv.org/abs/1905.08762}{{\ttfamily
  1905.08762}}].

\bibitem{Witten:1990hr}
E.~Witten, \emph{{Two-dimensional gravity and intersection theory on moduli
  space}}, \href{https://doi.org/10.4310/SDG.1990.v1.n1.a5}{\emph{Surveys Diff.
  Geom.} {\bfseries 1} (1991) 243}.

\bibitem{Jackiw:1992bw}
R.~Jackiw, \emph{{Gauge theories for gravity on a line}},
  \href{https://doi.org/10.1007/BF01017075}{\emph{Theor. Math. Phys.}
  {\bfseries 92} (1992) 979}
  [\href{https://arxiv.org/abs/hep-th/9206093}{{\ttfamily hep-th/9206093}}].

\bibitem{Mertens:2018fds}
T.~G. Mertens, \emph{{The Schwarzian theory \textemdash{} origins}},
  \href{https://doi.org/10.1007/JHEP05(2018)036}{\emph{JHEP} {\bfseries 05}
  (2018) 036} [\href{https://arxiv.org/abs/1801.09605}{{\ttfamily
  1801.09605}}].

\bibitem{Blommaert:2018oro}
A.~Blommaert, T.~G. Mertens and H.~Verschelde, \emph{{The Schwarzian Theory - A
  Wilson Line Perspective}},
  \href{https://doi.org/10.1007/JHEP12(2018)022}{\emph{JHEP} {\bfseries 12}
  (2018) 022} [\href{https://arxiv.org/abs/1806.07765}{{\ttfamily
  1806.07765}}].

\bibitem{Cotler:2019nbi}
J.~Cotler, K.~Jensen and A.~Maloney, \emph{{Low-dimensional de Sitter quantum
  gravity}}, \href{https://doi.org/10.1007/JHEP06(2020)048}{\emph{JHEP}
  {\bfseries 06} (2020) 048}
  [\href{https://arxiv.org/abs/1905.03780}{{\ttfamily 1905.03780}}].

\bibitem{Yang:2018gdb}
Z.~Yang, \emph{{The Quantum Gravity Dynamics of Near Extremal Black Holes}},
  \href{https://doi.org/10.1007/JHEP05(2019)205}{\emph{JHEP} {\bfseries 05}
  (2019) 205} [\href{https://arxiv.org/abs/1809.08647}{{\ttfamily
  1809.08647}}].

\bibitem{Kitaev:2018wpr}
A.~Kitaev and S.~J. Suh, \emph{{Statistical mechanics of a two-dimensional
  black hole}}, \href{https://doi.org/10.1007/JHEP05(2019)198}{\emph{JHEP}
  {\bfseries 05} (2019) 198}
  [\href{https://arxiv.org/abs/1808.07032}{{\ttfamily 1808.07032}}].

\bibitem{Mertens:2019tcm}
T.~G. Mertens and G.~J. Turiaci, \emph{{Defects in Jackiw-Teitelboim Quantum
  Gravity}}, \href{https://doi.org/10.1007/JHEP08(2019)127}{\emph{JHEP}
  {\bfseries 08} (2019) 127}
  [\href{https://arxiv.org/abs/1904.05228}{{\ttfamily 1904.05228}}].

\bibitem{Comtet:1984mm}
A.~Comtet and P.~Houston, \emph{{Effective Action on the Hyperbolic Plane in a
  Constant External Field}}, \href{https://doi.org/10.1063/1.526781}{\emph{J.
  Math. Phys.} {\bfseries 26} (1985) 185}.

\bibitem{Comtet:1986ki}
A.~Comtet, \emph{{On the Landau Levels on the Hyperbolic Plane}},
  \href{https://doi.org/10.1016/0003-4916(87)90098-4}{\emph{Annals Phys.}
  {\bfseries 173} (1987) 185}.

\bibitem{Pioline:2005pf}
B.~Pioline and J.~Troost, \emph{{Schwinger pair production in AdS(2)}},
  \href{https://doi.org/10.1088/1126-6708/2005/03/043}{\emph{JHEP} {\bfseries
  03} (2005) 043} [\href{https://arxiv.org/abs/hep-th/0501169}{{\ttfamily
  hep-th/0501169}}].

\bibitem{Deser:1988qn}
S.~Deser and R.~Jackiw, \emph{{Classical and Quantum Scattering on a Cone}},
  \href{https://doi.org/10.1007/BF01466729}{\emph{Commun. Math. Phys.}
  {\bfseries 118} (1988) 495}.

\bibitem{Carslaw1}
H.~S. Carslaw, \emph{{Some multiform solutions of the partial differential
  equations of physical mathematics and their applications}}, {\emph{Proc.
  London Math. Soc.} {\bfseries 30} (1899) 121}.

\bibitem{Carslaw2}
H.~S. Carslaw, \emph{{The Green's function for a wedge of any angle, and other
  problems in the conduction of heat}}, {\emph{Proc. London Math. Soc.}
  {\bfseries 8} (1910) 365}.

\bibitem{Carslaw3}
H.~S. Carslaw, \emph{{Diffraction of waves by a wedge of any angle}},
  {\emph{Proc. London Math. Soc.} {\bfseries 18} (1920) 291}.

\bibitem{Grosche:1998ff}
C.~Grosche, \emph{{On the path integral treatment for an Aharonov-Bohm field on
  the hyperbolic plane}},
  \href{https://doi.org/10.1023/A:1026629607510}{\emph{Int. J. Theor. Phys.}
  {\bfseries 38} (1999) 955}
  [\href{https://arxiv.org/abs/quant-ph/9808060}{{\ttfamily
  quant-ph/9808060}}].

\bibitem{Lisovyy:2007mj}
O.~Lisovyy, \emph{{Aharonov-Bohm effect on the Poincare disk}},
  \href{https://doi.org/10.1063/1.2738751}{\emph{J. Math. Phys.} {\bfseries 48}
  (2007) 052112} [\href{https://arxiv.org/abs/math-ph/0702066}{{\ttfamily
  math-ph/0702066}}].

\bibitem{Stanford:2020qhm}
D.~Stanford and Z.~Yang, \emph{{Finite-cutoff JT gravity and self-avoiding
  loops}},  \href{https://arxiv.org/abs/2004.08005}{{\ttfamily 2004.08005}}.

\bibitem{Kitaev:2017hnr}
A.~Kitaev, \emph{{Notes on $\widetilde{\mathrm{SL}}(2,\mathbb{R})$
  representations}},  \href{https://arxiv.org/abs/1711.08169}{{\ttfamily
  1711.08169}}.

\bibitem{ssolvable}
S.~Albeverio, F.~Gesztesy, P.~Hoegh-Krohn, H.~Holden and P.~Exner,
  \emph{{Solvable Models in Quantum Mechanics}}. Springer-Verlag, Berlin, 1988.

\bibitem{renyi1961}
A.~R\'enyi, \emph{On measures of entropy and information},  in
  \emph{Proceedings of the Fourth Berkeley Symposium on Mathematical Statistics
  and Probability, Volume 1: Contributions to the Theory of Statistics},
  (Berkeley, Calif.), pp.~547--561, University of California Press, 1961.

\bibitem{Calabrese:2004eu}
P.~Calabrese and J.~L. Cardy, \emph{{Entanglement entropy and quantum field
  theory}}, \href{https://doi.org/10.1088/1742-5468/2004/06/P06002}{\emph{J.
  Stat. Mech.} {\bfseries 0406} (2004) P06002}
  [\href{https://arxiv.org/abs/hep-th/0405152}{{\ttfamily hep-th/0405152}}].

\bibitem{Nishioka:2009un}
T.~Nishioka, S.~Ryu and T.~Takayanagi, \emph{{Holographic Entanglement Entropy:
  An Overview}}, \href{https://doi.org/10.1088/1751-8113/42/50/504008}{\emph{J.
  Phys. A} {\bfseries 42} (2009) 504008}
  [\href{https://arxiv.org/abs/0905.0932}{{\ttfamily 0905.0932}}].

\bibitem{Rangamani:2016dms}
M.~Rangamani and T.~Takayanagi, \emph{{Holographic Entanglement Entropy}},
  vol.~931. Springer, 2017,
  \href{https://doi.org/10.1007/978-3-319-52573-0}{10.1007/978-3-319-52573-0},
  [\href{https://arxiv.org/abs/1609.01287}{{\ttfamily 1609.01287}}].

\bibitem{Lewkowycz:2013nqa}
A.~Lewkowycz and J.~Maldacena, \emph{{Generalized gravitational entropy}},
  \href{https://doi.org/10.1007/JHEP08(2013)090}{\emph{JHEP} {\bfseries 08}
  (2013) 090} [\href{https://arxiv.org/abs/1304.4926}{{\ttfamily 1304.4926}}].

\bibitem{Lin:2018xkj}
J.~Lin, \emph{{Entanglement entropy in Jackiw-Teitelboim Gravity}},
  \href{https://arxiv.org/abs/1807.06575}{{\ttfamily 1807.06575}}.

\bibitem{Jafferis:2019wkd}
D.~L. Jafferis and D.~K. Kolchmeyer, \emph{{Entanglement Entropy in
  Jackiw-Teitelboim Gravity}},
  \href{https://arxiv.org/abs/1911.10663}{{\ttfamily 1911.10663}}.

\bibitem{Kitaev:2005dm}
A.~Kitaev and J.~Preskill, \emph{{Topological entanglement entropy}},
  \href{https://doi.org/10.1103/PhysRevLett.96.110404}{\emph{Phys. Rev. Lett.}
  {\bfseries 96} (2006) 110404}
  [\href{https://arxiv.org/abs/hep-th/0510092}{{\ttfamily hep-th/0510092}}].

\bibitem{Maloney:2007ud}
A.~Maloney and E.~Witten, \emph{{Quantum Gravity Partition Functions in Three
  Dimensions}}, \href{https://doi.org/10.1007/JHEP02(2010)029}{\emph{JHEP}
  {\bfseries 02} (2010) 029} [\href{https://arxiv.org/abs/0712.0155}{{\ttfamily
  0712.0155}}].

\bibitem{Shenker:2013pqa}
S.~H. Shenker and D.~Stanford, \emph{{Black holes and the butterfly effect}},
  \href{https://doi.org/10.1007/JHEP03(2014)067}{\emph{JHEP} {\bfseries 03}
  (2014) 067} [\href{https://arxiv.org/abs/1306.0622}{{\ttfamily 1306.0622}}].

\bibitem{Maldacena:2015waa}
J.~Maldacena, S.~H. Shenker and D.~Stanford, \emph{{A bound on chaos}},
  \href{https://doi.org/10.1007/JHEP08(2016)106}{\emph{JHEP} {\bfseries 08}
  (2016) 106} [\href{https://arxiv.org/abs/1503.01409}{{\ttfamily
  1503.01409}}].

\bibitem{Jahnke:2019gxr}
V.~Jahnke, K.-Y. Kim and J.~Yoon, \emph{{On the Chaos Bound in Rotating Black
  Holes}}, \href{https://doi.org/10.1007/JHEP05(2019)037}{\emph{JHEP}
  {\bfseries 05} (2019) 037}
  [\href{https://arxiv.org/abs/1903.09086}{{\ttfamily 1903.09086}}].

\bibitem{Shenker:2014cwa}
S.~H. Shenker and D.~Stanford, \emph{{Stringy effects in scrambling}},
  \href{https://doi.org/10.1007/JHEP05(2015)132}{\emph{JHEP} {\bfseries 05}
  (2015) 132} [\href{https://arxiv.org/abs/1412.6087}{{\ttfamily 1412.6087}}].

\bibitem{Benjamin:2020mfz}
N.~Benjamin, S.~Collier and A.~Maloney, \emph{{Pure Gravity and Conical
  Defects}}, \href{https://doi.org/10.1007/JHEP09(2020)034}{\emph{JHEP}
  {\bfseries 09} (2020) 034}
  [\href{https://arxiv.org/abs/2004.14428}{{\ttfamily 2004.14428}}].

\bibitem{McGough:2013gka}
L.~McGough and H.~Verlinde, \emph{{Bekenstein-Hawking Entropy as Topological
  Entanglement Entropy}},
  \href{https://doi.org/10.1007/JHEP11(2013)208}{\emph{JHEP} {\bfseries 11}
  (2013) 208} [\href{https://arxiv.org/abs/1308.2342}{{\ttfamily 1308.2342}}].

\bibitem{Maldacena:2017axo}
J.~Maldacena, D.~Stanford and Z.~Yang, \emph{{Diving into traversable
  wormholes}}, \href{https://doi.org/10.1002/prop.201700034}{\emph{Fortsch.
  Phys.} {\bfseries 65} (2017) 1700034}
  [\href{https://arxiv.org/abs/1704.05333}{{\ttfamily 1704.05333}}].

\bibitem{Kyono:2017jtc}
H.~Kyono, S.~Okumura and K.~Yoshida, \emph{{Deformations of the
  Almheiri-Polchinski model}},
  \href{https://doi.org/10.1007/JHEP03(2017)173}{\emph{JHEP} {\bfseries 03}
  (2017) 173} [\href{https://arxiv.org/abs/1701.06340}{{\ttfamily
  1701.06340}}].

\bibitem{VanRaamsdonk:2010pw}
M.~Van~Raamsdonk, \emph{{Building up spacetime with quantum entanglement}},
  \href{https://doi.org/10.1142/S0218271810018529}{\emph{Gen. Rel. Grav.}
  {\bfseries 42} (2010) 2323}
  [\href{https://arxiv.org/abs/1005.3035}{{\ttfamily 1005.3035}}].

\bibitem{Harlow:2018tqv}
D.~Harlow and D.~Jafferis, \emph{{The Factorization Problem in
  Jackiw-Teitelboim Gravity}},
  \href{https://doi.org/10.1007/JHEP02(2020)177}{\emph{JHEP} {\bfseries 02}
  (2020) 177} [\href{https://arxiv.org/abs/1804.01081}{{\ttfamily
  1804.01081}}].

\bibitem{Almheiri:2019hni}
A.~Almheiri, R.~Mahajan, J.~Maldacena and Y.~Zhao, \emph{{The Page curve of
  Hawking radiation from semiclassical geometry}},
  \href{https://doi.org/10.1007/JHEP03(2020)149}{\emph{JHEP} {\bfseries 03}
  (2020) 149} [\href{https://arxiv.org/abs/1908.10996}{{\ttfamily
  1908.10996}}].

\bibitem{Penington:2019kki}
G.~Penington, S.~H. Shenker, D.~Stanford and Z.~Yang, \emph{{Replica wormholes
  and the black hole interior}},
  \href{https://arxiv.org/abs/1911.11977}{{\ttfamily 1911.11977}}.

\bibitem{Almheiri:2019qdq}
A.~Almheiri, T.~Hartman, J.~Maldacena, E.~Shaghoulian and A.~Tajdini,
  \emph{{Replica Wormholes and the Entropy of Hawking Radiation}},
  \href{https://doi.org/10.1007/JHEP05(2020)013}{\emph{JHEP} {\bfseries 05}
  (2020) 013} [\href{https://arxiv.org/abs/1911.12333}{{\ttfamily
  1911.12333}}].

\bibitem{Susskind:1995da}
L.~Susskind, \emph{{Trouble for remnants}},
  \href{https://arxiv.org/abs/hep-th/9501106}{{\ttfamily hep-th/9501106}}.

\bibitem{Akers:2020pmf}
C.~Akers and G.~Penington, \emph{{Leading order corrections to the quantum
  extremal surface prescription}},
  \href{https://arxiv.org/abs/2008.03319}{{\ttfamily 2008.03319}}.

\bibitem{Streicher:2019wek}
A.~Streicher, \emph{{SYK Correlators for All Energies}},
  \href{https://doi.org/10.1007/JHEP02(2020)048}{\emph{JHEP} {\bfseries 02}
  (2020) 048} [\href{https://arxiv.org/abs/1911.10171}{{\ttfamily
  1911.10171}}].

\bibitem{Choi:2019bmd}
C.~Choi, M.~Mezei and G.~S\'arosi, \emph{{Exact four point function for large
  $q$ SYK from Regge theory}},
  \href{https://arxiv.org/abs/1912.00004}{{\ttfamily 1912.00004}}.

\bibitem{Hayden:2011ag}
P.~Hayden, M.~Headrick and A.~Maloney, \emph{{Holographic Mutual Information is
  Monogamous}}, \href{https://doi.org/10.1103/PhysRevD.87.046003}{\emph{Phys.
  Rev. D} {\bfseries 87} (2013) 046003}
  [\href{https://arxiv.org/abs/1107.2940}{{\ttfamily 1107.2940}}].

\bibitem{Hartman:2013qma}
T.~Hartman and J.~Maldacena, \emph{{Time Evolution of Entanglement Entropy from
  Black Hole Interiors}},
  \href{https://doi.org/10.1007/JHEP05(2013)014}{\emph{JHEP} {\bfseries 05}
  (2013) 014} [\href{https://arxiv.org/abs/1303.1080}{{\ttfamily 1303.1080}}].

\bibitem{Bao:2015bfa}
N.~Bao, S.~Nezami, H.~Ooguri, B.~Stoica, J.~Sully and M.~Walter, \emph{{The
  Holographic Entropy Cone}},
  \href{https://doi.org/10.1007/JHEP09(2015)130}{\emph{JHEP} {\bfseries 09}
  (2015) 130} [\href{https://arxiv.org/abs/1505.07839}{{\ttfamily
  1505.07839}}].

\bibitem{Shankar}
R.~Shankar, \emph{{Principles of quantum mechanics}}. Plenum, New York, NY,
  1980.

\bibitem{Sommerfeld}
A.~Sommerfeld, \emph{{Uber verzweigte Potentiale im Raum}}, {\emph{Proc. London
  Math. Soc.} {\bfseries 28} (1897) 417}.

\bibitem{Dowker:1977zj}
J.~Dowker, \emph{{Quantum Field Theory on a Cone}},
  \href{https://doi.org/10.1088/0305-4470/10/1/023}{\emph{J. Phys. A}
  {\bfseries 10} (1977) 115}.

\bibitem{De_Micheli_2006}
E.~De~Micheli, I.~Scorza and G.~A. Viano, \emph{Hyperbolic geometrical optics:
  Hyperbolic glass}, \href{https://doi.org/10.1063/1.2165796}{\emph{Journal of
  Mathematical Physics} {\bfseries 47} (2006) 023503}.

\end{thebibliography}\endgroup

\addcontentsline{toc}{section}{References}


\end{document}